\documentstyle[12pt,axodraw,epsfig]{article}
\newcommand{\eqnzero}{\setcounter{equation}{0}} 

\newcommand{\bq}{\begin{equation}}
\newcommand{\eq}{\end{equation}}
\newcommand{\ba}{\begin{eqnarray}}
\newcommand{\ea}{\end{eqnarray}}

\newcommand{\baa}[1]{\begin{array}{#1}}
\newcommand{\eaa}{\end{array}}

\newcommand{\nll}{\nonumber\\}

\newcommand{\Rw} {\mbox{$R_{\sss{W}}  $}}

\newcommand{\Rz} {\mbox{$R_{\sss{Z}}  $}}

\def\gz{\Gamma_{\sss{Z}}}

\def\Pgg{\Pi_{\gamma\gamma}}
\def\gfd{\gamma_5}

\def\nl{\nonumber \\}
\def\nn{\nonumber}

\newcommand{\bqa}{\begin{eqnarray}}
\newcommand{\eqa}{\end{eqnarray}}
\newcommand{\ds }{\displaystyle}
\newcommand{\sss}[1]{\scriptscriptstyle{#1}}
\newcommand{\QL}{\sss{QL}}
\newcommand{\LQ}{\sss{LQ}}
\newcommand{\zg}{{\sss{Z}}\ph}
\newcommand{\ph}{\gamma}
\newcommand{\ab}{A}
\newcommand{\zb}{Z}
\newcommand{\wb}{W}
\newcommand{\hb}{H}
\newcommand{\fe}{e}
\newcommand{\fbe}{{\bar{e}}}
\newcommand{\ff}{f}
\newcommand{\ffp}{f'}
\newcommand{\fep}{e^{+}}
\newcommand{\fem}{e^{-}}

\newcommand{\fu}{u}
\newcommand{\fd}{d}

\newcommand{\ft}{t}

\newcommand{\gap}{\gdp}
\newcommand{\gadi}[1]{\gamma_{#1}}
\newcommand{\cff}[6]{C_{0}\big( #1,#2,#3;#4,#5,#6 \big) }    
\newcommand{\mws}{M^2_{\sss{W}}}
\newcommand{\mzs}{M^2_{\sss{Z}}}

\newcommand{\mzf}{M^4_{\sss{Z}}}
\newcommand{\mhs}{M^2_{\sss{H}}}
\newcommand{\mts}{m^2_{t}}

\newcommand{\mf }{m^2_f}

\newcommand{\mtq }{m^4_{t}}
\newcommand{\mhl }{M_{\sss{H}}}

\newcommand{\mwl }{M_{\sss{W}}}
\newcommand{\mzl }{M_{\sss{Z}}}
\newcommand{\mfl }{m_f}
\newcommand{\mtl }{m_t}
\newcommand{\mbl }{m_b}

\newcommand{\mfpl}{m_{f'}}

\newcommand{\mfs }{m^2_f}

\newcommand{\uml}{ m   _{t}}

\newcommand{\wml}{ M  _{\sss{W}}}
\newcommand{\zml}{ M  _{\sss{Z}}}

\newcommand{\asums}[1]{\sum_{#1}}
\newcommand{\cf}{c_f}
\newcommand{\Nf}{N_f}     
\newcommand{\fbf}{{\bar{f}}}

\newcommand{\tpfi}{\lpar 2\pi\rpar^4\ib}

\newcommand{\Vverti}[3]{V_{#1}^{#2}\lpar{#3}\rpar}
\newcommand{\lpar}{\left(}                            
\newcommand{\rpar}{\right)}
\newcommand{\lrbr}{\left[}
\newcommand{\rrbr}{\right]}

\newcommand{\ib  }{i}
\newcommand{\qf  }{Q_f  }
\newcommand{\qfs }{Q^2_f}

\newcommand{\qe  }{Q_e  }

\newcommand{\gbc }{g^3}
\newcommand{\gdp}{\gamma_{+}}
\newcommand{\gdm}{\gamma_{-}}
\newcommand{\gdpm}{\gamma_{\pm}}

\newcommand{\gdmu}{\gamma_{\mu}}
\newcommand{\Gverti}[3]{G_{#1}^{{#2}}\lpar{#3}\rpar}
\newcommand{\Zverti}[3]{Z_{#1}^{{#2}}\lpar{#3}\rpar}
\newcommand{\vvertil}[3]{F^{#1}_{#2}\lpar{#3}\rpar}
\newcommand{\cvetril}[3]{{\cal{F}}^{#1}_{#2}\lpar{#3}\rpar}
\newcommand{\cvertil}[3]{{\cal{F}}^{#1}_{#2}\lpar{#3}\rpar}

\newcommand{\fverti}[2]{F^{#1}_{#2}}

\newcommand{\gadu}[1]{\gamma_{#1}}
\newcommand{\siw }{s_{\sss{W}}}           
\newcommand{\cow }{c_{\sss{W}}}
\newcommand{\siws}{s^2_{\sss{W}}}
\newcommand{\cows}{c^2_{\sss{W}}}

\newcommand{\cowsc}{c^6_{\sss{W}}}
\newcommand{\siwf}{s^4_{\sss{W}}}
\newcommand{\cowf}{c^4_{\sss{W}}}
\newcommand{\vpa}[2]{\sigma_{#1}^{#2}}
\newcommand{\vpae }{\sigma_e}
\newcommand{\vma}[2]{\delta_{#1}^{#2}}

\newcommand{\tcie}{I^{(3)}_e}
\newcommand{\tcif}{I^{(3)}_f}
\newcommand{\tcit}{I^{(3)}_t}
\newcommand{\saff}[1]{A_{#1}}             
                   
\newcommand{\sbff}[1]{B_{#1}}             
\newcommand{\sfbff}[1]{B^{F}_{#1}}
\newcommand{\bff}[4]{B_{#1}\big( #2;#3,#4\big)}             
\newcommand{\fbff}[4]{B^{F}_{#1}\big(#2;#3,#4\big)}        
\newcommand{\scff}[1]{C_{#1}}             
\newcommand{\sdff}[1]{D_{#1}}                 
\newcommand{\dffp}[6]{D_{0} \big( #1,#2,#3,#4,#5,#6;}       
\newcommand{\dffm}[4]{#1,#2,#3,#4 \big) }       
\newcommand{\delrho}[1]{{\Delta \rho}^{#1}}
\newcommand{\bdelrho}[1]{{\Delta\bar \rho}^{#1}}
\newcommand{\fbu}{{\overline{u}}}
\newcommand{\fbd}{{\overline{d}}}
\newcommand{\wbm}{W^{-}}
\newcommand{\wbp}{W^{+}}

\newcommand{\Lnrt}{\Lmmt}
\newcommand{\Lmmt}{L_\mu(\mts)}
\newcommand{\Lmmb}{L_\mu(\mbs)}
\newcommand{\Lmmz}{L_\mu(\mzs)}
\newcommand{\Lmmh}{L_\mu(\mhs)}
\newcommand{\Lmmw}{L_\mu(\mws)}

\newcommand{\rhw}{r_{_{\hb\wb}}}
\newcommand{\rhws}{r^2_{_{\hb\wb}}}
\newcommand{\rhzs}{r^2_{_{\hb\zb}}}

\newcommand{\Dz}[2]{{\cal{D}}_{\sss Z}^{#1}\lpar{#2}\rpar}
\newcommand{\Pzg}{\Pi_{\zg}}

    \newcommand{\tHs}{\mu}
    \newcommand{\tHss}{\mu^2}
    
    \newcommand{\stwl}{s_{\sss{W}}  }
    \newcommand{\ctwl}{c_{\sss{W}}  }
    \newcommand{\stws}{s^2_{\sss{W}}}
    \newcommand{\stwf}{s^4_{\sss{W}}}
    \newcommand{\ctws}{c^2_{\sss{W}}}
    \newcommand{\ctwf}{c^4_{\sss{W}}}
\newcommand{\qd }{ Q_b   }

  \newcommand{\vpau }{\sigma_{t}}
  \newcommand{\vmau }{\delta_{t}}
  \newcommand{\vmae }{\delta_{e}}

\newcommand{\bos}{\rm{bos}}
\newcommand{\fer}{\rm{fer}}
\newcommand{\pole}{\dlt}

\newcommand{\Trqf}{ \sum_f c_f Q^2_f }
\newcommand{\qum}{ \left| Q_{t} \right| }

\newcommand{\vd }{v_b}
\newcommand{\ad }{a_b}
  
\newcommand{\asum}[3]{\sum_{#1=#2}^{#3}}
\newcommand{\minds}[1]{m^2_{#1}}
\newcommand{\Minds}[1]{M^2_{#1}}

\newcommand{\pmom}{p}

\newcommand{\pmoms}{p^2}

\newcommand{\pmomi}[1]{p_{#1}}

\newcommand{\dlt}{\displaystyle{\frac{1}{\epsb}}}
\newcommand{\epsh}{\hat\varepsilon}
\newcommand{\epsb}{\bar\varepsilon}

\newcommand{\eqn}[1]{Eq.~(\ref{#1})}
\newcommand{\eqns}[2]{Eqs.~(\ref{#1})--(\ref{#2})}

\newcommand{\eqnsc}[2]{Eqs.~(\ref{#1}) and (\ref{#2})}

\newcommand{\tbn}[1]{Table~\ref{#1}}

\newcommand{\fig}[1]{Fig.~\ref{#1}}

\newcommand{\sman}{s}
\newcommand{\tman}{t}
\newcommand{\uman}{u}

\newcommand{\bPzga}[2]{{\bar\Pi}^{#1}_{_{\zb\gamma}}\lpar#2\rpar}

\newcommand{\qt}{Q_t}

\newcommand{\qu}{Q_t}

\newcommand{\au}{a_t}
\newcommand{\vu}{v_t}
\newcommand{\ruz}{r_{t{\sss Z}}}
\newcommand{\rtz}{r_{t{\sss Z}}}

\newcommand{\Longab}[3]{L_{ab} \lpar #1,#2,#3\rpar}        
\newcommand{\Longna}[3]{L_{na} \lpar #1,#2,#3\rpar}        
\newcommand{\LongHi}[3]{L_{Hi} \lpar #1,#2,#3\rpar}

\newcommand{\rtw }{r_{t{\sss W}}}

\newcommand{\ruw }{r_{t{\sss W}}}

\newcommand{\ruh }{r_{t{\sss H}}}
\newcommand{\rhz }{r_{\sss HZ}}

\newcommand{\hkg}{\phi}
\newcommand{\hkn}{\phi^{0}}                 
\newcommand{\fbfp}{{\overline{f}}'}
\newcommand{\Rxi}{R_{\gpar}}
\newcommand{\gpar}{\xi}
\newcommand{\hkp}{\phi^{+}}
\newcommand{\fpxm}{X^-}
\newcommand{\fpyZA}{Y_{\ssZ,\gamma}}
\newcommand{\ssZ}{{\sss{Z}}}
\newcommand{\fpxp}{X^+}
\newcommand{\btp}{\beta_t}
\newcommand{\hkm}{\phi^{-}}
\newcommand{\tpl}{t_+}
\newcommand{\tpls}{t^2_+}
\newcommand{\tmi}{t_-}
\newcommand{\tmis}{t^2_-}

\newcommand{\sz}{s_z}
\newcommand{\szs}{s^2_z}
\newcommand{\sdfit} {\Delta_{4r}}
\newcommand{\sdtit} {\Delta_{3r}}

\newcommand{\FQLt} {{\tilde F}_{\sss QL}\lpar s,t,u \rpar}
\newcommand{\FQLtc}{{\tilde F}^*_{\sss QL}\lpar s,t,u \rpar}
\newcommand{\FLQt} {{\tilde F}_{\sss LQ}\lpar s,t,u \rpar}
\newcommand{\FLQtc}{{\tilde F}^*_{\sss LQ}\lpar s,t,u \rpar}
\newcommand{\FQQt} {{\tilde F}_{\sss QQ}\lpar s,t,u \rpar}
\newcommand{\FQQtc}{{\tilde F}^*_{\sss QQ}\lpar s,t,u \rpar}
\newcommand{\FLLt}{{\tilde F}_{\sss LL}\lpar s,t,u \rpar}
\newcommand{\FLLtc}{{\tilde F}^*_{\sss LL}\lpar s,t,u \rpar}
\newcommand{\FLDt}{{\tilde F}_{\sss LD}\lpar s,t,u \rpar}

\newcommand{\FQDt}{{\tilde F}_{\sss QD}\lpar s,t,u \rpar}
\newcommand{\FQDtc}{{\tilde F}^*_{\sss QD}\lpar s,t,u \rpar}

\newcommand{\ip}[1]{u\lpar{#1}\rpar}             
\newcommand{\iap}[1]{{\bar{v}}\lpar{#1}\rpar}    
\newcommand{\op}[1]{{\bar{u}}\lpar{#1}\rpar}     
\newcommand{\oap}[1]{v\lpar{#1}\rpar}            
\newcommand{\qmomi}[1]{q_{#1}}
\newcommand{\rbw}{r_{b{\sss W}}}

\newcommand{\rD}{r^-_{tb}}
\newcommand{\rP}{r^+_{tb}}
\newcommand{\mbs}{m^2_{b}}

\begin{document}
\def\theequation{\arabic{section}.\arabic{equation}}
\newcommand{\chic}{\chi^*}
\def\href#1#2{#2}
\setcounter{page}{0}
\thispagestyle{empty}

\vspace*{-3cm}

\begin{flushright}
{\bf
CERN-TH/2001-308 \\
JINR E2--2000--292\\
{\tt hep-ph/0012080}\\
{revised version}\\
  November 2001 }
\end{flushright}
\vspace*{\fill}
\begin{center}

{\LARGE\bf
An electroweak library for the calculation of} \\[2.5mm] 
$\mbox{\hspace*{-3mm}\LARGE\bf EWRC to $e^+e^-\to f{\bar f}$ within the {\tt CalcPHEP} project}$

\vspace*{1.5cm}
{\bf 
\begin{tabular}[t]{c}
D.~Bardin,
L.~Kalinovskaya, and
G.~Nanava
\end{tabular}
}

\vspace*{3mm}
{\normalsize
{\it Laboratory for Nuclear Problems, JINR,
     ul. Joliot-Curie 6,\\

\vspace*{2mm}
     RU-141980 Dubna, Russia}}
\vspace*{2cm}

\end{center}

\begin{abstract}
\noindent
We present a description of calculations of the electroweak 
amplitude for $e^+e^-\to\ft\bar{t}$ process.
The calculations are done within the OMS (on-mass-shell) renormalization scheme
in two gauges: in $\Rxi$, which allows an explicit control of gauge invariance
by examining cancellation of gauge parameters
and search for gauge-invariant subsets of diagrams,
and in the unitary gauge as a cross-check.
The formulae we derived are realized
in a {\tt FORTRAN} code {\tt eeffLib}, which is being created within the framework
of the project {\tt CalcPHEP}.
We present a comprehensive comparison between {\tt eeffLib} results for
the light top with corresponding results of the well-known program {\tt ZFITTER} 
for the $\fu\bar{u}$ channel, as well as a preliminary comparison with results 
existing in the world literature.
\end{abstract}

\vfill

\begin{flushleft}
{\bf 
CERN-TH/2001--308\\
  November 2001 }
\end{flushleft}

\vspace*{1mm}
\bigskip
\footnoterule
\noindent
{\footnotesize \noindent
Work supported in part  by 
INTAS $N^{o}$ 00-00313.
\\
E-mails: bardin@nusun.jinr.ru, 
kalinov@nusun.jinr.ru, nanava@nusun.jinr.ru }
\clearpage
\tableofcontents
\clearpage
\listoffigures
\listoftables    
\clearpage
	\addcontentsline{toc}{section}{Introduction}
\section*{Introduction\label{eett_introduction}}
\eqnzero

The process $e^+e^-\to\ft\bar{t}$ has already been studied for about
ten years in connection with experiments at future linear colliders
(see, for instance, the review \cite{Beneke:2000hk}).

Actually, it is a six-fermion process (see \cite{Piccinini:2000ib}); 
one of the channels is shown in~\fig{six-fermion}.
\begin{figure}[!h]
\[
\begin{picture}(132,132)(0,0)
\ArrowLine(-40,22)(-2,66)
 \Text(-30,22)[lb]{$e^-$}
\ArrowLine(-2,66)(-40,110)
 \Text(-30,110)[lb]{$e^+$}
 \Photon(-2,66)(44,66){2}{7}
 \Text(12, 70)[lb]{$\gamma, Z$}
\ArrowLine(44,66)(66,88)
 \Text(45, 80)[lb]{$t$}
\ArrowLine(66,44)(44,66)
 \Text(45,47)[lb]{$\bar{t}$}
\Photon(66,44)(88,22){2}{5}
 \Text( 60, 99)[lb]{$W$}
\ArrowLine(88,22)(110,44)
\ArrowLine(110,0)(88,22)
 \Text(114,44)[lb]{$\bar{\nu}$}
 \Text(114,0)[lb]{$l$}
\Photon(66,88)(88,110){2}{5}
 \Text( 60,25)[lb]{$W$}
\ArrowLine(110,88)(88,110)
\ArrowLine(88,110)(110,132)
 \Text(114,132)[lb]{$f_1$}
 \Text(114, 88)[lb]{$\bar{f_2}$}
\ArrowLine(66,88)(90,68)
\ArrowLine(90,64)(66,44)
 \Text(85 , 80)[lb]{$b$}
 \Text(85 ,47)[lb]{$\bar{b}$}
\end{picture}
\]
\caption[The six-fermion $e^+e^-\to\ft\bar{t}$ process.]
{The six-fermion $e^+e^-\to\ft\bar{t}$ process.\label{six-fermion}}
\end{figure}

However, the cross-section of a hard subprocess, $\sigma(\fep\fem\to\ft\bar{t})$,
with tops on the mass shell is an ingredient in various approaches, such as 
DPA~\cite{Denner:2000bj} or the so-called Modified Perturbation Theory (MPT),
see~\cite{Nekrasov:1999cj}.

In this article, we present a brief description of a calculation 
of the electroweak part of the {\em amplitude} of the $e^+e^-\to\ft\bar{t}$ process.
This calculation is a part of the project {\tt CalcPHEP}~\cite{CalcPHEP:2000}
which started in Y2K after completion of the well-known project {\tt ZFITTER} 
\cite{Bardin:1999yd}. 
One of main goals of this paper is to cross-check the {\tt CalcPHEP} results
against the results obtained with the other existing codes. 

As before, we use the OMS renormalization scheme, a complete presentation 
of which was recently made in \cite{Bardin:1999ak}. However, for the first
time we performed calculations in two gauges: $\Rxi$ and the unitary gauge. 

  Note that there was wide experience of calculations in the $R_\xi$ gauge
for processes such as  $H\to\ff\bar{f},WW,ZZ,\gamma Z,\gamma\gamma$, or $e^+e^-\to ZH,WW$.
So, in \cite{Fleischer:1981ub} and \cite{Jegerlehner:1983bf}
a complete set of one-loop counterterms for the SM is given.
 Electromagnetic form factors for arbitrary $\xi$ are discussed in
\cite{Jegerlehner:1985ch} and \cite{Jegerlehner:1986ia}.
 Explicit expressions can be found in the CERN library program EEWW 
\cite{Fleischer:1987xa}.

 However, we are not aware of the existence of calculations in the $\Rxi$ gauge
for the $e^+e^-\to\ft\bar{t}$ process, although there are many studies in 
the $\xi=1$ gauge, see ~\cite{Beenakker:1991ca} -- \cite{Driesen:1996tn}.

\vspace*{2mm}

 Additional purposes of this study are: 

\begin{itemize}

\item[$-$] to explicitly control gauge invariance in $\Rxi$ by examining
cancellation of gauge parameters, and search for gauge-invariant subsets 
of diagrams;

\item[$-$] to offer a possibility to compare the results with those in the 
unitary gauge, as a cross-check;

\item[$-$] to present a self-contained list of results for one-loop amplitude
in terms of Passarino--Veltman functions $\saff{0},\;\sbff{0},\;\scff{0}$ 
and $\sdff{0}$ and their combinations in the spirit of the 
book~\cite{Bardin:1999ak}, where the process $e^+e^-\to\ft\bar{t}$ was
not covered; this article may thus be considered as an Annex to this book;

\item[$-$] to create a {\tt FORTRAN} code for the calculation of the improved
Born approximation (IBA) amplitudes and of the electroweak (EW) part of 
the cross-section of this process for a complementary study within the MPT framework;
 
\end{itemize}

This article consists of five sections.  \\

In Section 1, we present the Born amplitude of the process, basically to introduce our
notation and then define {\em the basis} in which the one-loop amplitude was 
calculated.
We explain {\em the splitting} between QED and EW corrections and between `dressed' 
$\ph$ and $\zb$ exchanges.

\vspace*{2mm}

Section 2 contains explicit expressions for all {\em the building blocks}: 
self-energies, vertices and EW boxes. Note that no diagram was computed by hand.
They are supplied by a new system, {\tt CalcPHEP}, which is being created at the site 
{\tt brg.jinr.ru}.
It roots back to dozens of supporting {\tt form} codes written by authors of ~\cite{Bardin:1999ak} 
while working on it. Later on, the idea came up to collect, 
order, unify and upgrade these codes up to the level of a `computer system'.
Its first phase will be described elsewhere~\cite{CalcPHEP:2000}.

\vspace*{2mm}

In Section 3, we describe the procedure of construction of 
{\em the scalar form factors}
of the one-loop amplitudes out of the building blocks. One of the aims of this section 
is to create a frame for a subsequent realization of this procedure within the 
{\tt CalcPHEP} project.

\vspace*{2mm}

Section 4 contains explicit expressions for the IBA cross-section.

\vspace*{2mm}

Finally, in Section 5 we present results of a comprehensive numerical comparison 
between {\tt eeffLib} and {\tt ZFITTER}.
We also discuss some preliminary results of a comparison between {\tt eeffLib} and 
the other available codes.

\clearpage

\section{Amplitudes\label{amplitudes}}
\subsection{Born amplitudes\label{B-amplitudes}}
We begin with the Born amplitudes for the process 
$\fep(\pmomi{+})\fem(\pmomi{-})\to\ft(\qmomi{-}){\bar t}(\qmomi{+})$, which 
is described by the two Feynman diagrams with $\ph$ and $\zb$ exchange.
The Born amplitudes are:
\bqa
A^{\sss{B}}_{\ph} &=&e\qe\,e\qt
                  \gadu{\mu} \otimes \gadu{\mu} \frac{-\ib}{Q^2}
               = -\ib\,4\pi\alpha(0)\frac{\qe\qt}{Q^2}
                  \gadu{\mu} \otimes \gadu{\mu}\,,
\\ 
\label{amplborn}
A^{\sss{B}}_{\sss{\zb}} &=&
       \frac{e}{2\stwl\ctwl}\,\frac{e}{ 2\stwl\ctwl}  
                         \gadu{\mu}
       \lrbr \tcie \gdp -2\qe \stws \rrbr 
                 \otimes \gadu{\mu}  
       \lrbr \tcit \gdp -2\qt \stws \rrbr 
       \frac{-\ib}{Q^2+\mzs}                                      
\nll &=& -                   
\ib e^2\frac{1}{4\stws\ctws(Q^2 +\mzs)}
\biggl[
  \tcie\tcit \gadu{\mu}\gdp\otimes\gadu{\mu}\gdp 
+\vmae \tcit \gadu{\mu}\otimes\gadu{\mu}\gdp 
\nll &&
+\tcie \vmau \gadu{\mu}\gdp\otimes\gadu{\mu}
+\vmae \vmau \gadu{\mu}\otimes\gadu{\mu} 
\biggr]\,,
\eqa
where $\;\gdpm=1\pm\gfd\;$ and the symbol $\otimes$ is used in the following 
short-hand notation:
\bq
\gadu{\mu}\lpar L_{1}\gdp+Q_{1} \rpar\otimes
\gadu{\nu}\lpar L_{2}\gdp+Q_{2} \rpar
=
\iap{\pmomi{+}}\gadu{\mu}\lpar L_{1}\gdp+Q_{1} \rpar\ip{\pmomi{-}} 
\op{\qmomi{-}}\gadu{\nu}\lpar L_{2}\gdp+Q_{2} \rpar\oap{\qmomi{+}};
\label{dlia_Lidy}
\eq
furthermore
\bqa
\delta_f &=& v_f - a_f = - 2 \qf \stws\,,\qquad f=e,t.
\eqa

Introducing the $LL$, $QL$, $LQ$, and $QQ$ structures, correspondingly
(see last \eqn{amplborn}), we have five structures to which the 
complete Born amplitude may be reduced: one for the $\ph$ exchange amplitude 
and four for the $\zb$ exchange amplitude. 
\subsection{One-loop amplitude for $e^+e^- \to\ft{\bar t}$}
 For the $e^+e^- \to\ft {\bar t}$ process at one loop,
it is possible to consider a gauge-invariant subset
of {\em electromagnetic corrections} separately: QED vertices, $\ph\ph$ and 
$\zb\ph$ boxes. Together with QED bremsstrahlung diagrams, it is free of 
infrared divergences. 
The contribution of QED diagrams is considered elsewhere~\cite{part2:2001}. 
Here we keep in mind only the 
remaining one-loop diagrams forming {\it electroweak corrections}.
The total electroweak amplitude is a sum of `dressed' $\gamma$ and $Z$
exchange amplitudes, plus the contribution from {\it the weak box} diagrams
($WW$ and $ZZ$ boxes).

 Contrary  to the Born amplitude, the one-loop amplitude may be parametrized
by 6 form factors, a number equal to the number of independent helicity 
amplitudes for this process.

 We work in the so-called $LQD$ basis, which naturally arises if the 
final-state fermion masses are not ignored
\footnote{
If the initial-state masses were not ignored too, we would have ten independent helicity
amplitudes, ten structures and ten scalar form factors.
}.
Then the amplitude may be schematically represented as:
\bqa
  \lrbr i \gadu{\mu}\gdp \vvertil{e}{\sss{L}}{\sman} 
       +i \gadu{\mu}                   \vvertil{e}{\sss{Q}}{\sman} \rrbr
  \otimes
  \lrbr 
        i \gadu{\mu}\gdp \vvertil{t}{\sss{L}}{\sman} 
       +i \gadu{\mu}                   \vvertil{t}{\sss{Q}}{\sman} 
        + m_t I D_\mu                  \vvertil{t}{\sss{D}}{\sman} \rrbr,
\eqa
with 
\bq
D_\mu=(\qmomi{+}-\qmomi{-})_{\mu}\,.
\label{difference}
\eq

 Every form factor in the $\Rxi$ gauge could be represented as a sum of two terms:
\bq
\vvertil{\xi}{\sss{L,Q,D}}{\sman}=
\vvertil{(1)}{\sss{L,Q,D}}{\sman} + \vvertil{\rm add}{\sss{L,Q,D}}{\sman}.
\eq
The first term corresponds to the $\xi=1$ gauge and the second contains
all $\xi$ dependences and vanishes for $\xi=1$ by construction. 

 The $LQD$ basis was found to be particularly convenient to 
explicitly demonstrate the cancellation of all $\xi$-dependent terms.
We checked the cancellation of these terms in several groups of diagrams
separately:
the so-called $\ph$, $\zb$, and $\hb$ clusters, defined below; 
the $\wb$ cluster together with the self-energies and the $\wb\wb$ box; 
and the $\zb\zb$ boxes. Therefore, for our process we found seven separately 
gauge-invariant subgroups of diagrams: three in the QED sector, and four in the EW 
sector.

 The `dressed' $\ph$ exchange amplitude is
\bq
A^{\sss{\rm{IBA}}}_{\ph} = \ib\frac{4\pi\qe\qf}{\sman}
\alpha(\sman) \gadu{\mu} \otimes \gadu{\mu}\,,
\label{Born_modulo-old}
\eq
which is identical to the Born amplitude of \eqn{amplborn} 
modulo the replacement of $\alpha(0)$ by the
running electromagnetic coupling $\alpha(\sman)$:
\bq
\alpha(\sman)=\frac{\alpha}
{\ds{1-\frac{\alpha}{4\pi}\Bigl[\Pgg^{\fer}(\sman)-\Pgg^{\fer}(0)\Bigr]}}\,.
\label{alpha_fer-old}
\eq

 In the $LQD$ basis the $\zb$ exchange amplitude has the following Born-like 
structure in terms of six ($LL$, $QL$, $LQ$, $QQ$, $LD$ and $QD$) form factors:
\bqa
{\cal A}^{\sss{\rm{IBA}}}_{\sss{\zb}}&=&
\ib\,e^2
\frac{\chi_{\sss{Z}}(\sman)}{\sman}
   \Biggl\{\gadu{\mu} {\gdp } \otimes
       \gadu{\mu} {\gdp } \tilde\vvertil{}{\sss{LL}}{\sman,\tman}     
+\gadu{\mu}     
      \otimes \gadu{\mu} {\gdp} \tilde\vvertil{}{\sss{QL}}{\sman,\tman} 
\nll &&
+\gadu{\mu}{\gdp}\otimes\gadu{\mu} \tilde\vvertil{}{\sss{LQ}}{\sman,\tman}
+\gadu{\mu}\otimes\gadu{\mu}
 \tilde\vvertil{}{\sss{QQ}}{\sman,\tman}
\nll &&
+\gadu{\mu}{\gdp}\otimes\lpar 
     -i m_t D_{\mu} \rpar  \tilde\vvertil{}{\sss{LD}}{\sman,\tman}
+\gadu{\mu} \otimes \lpar  
     -i m_t D_{\mu} \rpar  \tilde\vvertil{}{\sss{QD}}{\sman,\tman}
\Biggr\},\qquad
\label{structures-old}
\eqa
where we introduce the notation for $\tilde\vvertil{}{ij}{\sman,\tman}$ :
\bqa
\tilde\vvertil{}{\sss{LL}}{\sman,\tman} &=& \tcie\tcit\vvertil{}{\sss{LL}}{\sman,\tman},
\nll     
\tilde\vvertil{}{\sss{QL}}{\sman,\tman} &=& \vmae\tcit\vvertil{}{\sss{QL}}{\sman,\tman},
\nll
\tilde\vvertil{}{\sss{LQ}}{\sman,\tman} &=& \tcie\vmau\vvertil{}{\sss{LQ}}{\sman,\tman},
\nll
\tilde\vvertil{}{\sss{QQ}}{\sman,\tman} &=& \vmae\vmau\vvertil{}{\sss{QQ}}{\sman,\tman},
\nll
\tilde\vvertil{}{\sss{LD}}{\sman,\tman} &=& \tcie\tcit\vvertil{}{\sss{LD}}{\sman,\tman},
\nll
\tilde\vvertil{}{\sss{QD}}{\sman,\tman} &=& \vmae\tcit\vvertil{}{\sss{QD}}{\sman,\tman}.
\eqa
Note that {\it tilded} form factors absorb couplings, which leads to a compactification 
of formulae for the amplitude and IBA cross-section, while explicit expressions will be 
given for {\it untilded} quantities.
The representation of \eqn{structures-old} is very convenient for the subsequent 
discussion of one-loop amplitudes.
 
Furthermore, in \eqn{structures-old} we 
use the $\zb/\ph$ propagator ratio with an $\sman$-dependent (or constant) $\zb$ width:   
\bqa
\chi_{\sss{Z}}(\sman)&=&\frac{1}{4\siws\cows}\frac{\sman}
{\ds{\sman - \mzs + \ib\frac{\gz}{\mzl}\sman}}\,.
\label{propagators}
\eqa

\clearpage

\section{ Building Blocks in the OMS Approach\label{building-blocks}}
\eqnzero
We start our discussion by presenting various {\it building blocks}, used 
to construct the one-loop form factors of the processes $\fep\fem\to\ff\fbf$ 
in terms of the $\saff{0},\;\sbff{0},\;\scff{0}$ and $\sdff{0}$ functions. 
They are shown in order of increasing complexity: self-energies, 
vertices, and boxes.
 
\subsection{Bosonic self-energies \label{bosonicse}}
\subsubsection{$\zb,\ph$ bosonic self-energies and $\zb$--$\ph$ transition
\label{bsetran}}
\begin{figure}[h]
\vspace*{-10mm}
\[
\baa{ccccccc}
\begin{picture}(75,20)(0,8)
  \Photon(0,10)(25,10){3}{5}
    \GCirc(37.5,10){12.5}{0.5}
  \Photon(50,10)(75,10){3}{5}
\Text(0,18)[bl]{$\zb,\gamma$}
\Text(12.5, 2)[tc]{$\mu$}
\Text(75,18)[br]{$\zb,\gamma$}
\Text(62.5, 2)[tc]{$\nu$}
\end{picture}
&=&
\begin{picture}(75,20)(0,8)
  \Photon(0,10)(25,10){3}{5}
  \ArrowArcn(37.5,10)(12.5,0,180)
  \ArrowArcn(37.5,10)(12.5,180,0)
    \Vertex(25,10){2.5}
    \Vertex(50,10){2.5}
  \Photon(50,10)(75,10){3}{5}
\Text(37.5,-8)[tc]{$\fbu,\fbd$}
\Text(37.5,30)[bc]{$\fu,\fd$}
\Text(0,-8)[lt]{$(1)$}
\end{picture}
&+&
\begin{picture}(75,20)(0,8)
  \Photon(0,10)(25,10){3}{5}
  \PhotonArc(37.5,10)(12.5,0,180){3}{7}
  \PhotonArc(37.5,10)(12.5,180,0){3}{7}
    \Vertex(25,10){2.5}
    \Vertex(50,10){2.5}
  \Photon(50,10)(75,10){3}{5}
    \ArrowLine(37.6,22.2)(38.3,24.2)
    \ArrowLine(37.5,-2.2)(36.7,-4)
\Text(37.5,-8)[tc]{$\wbm$}
\Text(37.5,30)[bc]{$\wbp$}
\Text(0,-8)[lt]{$(2)$}
\end{picture}
&+&
\begin{picture}(75,20)(0,8)
  \Photon(0,10)(25,10){3}{5}
  \PhotonArc(37.5,10)(12.5,0,180){3}{7}
  \DashCArc(37.5,10)(12.5,180,0){3.}
    \Vertex(25,10){2.5}
    \Vertex(50,10){2.5}
  \Photon(50,10)(75,10){3}{5}
\Text(37.5,-8)[tc]{$\hb$}
\Text(37.5,30)[bc]{$\zb$}
\Text(0,-8)[lt]{$(3)$}
\end{picture}
\nl \nl \nl \nl \nl
&+&
\begin{picture}(75,20)(0,8)
  \Photon(0,10)(25,10){3}{5}
  \PhotonArc(37.5,10)(12.5,0,180){3}{7}
  \DashArrowArcn(37.5,10)(12.5,0,180){3.}
    \Vertex(25,10){2.5}
    \Vertex(50,10){2.5}
  \Photon(50,10)(75,10){3}{5}
    \ArrowLine(37.6,22.2)(38.3,24.2)
\Text(37.5,-8)[tc]{$\hkm$}
\Text(37.5,30)[bc]{$\wbp$}
\Text(0,-8)[lt]{$(4)$}
\end{picture}
&+&
\begin{picture}(75,20)(0,8)
  \Photon(0,10)(25,10){3}{5}
  \DashArrowArcn(37.5,10)(12.5,180,0){3.}
  \PhotonArc(37.5,10)(12.5,180,0){3}{7}
    \Vertex(25,10){2.5}
    \Vertex(50,10){2.5}
  \Photon(50,10)(75,10){3}{5}
    \ArrowLine(37.5,-2.2)(36.7,-4)
\Text(37.5,-8)[tc]{$\wbm$}
\Text(37.5,30)[bc]{$\hkp$}
\Text(0,-8)[lt]{$(5)$}
\end{picture}
&&
\nl \nl \nl \nl \nl
&+&
\begin{picture}(75,20)(0,8)
  \Photon(0,10)(25,10){3}{5}
  \DashCArc(37.5,10)(12.5,0,180){3.}
  \DashCArc(37.5,10)(12.5,180,0){3.}
    \Vertex(25,10){2.5}
    \Vertex(50,10){2.5}
  \Photon(50,10)(75,10){3}{5}
\Text(37.5,-8)[tc]{$\hb$}
\Text(37.5,30)[bc]{$\hkn$}
\Text(0,-8)[lt]{$(6)$}
\end{picture}
&+&
\begin{picture}(75,20)(0,8)
  \Photon(0,10)(25,10){3}{5}
  \DashArrowArcn(37.5,10)(12.5,180,0){3.}
  \DashArrowArcn(37.5,10)(12.5,0,180){3.}
    \Vertex(25,10){2.5}
    \Vertex(50,10){2.5}
  \Photon(50,10)(75,10){3}{5}
\Text(37.5,-8)[tc]{$\hkm$}
\Text(37.5,30)[bc]{$\hkp$}
\Text(0,-8)[lt]{$(7)$}
\end{picture}
& &
\nl \nl \nl \nl \nl
&+&
\begin{picture}(75,20)(0,8)
  \Photon(0,10)(25,10){3}{5}
\SetWidth{1.}
  \DashArrowArcn(37.5,10)(12.5,0,180){1.5}
  \DashArrowArcn(37.5,10)(12.5,180,0){1.5}
\SetWidth{.5}
    \Vertex(25,10){2.5}
    \Vertex(50,10){2.5}
  \Photon(50,10)(75,10){3}{5}
\Text(37.5,-8)[tc]{$\fpxm$}
\Text(37.5,30)[bc]{$\fpxm$}
\Text(0,-8)[lt]{$(8)$}
\end{picture}
&+&
\begin{picture}(75,20)(0,8)
  \Photon(0,10)(25,10){3}{5}
\SetWidth{1.}
  \DashArrowArcn(37.5,10)(12.5,0,180){1.5}
  \DashArrowArcn(37.5,10)(12.5,180,0){1.5}
\SetWidth{.5}
    \Vertex(25,10){2.5}
    \Vertex(50,10){2.5}
  \Photon(50,10)(75,10){3}{5}
\Text(37.5,-8)[tc]{$\fpxp$}
\Text(37.5,30)[bc]{$\fpxp$}
\Text(0,-8)[lt]{$(9)$}
\end{picture}
& &
\nl \nl \nl \nl \nl \nl \nl
&+&
\begin{picture}(75,20)(0,8)
  \Photon(0,10)(75,10){3}{15}
  \PhotonArc(37,28)(12.5,17,197){3}{8}
  \PhotonArc(37,28)(12.5,197,17){3}{8}
     \Vertex(37,13.){2.5}
\Text(36.26,50)[bc]{$\wb$}
\Text(0,48)[lb]{$(10)$}
\end{picture}
&+&
\begin{picture}(75,20)(0,8)
  \Photon(0,10)(75,10){3}{15}
  \DashCArc(37,26)(12.5,0,180){3}
  \DashCArc(37,26)(12.5,180,0){3}
    \Vertex(37,13){2.5}
\Text(36.26,50)[bc]{$\hb$}
\Text(0,48)[lb]{$(11)$}
\end{picture}
&&
\nl \nl \nl \nl \nl 
&+&
\begin{picture}(75,20)(0,8)
  \Photon(0,10)(75,10){3}{15}
  \DashArrowArcn(37,26)(12.5,0,180){3}
  \DashArrowArcn(37,26)(12.5,180,0){3}
    \Vertex(37,13.){2.5}
\Text(36.26,45)[bc]{$\hkp$}
\Text(0,45)[lb]{$(12)$}
\end{picture}
&+&
\begin{picture}(75,20)(0,8)
  \Photon(0,10)(75,10){3}{15}
  \DashCArc(37,26)(12.5,0,180){3}
  \DashCArc(37,26)(12.5,180,0){3}
    \Vertex(37,13.){2.5}
\Text(36.26,45)[bc]{$\hkn$}
\Text(0,45)[lb]{$(13)$}
\end{picture}
\nl \nl \nl 
&+&
\begin{picture}(75,20)(0,8)
  \Photon(0,10)(75,10){3}{15}
    \Vertex(37.5,10.){2.5}
\Text(37.5,20)[bc]{$\btp$}
\Text(0,22)[lb]{$(14)$}
\Text(53,20)[lb]{$(\zb)$}
\end{picture}
& &
\eaa
\]
\vspace*{5mm}
\caption{$(\zb,\gamma)$-boson self-energy;~~$\zb$--$\gamma$~~transition.
\label{se_zgamma}}
\vspace*{-20mm}
\end{figure}
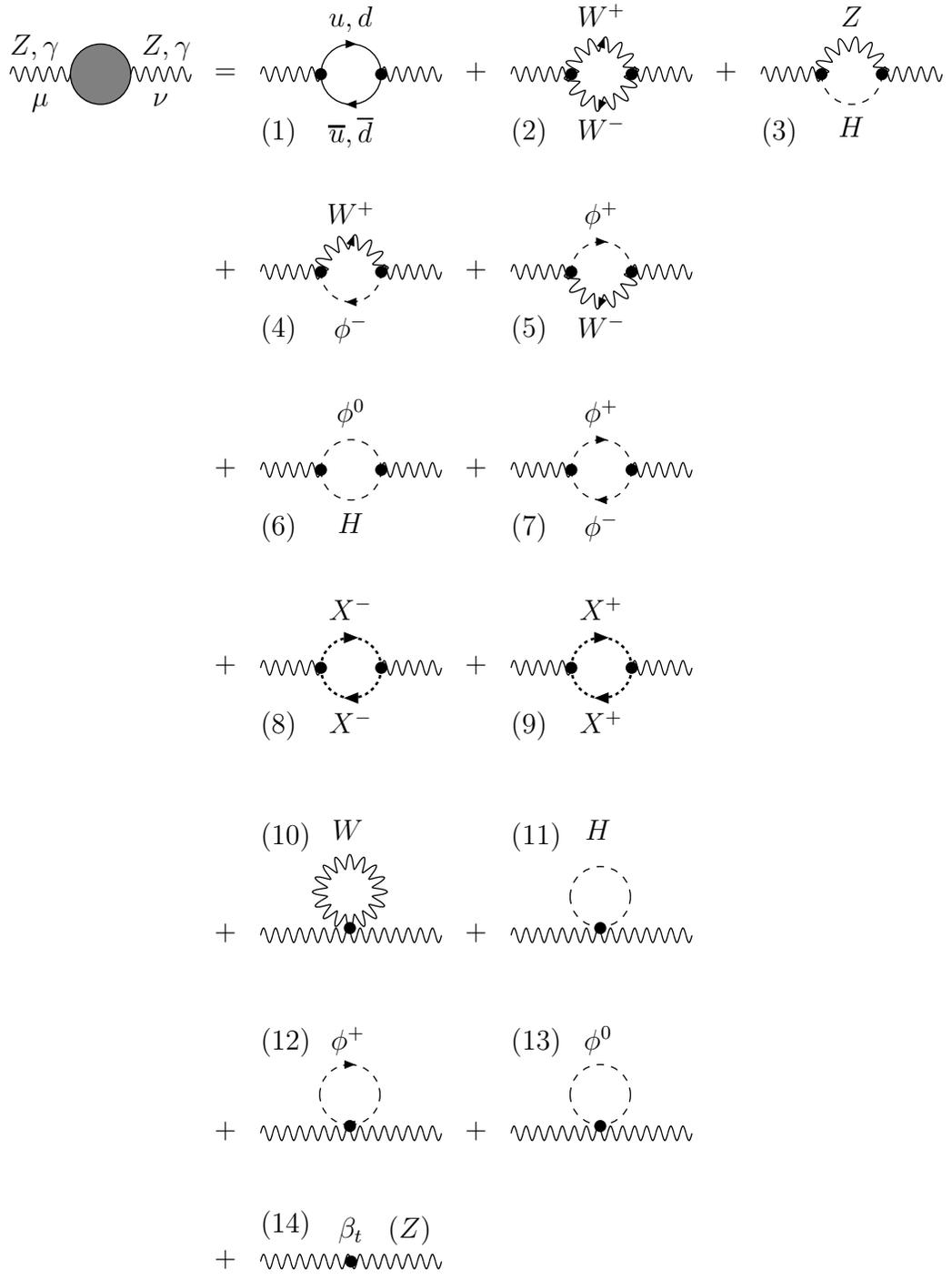

In the $\Rxi$ gauge there are 14 diagrams that contribute to the
{\em total} $\zb$ and $\ph$ bosonic self-energies and to the $\zb$--$\ph$ 
transition. They are shown in \fig{se_zgamma}.

With $S_{\sss{\zb\zb}}$, $S_{\zg}$ and $S_{\ph\ph}$ standing for the sum of all
diagrams, depicted by a grey circle in \fig{se_zgamma}, we define the three 
corresponding self-energy functions $\Sigma_{\sss{AB}}$:
\ba
S_{\sss{\zb\zb}}&=&(2\pi)^4\ib\frac{g^2}{16\pi^2\cows}\Sigma_{\sss{\zb\zb}}\,,
\\
S_{\zg}   &=&(2\pi)^4\ib\frac{g^2\siw}{16\pi^2\cow}\Sigma_{\zg}\,,
\\
S_{\ph\ph}   &=&(2\pi)^4\ib\frac{g^2\siws}{16\pi^2}\Sigma_{\ph\ph}\,.
\label{sefunct}
\ea
All {\bf bosonic} self-energies and transitions may be naturally split into
{\em bosonic} and {\em fermionic} components.
\begin{itemize}
\item {Bosonic components of $\zb,\ph$ self-energies and
        $\zb$--$\ph$ transitions} (see diagrams \fig{se_zgamma})
\end{itemize}
\bqa
\Sigma^{\bos}_{\sss{\zb\zb}}(\sman)  &=& 
 \mzs \Biggl\{
 \frac{ 1}{3}\frac{1}{\Rz}  \lpar  \frac{1}{2} - \ctws - 9 \ctwf \rpar
\\ && 
   - \frac{3}{2} \lrbr \lpar 1 + 2 \ctwf \rpar\frac{1}{\rhz}
   - \frac{1}{2} - \ctws + \frac{8}{3}\ctwf + \frac{1}{2} \rhz \rrbr 
      \Biggr\}
\pole + \Sigma^{{\bos},F}_{\sss{\zb\zb}}(\sman),
\nonumber
\eqa
\bqa
\label{self_zz_finite}
\Sigma^{{\bos},F}_{\sss{\zb\zb}}(\sman)  &=& 
\frac{\mzs}{12}
\Biggl\{
 \lrbr 4 \ctws \lpar 5 - 8 \ctws - 12\ctwf \rpar
              +\lpar 1 - 4 \ctws - 36\ctwf \rpar \frac{1}{\Rz} \rrbr
 \fbff{0}{-\sman}{\mwl}{\mwl}
\nll &&
 +\lrbr \frac{1}{\Rz} + 10 - 2 \rhz + \lpar \rhz-1 \rpar^2 \Rz \rrbr                  
 \fbff{0}{-\sman}{\mhl}{\mzl}
\nll &&
  +  \lrbr \frac{18}{\rhz} + 1 +  \lpar 1-\rhz \rpar \Rz \rrbr   \Lmmz
  +  \rhz \Big[ 7 - \lpar 1 - \rhz \rpar \Rz \Big] \Lmmh
\nll &&
  +  2 \ctws \lpar \frac{18}{\rhw} + 1 +  8 \ctws -24 \ctwf \rpar \Lmmw
\\ &&
  + \frac{4}{3}
 \lpar 1 - 2\ctws \rpar \frac{1}{\Rz} 
        - 6 \lpar 1+2\ctwf \rpar \frac{1}{\rhz}
        - 3 (1+2 \ctws) - 9\rhz - \lpar 1-\rhz \rpar^2 \Rz 
  \Biggr\}.
\nonumber
\eqa
\clearpage

\clearpage

 Here $L_\mu(M^2)$ denotes the log containing the 't Hooft scale $\mu$:
\bqa
L_\mu(M^2)=\ln\frac{M^2}{\mu^2}\;,
\eqa
and it should be understood that, contrary to the one used in 
\cite{Bardin:1999ak}, we define here
\bqa
 \bff{0}{-\sman}{M_1}{M_2}=\pole+\fbff{0}{-\sman}{M_1}{M_2},
\label{b0_finite}
\eqa
meaning that $B_0^F$ also depends on the scale $\mu$.
We will not explicitly maintain $\mu$ in the arguments list 
of $L_\mu$ and $B_0^F$. Leaving $\tHs$ unfixed, we retain an opportunity 
to control $\tHs$ ~independence (and therefore UV finiteness) in numerical
realization of one-loop form factors, providing thereby an additional 
cross-check.

Next, it is convenient to introduce the dimensionless quantities
$\Pzg^{\bos}(\sman)$ and $\Pgg^{\bos}(\sman)$ (vacuum polarizations):
\bqa
\Sigma^{\bos}_{{\sss \zb}\ph}(\sman) &=& - \sman \Pzg^{\bos}(\sman),
\label{sig_Zg} 
\\[2mm]
\Sigma^{\bos}_{\ph\ph}(\sman) &=& - \sman \Pgg^{\bos}(\sman).
\label{sig_gg} 
\eqa

In \eqnsc{self_zz_finite}{b0_finite}
and below, the following abbreviations are used:
\bqa
\cows=\frac{\mws}{\mzs}\,,
\qquad
r_{ij}=\frac{m^2_i}{m^2_j}\,,
\qquad
\Rw=\frac{\mws}{\sman}\,,
\qquad
\Rz=\frac{\mzs}{\sman}\,.
\qquad
\label{abbrev-old}
\eqa
 Since only finite parts will contribute to resulting expressions for
physical amplitudes, which should be free from ultraviolet poles, 
it is convenient to split every divergent function into singular and 
finite parts:
\bqa
\Pgg^{\bos}(\sman) & = &  3 \pole + \Pgg^{{\bos},F}(\sman),
\label{pigg_pf}
\\[2mm]   
\label{pigg_pf1}
\Pgg^{{\bos},F}(\sman)&=& 
 \lpar 3 + 4 \Rw \rpar \fbff{0}{-s}{\mwl}{\mwl} + 4 \Rw \Lmmw,
\eqa
and
\bqa
\Pzg^{\bos}(\sman) & = &  \lpar \frac{1}{6} + 3 \ctws + 2\Rw\rpar \pole
                           + \Pzg^{{\bos},F}(\sman),
\label{pizg_pf}
\\   
\Pzg^{{\bos},F}(\sman)  &=&
   \lrbr
  \frac{1}{6} + 3 \cows 
 + 4 \lpar \frac{1}{3}+\cows \rpar \Rw
\rrbr 
\fbff{0}{-s}{\mwl}{\mwl}
\nll &&
 +\frac{1}{9}
-\lpar \frac{2}{3} - 4 \ctws \rpar \Rw \Lmmw.
\label{pizg_f}
\eqa

With the $\zb$ boson self-energy, $\Sigma_{\sss{\zb\zb}}$, we construct
a useful ratio:
\bqa
\Dz{}{\sman}&=&\frac{1}{\cows}
            \frac{\Sigma^{}_{\sss{\zb\zb}}\lpar s\rpar
                 -\Sigma^{}_{\sss{\zb\zb}}\lpar \mzs\rpar}{\mzs-\sman}\,,
\label{usefulratio}
\eqa
which also has bosonic and fermionic parts.
The bosonic component is:
\bqa
\Dz{\bos}{\sman}&=&
 \frac{1}{\ctws} \lpar - \frac{1}{6} + \frac{1}{3}\ctws + 3\ctwf \rpar 
                  \pole
+\Dz{{\bos},F}{\sman},
\\
\Dz{{\bos},F}{\sman}&=&
 \frac{1}{\cows}
\Biggl\{
     \lpar \frac{1}{12} + \frac{4}{3} \cows -
                        \frac{17}{3} \cowf - 4 \cowsc \rpar 
\\ &&
\times \frac{\mzs}{\mzs-s}
\lrbr 
\fbff{0}{-s}{\mwl}{\mwl} - \fbff{0}{-\mzs}{\mwl}{\mwl} \rrbr
\nll &&
  -\lpar \frac{1}{12} - \frac{1}{3}\cows - 3 \cowf \rpar 
      \fbff{0}{-s}{\mwl}{\mwl}
\nll &&
  + \lpar  1 - \frac{1}{3} \rhz + \frac{1}{12} \rhzs \rpar  
    \frac{\mzs}{\mzs-s} 
    \lrbr \fbff{0}{-s}{\mhl}{\mzl} - \fbff{0}{-\mzs}{\mhl}{\mzl} \rrbr 
\nll &&
  -   \frac{1}{12}
  \lrbr 1 - \lpar 1 - \rhz \rpar^2 \Rz  \rrbr 
\fbff{0}{-s}{\mhl}{\mzl}
\nll &&
 -\frac{1}{12} \Rz
   \lpar  1 - \rhz \rpar 
   \lrbr  \rhz \lpar \Lmmh -1 \rpar - \Lmmz + 1 \rrbr 
  - \frac{1}{9}\lpar 1 - 2 \cows  \rpar 
\Biggr\}.
\nonumber
\label{Dz_bF}
\eqa
\begin{itemize}
 \item{Fermionic components of the $\zb$ and $\ph$ bosonic self-energies and
of the $\zb$--$\ph$ transition}
\end{itemize}
These are represented as sums over all fermions of
the theory, $\sum_f$. They, of course, depend on vector and axial couplings
of fermions to the $\zb$ boson, $v_f$ and $a_f$, and to the photon, 
electric charge $e Q_f$,
as well as on the colour factor $c_f$ and fermion mass $\mfl$.
The couplings are defined as usual:
\ba
v_f=I^{(3)}_f-2Q_f\siws\,,
\qquad
a_f=I^{(3)}_f,
\ea
with weak isospin $I^{(3)}_f$, and
\ba
Q_f&=&-1\quad\mbox{for leptons},\quad+\frac{2}{3}\quad\mbox{for up quarks},
\quad-\frac{1}{3}\quad\mbox{for down quarks},
\\
c_f&=&1\quad\mbox{for leptons},\quad 3\quad\mbox{for quarks}.
\ea

The three main self-energy functions are: 
\bqa
\Sigma^{\fer}_{\sss{\zb\zb}}(\sman) &=& 
\sum_f c_f \Biggl[
-\lpar v^2_f+a^2_f \rpar\sman  
\bff{f}{-\sman}{\mfl}{\mfl}
-2a^2_f\mf\bff{0}{-\sman}{\mfl}{\mfl}
           \Biggr],
\label{Sigma_zz_fer}
\\
\Sigma^{\fer}_{\ph\ph}(\sman)   &=& -\sman \Pgg^{\fer}(\sman),
\label{Sgg_Pgg}
\\[2mm]
\Sigma^{\fer}_{\zg}(\sman)&=& -\sman \Pzg^{\fer}\lpar\sman\rpar. 
\label{Szg_Pzg}
\eqa

The quantities $\Pgg^{\fer}$ and $\Pzg^{\fer}$ are different according to
different couplings, but proportional to one function $\sbff{f}$
(see Eq. (5.252) of \cite{Bardin:1999ak} for its definition):
\bqa
\Pgg^{\fer}(\sman) &=& 4 \sum_f c_f\,Q^2_f\,   \bff{f}{-\sman}{\mfl}{\mfl}\,,
\label{Pi_gg_fer}
\\
\Pzg^{\fer}(\sman) &=& 2 \sum_f c_f\,Q_f\,v_f\,\bff{f}{-\sman}{\mfl}{\mfl}\,.
\label{Pi_zg_fer}
\eqa
As usual, we subdivided them into singular and finite parts:
\bqa
\label{fermionic-split}
\Pzg^{\fer}(\sman) &=&
  -  \frac{1}{3} 
     \Biggl(\frac{1}{2}\Nf-4\siws\asums{\ff}\cf\qfs\Biggr)
     \pole
  +  \Pzg^{{\fer},F}(\sman),
\\
\Sigma^{\fer}_{\sss{\zb\zb}}(\sman)&=&
\Biggl\{-\frac{1}{2}\asums{\ff}\cf\mfs + \frac{\sman}{3}
\Biggl[\lpar\frac{1}{2}-\siws\rpar\Nf+4\siwf\asums{\ff}\cf\qfs\Biggr]
\Biggr\} \pole
 + \Sigma^{{\fer},F}_{\sss{\zb\zb}}(\sman).
\nonumber
\eqa
In \eqn{fermionic-split}, $\Nf=24$ is the total number of fermions in the SM.
We do not show explicit expressions for finite parts, marked with superscript 
$F$, because these might be trivially derived from 
\eqn{Sigma_zz_fer} and 
Eqs.~(\ref{Pi_gg_fer}),~(\ref{Pi_zg_fer}) by replacing complete 
expressions for $\sbff{f}$ and $\sbff{0}$ with their finite parts
$\sfbff{f}$ and $\sfbff{0}$, correspondingly.

\subsubsection{$\wb$ boson self-energy}

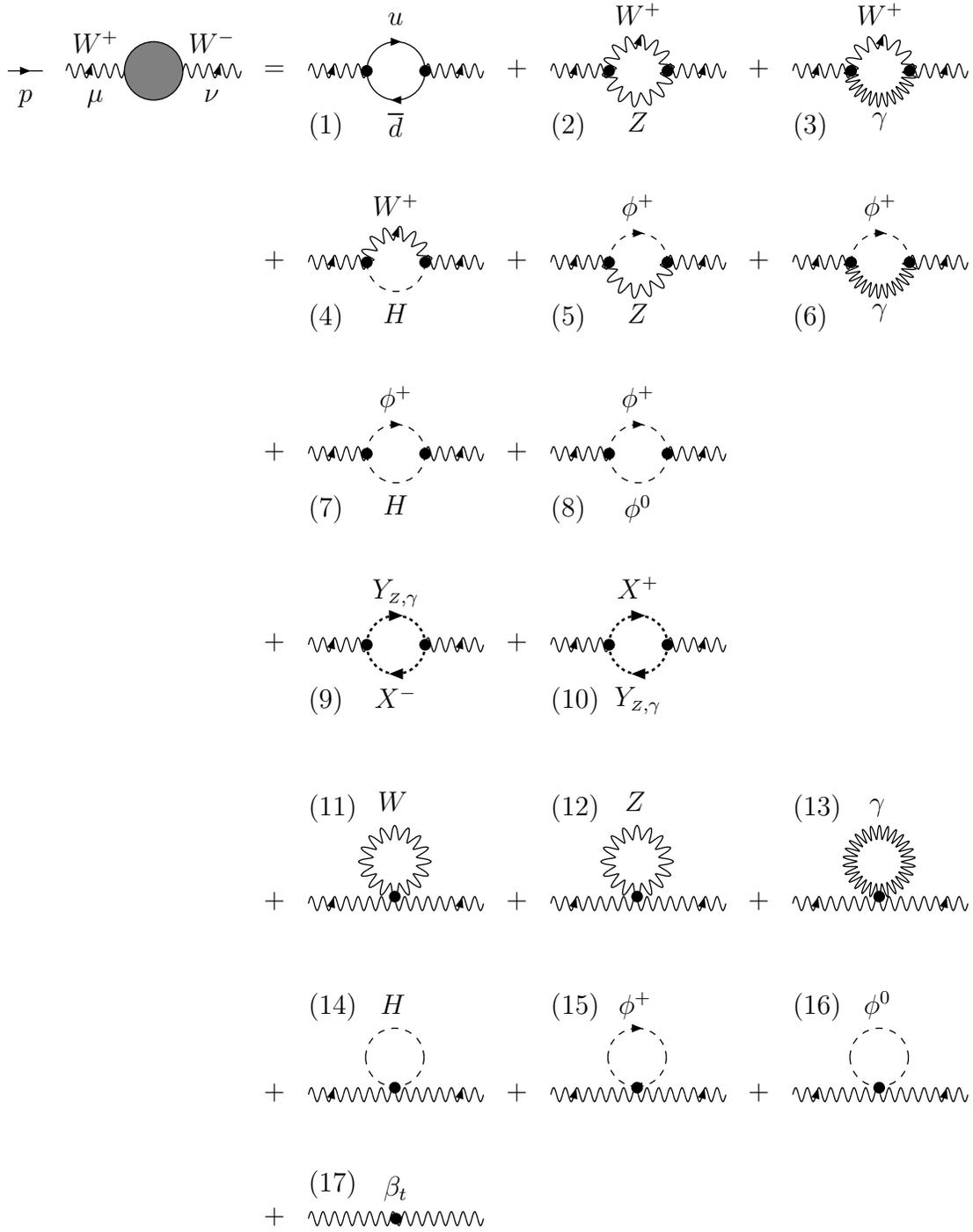
\begin{figure}[t]
\vspace*{-20mm}
\[
\baa{ccccccc}
\begin{picture}(75,20)(0,8)
  \Photon(0,10)(25,10){3}{5}
\ArrowLine(-25,10)(-10,10)
\Text(-17.5,2)[tc]{$\pmom$}
    \GCirc(37.5,10){12.5}{0.5}
  \Photon(50,10)(75,10){3}{5}
    \ArrowLine(10,9.6)(10.55,11.6)
    \ArrowLine(65,9.6)(65.55,11.6)
\Text(12.5,18)[bc]{$\wbp$}
\Text(12.5, 2)[tc]{$\mu$}
\Text(62.5,18)[bc]{$\wbm$}
\Text(62.5, 2)[tc]{$\nu$}
\end{picture}
&=&
\begin{picture}(75,20)(0,8)
  \Photon(0,10)(25,10){3}{5}
  \ArrowArcn(37.5,10)(12.5,0,180)
  \ArrowArcn(37.5,10)(12.5,180,0)
    \Vertex(25,10){2.5}
    \Vertex(50,10){2.5}
  \Photon(50,10)(75,10){3}{5}
    \ArrowLine(10,9.6)(10.55,11.6)
    \ArrowLine(65,9.6)(65.55,11.6)
\Text(37.5,-8)[tc]{$\fbd$}
\Text(37.5,30)[bc]{$\fu$}
\Text(0,-8)[lt]{$(1)$}
\end{picture}
&+&
\begin{picture}(75,20)(0,8)
  \Photon(0,10)(25,10){3}{5}
  \PhotonArc(37.5,10)(12.5,0,180){3}{7}
  \PhotonArc(37.5,10)(12.5,180,0){3}{7}
    \Vertex(25,10){2.5}
    \Vertex(50,10){2.5}
  \Photon(50,10)(75,10){3}{5}
    \ArrowLine(10,9.6)(10.55,11.6)
    \ArrowLine(65,9.6)(65.55,11.6)
    \ArrowLine(37.6,22.2)(38.3,24.2)
\Text(37.5,-8)[tc]{$\zb$}
\Text(37.5,30)[bc]{$\wbp$}
\Text(0,-8)[lt]{$(2)$}
\end{picture}
&+&
\begin{picture}(75,20)(0,8)
  \Photon(0,10)(25,10){3}{5}
  \PhotonArc(37.5,10)(12.5,0,180){3}{7}
  \PhotonArc(37.5,10)(12.5,180,0){3}{13}
    \Vertex(25,10){2.5}
    \Vertex(50,10){2.5}
  \Photon(50,10)(75,10){3}{5}
    \ArrowLine(10,9.6)(10.55,11.6)
    \ArrowLine(65,9.6)(65.55,11.6)
    \ArrowLine(37.6,22.2)(38.3,24.2)
\Text(37.5,-8)[tc]{$\gamma$}
\Text(37.5,30)[bc]{$\wbp$}
\Text(0,-8)[lt]{$(3)$}
\end{picture}
\nl \nl \nl \nl \nl
&+&
\begin{picture}(75,20)(0,8)
  \Photon(0,10)(25,10){3}{5}
  \PhotonArc(37.5,10)(12.5,0,180){3}{7}
  \DashCArc(37.5,10)(12.5,180,0){3.}
    \Vertex(25,10){2.5}
    \Vertex(50,10){2.5}
  \Photon(50,10)(75,10){3}{5}
    \ArrowLine(10,9.6)(10.55,11.6)
    \ArrowLine(65,9.6)(65.55,11.6)
    \ArrowLine(37.6,22.2)(38.3,24.2)
\Text(37.5,-8)[tc]{$\hb$}
\Text(37.5,30)[bc]{$\wbp$}
\Text(0,-8)[lt]{$(4)$}
\end{picture}
&+&
\begin{picture}(75,20)(0,8)
  \Photon(0,10)(25,10){3}{5}
  \DashArrowArcn(37.5,10)(12.5,180,0){3.}
  \PhotonArc(37.5,10)(12.5,180,0){3}{7}
    \Vertex(25,10){2.5}
    \Vertex(50,10){2.5}
  \Photon(50,10)(75,10){3}{5}
    \ArrowLine(10,9.6)(10.55,11.6)
    \ArrowLine(65,9.6)(65.55,11.6)
\Text(37.5,-8)[tc]{$\zb$}
\Text(37.5,30)[bc]{$\hkp$}
\Text(0,-8)[lt]{$(5)$}
\end{picture}
&+&
\begin{picture}(75,20)(0,8)
  \Photon(0,10)(25,10){3}{5}
  \DashArrowArcn(37.5,10)(12.5,180,0){3.}
  \PhotonArc(37.5,10)(12.5,180,0){3}{13}
    \Vertex(25,10){2.5}
    \Vertex(50,10){2.5}
  \Photon(50,10)(75,10){3}{5}
    \ArrowLine(10,9.6)(10.55,11.6)
    \ArrowLine(65,9.6)(65.55,11.6)
\Text(37.5,-8)[tc]{$\gamma$}
\Text(37.5,30)[bc]{$\hkp$}
\Text(0,-8)[lt]{$(6)$}
\end{picture}
\nl \nl \nl \nl \nl
&+&
\begin{picture}(75,20)(0,8)
  \Photon(0,10)(25,10){3}{5}
  \DashArrowArcn(37.5,10)(12.5,180,0){3.}
  \DashCArc(37.5,10)(12.5,180,0){3.}
    \Vertex(25,10){2.5}
    \Vertex(50,10){2.5}
  \Photon(50,10)(75,10){3}{5}
    \ArrowLine(10,9.6)(10.55,11.6)
    \ArrowLine(65,9.6)(65.55,11.6)
\Text(37.5,-8)[tc]{$\hb$}
\Text(37.5,30)[bc]{$\hkp$}
\Text(0,-8)[lt]{$(7)$}
\end{picture}
&+&
\begin{picture}(75,20)(0,8)
  \Photon(0,10)(25,10){3}{5}
  \DashArrowArcn(37.5,10)(12.5,180,0){3.}
  \DashCArc(37.5,10)(12.5,180,0){3.}
    \Vertex(25,10){2.5}
    \Vertex(50,10){2.5}
  \Photon(50,10)(75,10){3}{5}
    \ArrowLine(10,9.6)(10.55,11.6)
    \ArrowLine(65,9.6)(65.55,11.6)
\Text(37.5,-8)[tc]{$\hkn$}
\Text(37.5,30)[bc]{$\hkp$}
\Text(0,-8)[lt]{$(8)$}
\end{picture}
& &
\nl \nl \nl \nl \nl
&+&
\begin{picture}(75,20)(0,8)
  \Photon(0,10)(25,10){3}{5}
\SetWidth{1.}
  \DashArrowArcn(37.5,10)(12.5,0,180){1.5}
  \DashArrowArcn(37.5,10)(12.5,180,0){1.5}
\SetWidth{.5}
    \Vertex(25,10){2.5}
    \Vertex(50,10){2.5}
  \Photon(50,10)(75,10){3}{5}
    \ArrowLine(10,9.6)(10.55,11.6)
    \ArrowLine(65,9.6)(65.55,11.6)
\Text(37.5,-8)[tc]{$\fpxm$}
\Text(37.5,30)[bc]{$\fpyZA$}
\Text(0,-8)[lt]{$(9)$}
\end{picture}
&+&
\begin{picture}(75,20)(0,8)
  \Photon(0,10)(25,10){3}{5}
\SetWidth{1.}
  \DashArrowArcn(37.5,10)(12.5,0,180){1.5}
  \DashArrowArcn(37.5,10)(12.5,180,0){1.5}
\SetWidth{.5}
    \Vertex(25,10){2.5}
    \Vertex(50,10){2.5}
  \Photon(50,10)(75,10){3}{5}
    \ArrowLine(10,9.6)(10.55,11.6)
    \ArrowLine(65,9.6)(65.55,11.6)
\Text(37.5,-8)[tc]{$\fpyZA$}
\Text(37.5,30)[bc]{$\fpxp$}
\Text(0,-8)[lt]{$(10)$}
\end{picture}
& &
\nl \nl \nl \nl \nl \nl \nl
&+&
\begin{picture}(75,20)(0,8)
  \Photon(0,10)(75,10){3}{15}
  \PhotonArc(37,28)(12.5,17,197){3}{8}
  \PhotonArc(37,28)(12.5,197,17){3}{8}
     \Vertex(37,13.){2.5}
    \ArrowLine(10,9.6)(10.55,11.6)
    \ArrowLine(65,9.6)(65.55,11.6)
\Text(36.26,50)[bc]{$\wb$}
\Text(0,48)[lb]{$(11)$}
\end{picture}
&+&
\begin{picture}(75,20)(0,8)
  \Photon(0,10)(75,10){3}{15}
  \PhotonArc(37,28)(12.5,17,197){3}{8}
  \PhotonArc(37,28)(12.5,197,17){3}{8}
     \Vertex(37,13.){2.5}
    \ArrowLine(10,9.6)(10.55,11.6)
    \ArrowLine(65,9.6)(65.55,11.6)
\Text(36.26,50)[bc]{$\zb$}
\Text(0,48)[lb]{$(12)$}
\end{picture}
&+&
\begin{picture}(75,20)(0,8)
  \Photon(0,10)(75,10){3}{15}
  \PhotonArc(37,28)(12.5,3,183){3}{13}
  \PhotonArc(37,28)(12.5,183,3){3}{13}
     \Vertex(37,13.){2.5}
    \ArrowLine(10,9.6)(10.55,11.6)
    \ArrowLine(65,9.6)(65.55,11.6)
\Text(36.26,50)[bc]{$\gamma$}
\Text(0,48)[lb]{$(13)$}
\end{picture}
\nl \nl \nl \nl \nl 
&+&
\begin{picture}(75,20)(0,8)
  \Photon(0,10)(75,10){3}{15}
  \DashCArc(37,26)(12.5,0,180){3}
  \DashCArc(37,26)(12.5,180,0){3}
    \Vertex(37,13){2.5}
    \ArrowLine(10,9.6)(10.55,11.6)
    \ArrowLine(65,9.6)(65.55,11.6)
\Text(36.26,45)[bc]{$\hb$}
\Text(0,45)[lb]{$(14)$}
\end{picture}
&+&
\begin{picture}(75,20)(0,8)
  \Photon(0,10)(75,10){3}{15}
  \DashArrowArcn(37,26)(12.5,0,180){3}
  \DashArrowArcn(37,26)(12.5,180,0){3}
    \Vertex(37,13.){2.5}
    \ArrowLine(10,9.6)(10.55,11.6)
    \ArrowLine(65,9.6)(65.55,11.6)
\Text(36.26,45)[bc]{$\hkp$}
\Text(0,45)[lb]{$(15)$}
\end{picture}
&+&
\begin{picture}(75,20)(0,8)
  \Photon(0,10)(75,10){3}{15}
  \DashCArc(37,26)(12.5,0,180){3}
  \DashCArc(37,26)(12.5,180,0){3}
    \Vertex(37,13.){2.5}
    \ArrowLine(10,9.6)(10.55,11.6)
    \ArrowLine(65,9.6)(65.55,11.6)
\Text(36.26,45)[bc]{$\hkn$}
\Text(0,45)[lb]{$(16)$}
\end{picture}
\nl \nl \nl 
&+&
\begin{picture}(75,20)(0,8)
  \Photon(0,10)(75,10){3}{15}
    \Vertex(37.5,10.){2.5}
\Text(37.5,20)[bc]{$\btp$}
\Text(0,22)[lb]{$(17)$}
\end{picture}
\eaa
\]
\vspace*{5mm}
\caption{$\wb$ boson self-energy.\label{se_wb}}
\vspace*{-20mm}
\end{figure}

Next we consider the $\wb$ boson self-energy, which
is described, in the $\Rxi$  gauge, by 16 diagrams, shown in \fig{se_wb}.

First, we present an explicit expression for its bosonic component:
\bq
\Sigma^{\bos}_{_{\wb\wb}}(\sman)=
\mws\Biggl\{ 
  - \frac{19}{6} \frac{1}{\Rw} 
  - \frac{1}{4} \biggl[ 
 \frac{6}{\rhw} \lpar \frac{1}{\ctwf} + 2 \rpar 
  - \frac{3}{\ctws}
 + 10 + 3 \rhw \biggr] \Biggr\}
          \pole
      + \Sigma^{{\bos},F}_{_{\wb\wb}}(\sman),
\eq
where
\bqa
\Sigma^{{\bos},F}_{_{\wb\wb}}(\sman)&=&
  \frac{\mws}{12} 
        \Biggl\{
        \biggl[ \lpar 1 - 40\ctws \rpar \frac{1}{\Rw} 
         + 2 \lpar \frac{5}{\ctws} - 27 - 8\ctws \rpar 
\nll &&
    +\frac{\stwf}{\ctws} 
    \lpar\frac{1}{\ctws} + 8 \rpar \Rw \biggr] 
\bff{0}{-\sman}{\mwl}{\mzl}
\nll && 
  + \lrbr \frac{1}{\Rw}+2 \lpar 5-\rhw \rpar + \lpar 1-\rhw \rpar^2 \Rw \rrbr 
\bff{0}{-\sman}{\mwl}{\mhl}
\nll &&
  - 8\stws \lpar \frac{5}{\Rw} + 2 - \Rw \rpar
\bff{0}{-\sman}{\mwl}{0}
\nll &&
  + \rhw \Big[ 7 - \lpar 1 - \rhw \rpar \Rw \Big] \Lmmh 
\nll &&
  + \frac{1}{\ctws}\lrbr  \frac{18}{\rhw} \frac{1}{\ctws} 
  + 1 - 16 \ctws 
  + \stws \lpar \frac{1}{\ctws}+8 \rpar \Rw \rrbr \Lmmz 
\nll &&
  + \lrbr 2 \lpar\frac{18}{\rhw}-7\rpar-\lpar\frac{1}{\ctws}-2+\rhw\rpar\Rw\rrbr\Lmmw
\nll &&
  - \frac{4}{3}\frac{1}{\Rw}
  - 12 \lpar   \frac{1}{2}\frac{1}{\ctwf} + 1 \rpar \frac{1}{\rhw}
   - 3 \lpar   \frac{1}{\ctws}            + 2 \rpar - 9\rhw   
\nll &&
  - \lrbr\lpar \frac{1}{\ctws} + 6\stws \rpar \frac{1}{\ctws}
   - \rhw \lpar 2 - \rhw \rpar \rrbr \Rw 
      \Biggr\}. 
\label{sig_ww}
\eqa
\clearpage

\clearpage

Secondly, we give its fermionic component:
\bqa
\label{sig_wwf}
\Sigma^{\fer}_{_{\wb\wb}}(\sman)&=&
    -\sman\asum{f}{d}{} c_f\bff{f}{-\sman}{\mfpl}{\mfl} 
         +\asums{\ff} c_f\minds{\ff}\bff{1}{-\sman}{\mfpl}{\mfl},
\eqa
where summation in the first term extends to all {\em doublets} of the SM.

\subsubsection{Bosonic self-energies and counterterms\label{bosonic_ct}}
Bosonic self-energies and transitions enter one-loop amplitudes either
directly through the functions $\Dz{}{\sman}$, $\Pgg^{}(\sman)$ and $\Pzg^{}(\sman)$,
or by means of bosonic counterterms, which are made of self-energy functions at
zero argument, owing to {\em electric charge renormalization}, or  
at $\pmoms=-\Minds{}$, that is on a mass shell, owing to {\em 
on-mass-shell renormalization} (OMS scheme).
\begin{itemize}
\item Electric charge renormalization
\end{itemize}

The electric charge renormalization introduces the quantity $z_{\ph}-1$:
\bq
\lpar z_{\ph}-1 \rpar = 
    \stws   \bigg[ \Pgg(0)-\frac{2}{\mws}{\overline{\Sigma}}_{3Q}(0)\bigg],
\eq
with bosonic (see Eq. (6.161) of \cite{Bardin:1999ak}):
\bq
\lpar z_{\ph}-1 \rpar^{\bos}=
\stws   \bigg[ 3 \bigg(\pole-\Lmmw \bigg) + \frac{2}{3}\bigg]\,,  
\eq
and fer\-mio\-nic
\bqa
\lpar z_{\ph}-1 \rpar^{\fer}&=&
\stws   \bigg[ \bigg( -\frac{4}{3}\Trqf \bigg) \pole + \Pgg^{{\fer},F}(0) \bigg]
\label{elcharen}
\eqa
components.

\begin{itemize}
\item $\rho$-parameter
\end{itemize}

Finally, two self-energy functions enter Veltman's parameter $\Delta\rho$,
a gauge-invariant combination of self-energies, which naturally appears 
in the one-loop calculations:
\bqa
\Delta \rho &=&
\frac{1}{\mws}
\lrbr \Sigma_{_{WW}}(\mws) - \Sigma_{_{ZZ}}(\mzs) \rrbr,
\label{rhodef-old}
\eqa
with individual components where we explicitly show the pole parts: 
\bqa
\Delta \rho^{\bos}
&=& \lpar-\frac{1}{ 6\ctws} - \frac{41}{6} + 7 \ctws \rpar
      \pole + \Delta \rho^{{\bos},F},
\\
\Delta \rho^{\fer} &=& \frac{1}{3} \frac{\stws}{\ctws} 
\Biggl( \frac{1}{2}  N_f -4 \stws  \Trqf \Biggr)
         \pole + \Delta \rho^{{\fer},F}.\qquad
\eqa
The finite part of $\Delta \rho^{\bos}$ is given explicitly by
\bqa
\Delta \rho^{{\bos},F}
     &=&
   \lpar \frac{1}{ 12 \cowf}+\frac{4}{3\cows}
              -\frac{17}{3}-4\cows    \rpar
\biggl[  \fbff{0}{-\mws}{\mwl}{\mzl}
                  -\ctws       \fbff{0}{-\mzs}{\mwl}{\mwl} \biggr]
\nll &&
+\lpar 1-\frac{1}{3}\rhw
 +\frac{1}{12} \rhws  
 \rpar                                       \fbff{0}{-\mws}{\mwl}{\mhl}
\nll &&
-\lpar 1 - \frac{1}{3} \rhz
         + \frac{1}{12}\rhzs \rpar \frac{1}{\cows}
                                            \fbff{0}{-\mzs}{\mzl}{\mhl} 
 -4 \stws                                    \fbff{0}{-\mws}{\mwl}{0}
\nll && 
 + \frac{1}{12} \Biggl[ 
          \lpar 
 \frac{1}{\cowf} + \frac{6}{\cows} - 24 + \rhw \rpar  \Lmmz 
       + \stws \rhws \lrbr  \Lmmh - 1 \rrbr
\nll &&
    -\lpar \frac{1}{\ctws} + 14 + 16 \ctws - 48 \ctwf + \rhw \rpar \Lmmw 
    -\frac{1}{\cowf} - \frac{19}{3 \cows} + \frac{22}{3}
         \Biggr],
\eqa
while the finite part of $\Delta \rho^{\fer}$ is not shown, since  
it is trivially derived from the defining equation 
(\ref{rhodef-old}) by replacing the total self-energies with their finite parts.
\subsection{Fermionic self-energies}
\subsubsection{Fermionic self-energy diagrams}
\begin{figure}[th]
\[
\baa{ccccccc}
\begin{picture}(75,20)(0,7.5)
\Text(12.5,13)[bc]{$\ff$}
\Text(62.5,13)[bc]{$\ff$}
  \ArrowLine(0,10)(25,10)
  \GCirc(37.5,10){12.5}{0.5}
  \ArrowLine(50,10)(75,10)
\end{picture}
&=&
\begin{picture}(75,20)(0,7.5)
\Text(12.5,13)[bc]{$\ff$}
\Text(62.5,13)[bc]{$\ff$}
\Text(37.5,26)[bc]{$\ff$}
\Text(37.5,-8)[tc]{$\gamma$}
\Text(0,-7)[lt]{$(1)$}
  \ArrowLine(0,10)(25,10)
  \ArrowArcn(37.5,10)(12.5,180,0)
  \PhotonArc(37.5,10)(12.5,180,0){3}{15}
  \Vertex(25,10){2.5}
  \Vertex(50,10){2.5}
  \ArrowLine(50,10)(75,10)
\end{picture}
&+&
\begin{picture}(75,20)(0,7.5)
\Text(12.5,13)[bc]{$\ff$}
\Text(62.5,13)[bc]{$\ff$}
\Text(37.5,26)[bc]{$\ff$}
\Text(37.5,-8)[tc]{$\zb$}
\Text(0,-7)[lt]{$(2)$}
  \ArrowLine(0,10)(25,10)
  \ArrowArcn(37.5,10)(12.5,180,0)
  \PhotonArc(37.5,10)(12.5,180,0){3}{9}
  \Vertex(25,10){2.5}
  \Vertex(50,10){2.5}
  \ArrowLine(50,10)(75,10)
\end{picture}
&+&
\begin{picture}(75,20)(0,7.5)
\Text(12.5,13)[bc]{$\ff$}
\Text(62.5,13)[bc]{$\ff$}
\Text(37.5,26)[bc]{$\ffp$}
\Text(37.5,-8)[tc]{$\wb$}
\Text(0,-7)[lt]{$(3)$}
  \ArrowLine(0,10)(25,10)
  \ArrowArcn(37.5,10)(12.5,180,0)
  \PhotonArc(37.5,10)(12.5,180,0){3}{7}
  \Vertex(25,10){2.5}
  \Vertex(50,10){2.5}
  \ArrowLine(50,10)(75,10)
\end{picture}
\nl \nl \nl \nl
&+&
\begin{picture}(75,20)(0,7.5)
\Text(12.5,13)[bc]{$\ff$}
\Text(62.5,13)[bc]{$\ff$}
\Text(37.5,26)[bc]{$\ff$}
\Text(37.5,-8)[tc]{$\hb$}
\Text(0,-7)[lt]{$(4)$}
  \ArrowLine(0,10)(25,10)
  \ArrowArcn(37.5,10)(12.5,180,0)
  \DashCArc(37.5,10)(12.5,180,0){3}
  \Vertex(25,10){2.5}
  \Vertex(50,10){2.5}
  \ArrowLine(50,10)(75,10)
\end{picture}
&+&
\begin{picture}(75,20)(0,7.5)
\Text(12.5,13)[bc]{$\ff$}
\Text(62.5,13)[bc]{$\ff$}
\Text(37.5,26)[bc]{$\ff$}
\Text(37.5,-8)[tc]{$\hkn$}
\Text(0,-7)[lt]{$(5)$}
  \ArrowLine(0,10)(25,10)
  \ArrowArcn(37.5,10)(12.5,180,0)
  \DashCArc(37.5,10)(12.5,180,0){3}
  \Vertex(25,10){2.5}
  \Vertex(50,10){2.5}
  \ArrowLine(50,10)(75,10)
\end{picture}
&+&
\begin{picture}(75,20)(0,7.5)
\Text(12.5,13)[bc]{$\ff$}
\Text(62.5,13)[bc]{$\ff$}
\Text(37.5,26)[bc]{$\ffp$}
\Text(37.5,-8)[tc]{$\hkg$}
\Text(0,-7)[lt]{$(6)$}
  \ArrowLine(0,10)(25,10)
  \ArrowArcn(37.5,10)(12.5,180,0)
  \DashCArc(37.5,10)(12.5,180,0){3}
  \Vertex(25,10){2.5}
  \Vertex(50,10){2.5}
  \ArrowLine(50,10)(75,10)
\end{picture}
\label{fsesm}
\eaa
\]
\vspace*{5mm}
\caption{Fermionic self-energy diagrams. \label{f_se}}
\vspace*{5mm}
\end{figure}
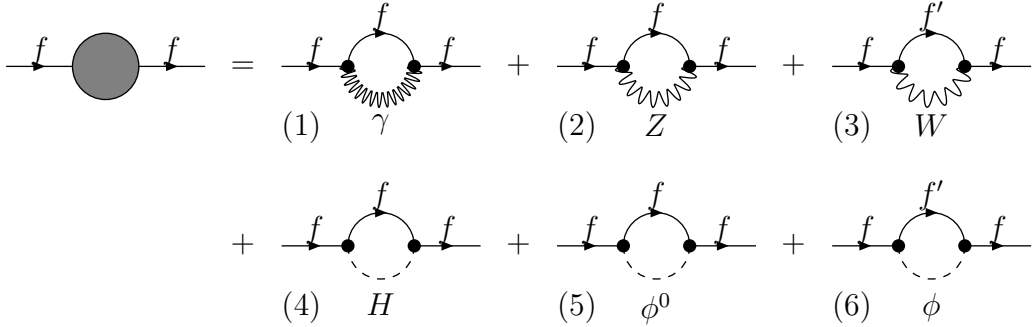

 The total self-energy function of a fermion 
in the $\Rxi$ gauge is described by the six diagrams of \fig{f_se}.

 Calculating derivatives straightforwardly and substituting the $a_i$'s, we
obtain (see \cite{Bardin:1999yd}) explicit expressions for
the wave-function renormalization factor $\sqrt{z_{_{L,R}}}$. 

It is convenient to distinguish the electromagnetic components
\bqa 
\lpar \sqrt{z_{_{L}}} - I \rpar^{em}_f &=& 
          \lpar \sqrt{z_{_{R}}} - I \rpar^{em}_f
        = \stws Q^2_f \lpar
        - \frac{1}{2\epsb} + \frac{1}{\epsh}
        + \frac{3}{2} \ln \frac{\mf}{\tHss} - 2 \rpar
\label{wfren_em}
\eqa
and the weak components
\bqa
 \Big| \sqrt{z_{\sss L,R}} \Big| - I   = 
       \lpar w_v \pm w_a \rpar,                   
\eqa
where 
\bqa
w_v^{\sss Z} &=&
- \frac{1}{8}\frac{1}{\ctws}
\Biggl\{ 
  \lpar v_t^2 + a_t^2 + 2 a_t^2 \rtz \rpar \pole
+ \lpar v_t^2 + a_t^2 \rpar
               \biggl[ 
     \frac{1}{\rtz} \lpar \fbff{0}{- \mts}{\mzl}{\mtl}+\Lmmz-1 \rpar
\nll &&
    + 2 \lpar 1 + 2 \rtz \rpar \mzs \bff{0p}{-\mts}{\mtl}{\mzl} 
    - \Lmmt     \biggr]
  +2 a_t^2 \rtz \biggl[
    \frac{1}{\rtz} \Bigl( \fbff{0}{-\mts}{\mzl}{\mtl} 
\nll &&
+ \Lmmz - 1 \Bigr)
    - 6 \mzs \bff{0p}{ -\mts}{\mtl}{\mzl}
    - \Lmmt + 1  \biggr]   
\Biggr\}, 
\\ 
w_v^{\sss W} &=&
  -\frac{1}{16} \lpar 2 + \rtw \rpar 
    \Biggl\{  \pole
   +    \frac{1 + \rtw}{\rtw}   \lrbr \fbff{0}{-\mts}{\mwl}{0} - 1 \rrbr 
\nll &&
   + 2 \lpar 1 - \rtw \rpar \mws  \bff{0p}{ -\mts}{0}{\mwl}
         +\frac{1}{4 \rtw} \Lmmw \Biggr\} + \rtw\,,
\\
w_v^{\sss H} &=&
  - \frac{1}{16}\rtw      \Biggl\{
         \pole           
 + r_{{\sss H}t} \lrbr \fbff{0}{-\mts}{\mhl}{\mtl} + \Lmmh - 1 \rrbr 
\nll &&
 -  2 \lpar 4 r_{t{\sss H}} 
       - 1 \rpar \mhs \bff{0p}{ -\mts}{\mtl}{\mhl} 
 -  \Lmmt + 1 
                         \Biggr\},
\\
w_a^{\sss Z} &=&
- \frac{1}{4\ctws} v_t a_t 
            \Biggl\{
         \pole
  - \frac{1}{\rtz}  \lrbr \fbff{0}{-\mts}{\mzl}{\mtl}
 + \Lmmz - 1 \rrbr
\nll &&
  + 2 \fbff{0}{-\mts}{\mzl}{\mtl} + \Lnrt - 2 \Biggr\},
\\
w_a^{\sss W} &=&
   -\frac{1}{16} \Biggl\{
         \lpar  2 - \rtw        \rpar  \pole
       - \lpar \frac{2}{\rtw} - 3 \rpar
            \lrbr  \fbff{0}{-\mts}{\mwl}{0} - 1 \rrbr
\nll &&
       -  \rtw  \fbff{0}{-\mts}{\mwl}{0} 
      - \lpar \frac{2}{\rtw} - 1 \rpar \Lmmw
               \Biggr\}.
\eqa
\subsection{The $\zb\ff\fbf$ and $\ph\ff\fbf$ vertices}
 Consider now the sum of all vertices and corresponding counterterms 
whose contribution originates from the fermionic self-energy diagrams of 
Fig.~\ref{f_se}. This sum is shown in Fig.~\ref{zavert1}.

\begin{figure}[!h]
\vspace*{-20mm}
\[
\begin{array}{cccccccc}
& \begin{picture}(125,86)(20,40)
  \Vertex(100,43){12.5}
\SetScale{2.}
  \Photon(13,22)(50,22){1.5}{15}
  \ArrowLine(50,21.5)(62.5,0)
  \ArrowLine(62.5,43)(50,22.5)
  \Text(108,74)[lb]{$\fbf$}
  \Text(62.5,50)[bc]{$(\ph,\zb)$}
  \Text(108,12)[lt]{$\ff$}
\end{picture}
&=&
\begin{picture}(125,86)(20,40)
  \Photon(25,43)(100,43){3}{15}
  \GCirc(100,43){12.5}{0.5}
  \ArrowLine(125,86)(107,53)
  \ArrowLine(106,31)(125,0)
  \Text(108,74)[lb]{$\fbf$}
  \Text(62.5,50)[bc]{$(\ph,\zb)$}
  \Text(108,12)[lt]{$\ff$}
\end{picture}
&+&
\begin{picture}(125,86)(20,40)
  \Photon(25,43)(100,43){3}{15}
\SetScale{2.0}
  \Line(45.25,16.75)(54.75,26.25)
  \Line(45.25,26.25)(54.75,16.75)
\SetScale{1.0}
  \ArrowLine(125,86)(100,43)
  \Vertex(100,43){2.5}
  \ArrowLine(100,43)(125,0)
  \Text(108,74)[lb]{$\fbf$}
  \Text(62.5,50)[bc]{$(\ph,\zb)$}
  \Text(108,12)[lt]{$\ff$}
\end{picture}
\end{array}
\]
\vspace{10mm}
\caption[$\zb\ff\fbf$ and $\ph\ff\fbf$ vertices with counterterms.]
{\it
$\zb\ff\fbf$ and $\ph\ff\fbf$ vertices with fermionic counterterms.
\label{zavert1}}
\end{figure}
%

The formulae which determine the counterterms are:
\bqa
 F^{\gamma,ct}_{\sss Q} & = &  2 \lpar\sqrt{z_{\sss R}}-I \rpar, \\
 F^{\gamma,ct}_{\sss L} & = &  \lpar\sqrt{z_{\sss L}}-I \rpar
                       - \lpar\sqrt{z_{\sss R}}-I \rpar,         \\
 F^{z,ct}_{\sss Q} & = &  \delta^2_f
                         \lpar\sqrt{z_{\sss R}}-I \rpar,   \\
 F^{z,ct}_{\sss L} & = &  \sigma^2_f
                          \lpar\sqrt{z_{\sss R}}-I \rpar
                        -\delta^2_f
                          \lpar\sqrt{z_{\sss R}}-I \rpar,    
\eqa
where
\bqa
\delta_f=v_f-a_f,\qquad
\sigma_f=v_f+a_f.
\eqa
 For the sum of all $\ph\to\ff\fbf$ and $\zb\to\ff\fbf$ vertices 
(the {\em total} $\ph(\zb)\ff\fbf$ vertex depicted by a grey circle
in \fig{zavert1}) 
we use the standard
normalization
\bq
\ib\pi^2=\tpfi\frac{1}{16\pi^2}\,,
\eq
and define
\bqa
\Vverti{\mu}{\ph}{\sman} &=&
\tpfi\frac{1}{16\pi^2}\Gverti{\mu}{ }{\sman},
\\
\Vverti{\mu}{\zb}{\sman} &=&
\tpfi\frac{1}{16\pi^2}\Zverti{\mu}{ }{\sman},
\eqa
while we denote the individual vertices as follows:
\bqa
\Gverti{\mu}{}{\sman}&=&
 \Gverti{\mu}{\ph}{\sman}+\Gverti{\mu}{\zb}{\sman}
+\Gverti{\mu}{\wb}{\sman}+\Gverti{\mu}{\hb}{\sman},
\\
\Zverti{\mu}{}{\sman}&=&
 \Zverti{\mu}{\ph}{\sman}+\Zverti{\mu}{\zb}{\sman}
+\Zverti{\mu}{\wb}{\sman}+\Zverti{\mu}{\hb}{\sman}.
\eqa
All vertices  have three components in our $LQD$  basis.

\subsubsection{Scalar form factors}
Now we construct the $24=(4:\ab,\zb,\hb,\wb\,\mbox{-virtual})
\otimes(3:L,Q,D)\otimes(2:\ph,\zb\,\mbox{-incoming})$
scalar form factors, originating from the diagrams of \fig{zavert1}. 
They are derived from the following six equations --- three projections 
for $\ph\ff\fbf$ vertices:
\bqa
\vvertil{\gamma {\sss B}}{\sss L}{\sman}
&=&\frac{2}{\siw \tcif}
\Bigl\{ \Gverti{\mu}{\sss B}{\sman}[\ib\gbc \gadu{\mu}\gap]
+\siw\qf\fverti{\gamma,ct}{\sss L} \Bigr\},
\\
\vvertil{\gamma {\sss B}}{\sss Q}{\sman}
&=&\frac{1}{\siw\qf}
\Bigl\{\Gverti{\mu}{\sss B}{\sman}[\ib\gbc \gadu{\mu}]    
+\siw\qf
\fverti{\gamma,ct}{\sss Q} \Bigr\},
\\ 
\vvertil{\gamma {\sss B}}{\sss D}{\sman}
&=&
 \frac{2}{\stwl I^{(3)}_t} 
  \Gverti{\mu}{\sss B}{\sman}[\gbc \mtl I D_\mu]\,,
\eqa
and three projections for $\zb\ff\fbf$ vertices:
\bqa
\vvertil{z{\sss B}}{\sss L}{\sman}&=&\frac{2\cow}{\tcif}
\Biggl\{\Zverti{\mu}{\sss B}{\sman}[\ib \gbc \gadu{\mu} \gap]
+\frac{1}{\cow}\fverti{z,ct}{\sss L}
\Biggr\},
\\
\vvertil{z{\sss B}}{\sss Q}{\sman}&=&
\frac{2\cow}{\delta_f}
\Biggl\{
\Zverti{\mu}{\sss B}{\sman}[\ib \gbc \gadu{\mu}]
+\frac{1}{\cow}\fverti{z,ct}{\sss Q}
\Biggr\},
\\
\vvertil{z{\sss B}}{\sss D}{\sman}
& = & \frac{2\ctwl}{I^{(3)}_t } 
 \Zverti{\mu}{\sss B}{\sman}[\gbc \mtl I D_\mu]\,.
\eqa
Here we have $ f=t,e $, and $ B = \ab,\zb,\wb,\hb $, 
and we introduce the symbol $[\dots]$ for the definition
of the procedure of the projection of $\Gverti{\mu}{}{\sman}$
and $\Zverti{\mu}{}{\sman}$ to our basis.
It has the same meaning as in {\tt form} language \cite{Vermaseren:2000f},
namely, e.g. $\Gverti{\mu}{\sss B}{\sman}[\ib\gbc \gadu{\mu}\gap]$
means that only the coefficient of $[\ib\gbc \gadu{\mu}\gap]$ of
the whole expression $\Gverti{\mu}{\sss B}{\sman}$ is taken
({\em projected}).

 The factors $ 1/\bigl(\qf \siw \bigr)$, $ 2/\bigl(\siw\tcit\bigr)$ 
and $ 2 \ctws /\bigl(\stwl I^{(3)}_t \bigr)$
for $\ph\ff\fbf$ vertices, and the 
factors $2\cow/\tcit$, $2\cow/\delta_f$
and  $ 2\ctwl/ I^{(3)}_t $ 
for $\zb\ff\fbf$ are due to the form factor definitions of
\eqn{structures-old}.

The total $\gamma\ft\bar{t}$ and $\zb\ft\bar{t}$ form factors are sums over 
three bosonic contributions $B =\zb,\wb,\hb$
since we separated out the contribution of the diagram with virtual $\ph\equiv\ab$:
\bqa
\vvertil{\gamma tt}{\sss{L}}{\sman}  &=&
  \vvertil{\gamma{\sss Z}}{\sss{L}}{\sman} 
+ \vvertil{\gamma{\sss W}}{\sss{L}}{\sman},
\nll
\vvertil{\gamma tt}{\sss{Q}}{\sman}  &=&
  \vvertil{\gamma{\sss Z}}{\sss{Q}}{\sman} 
+ \vvertil{\gamma{\sss W}}{\sss{Q}}{\sman}
+ \vvertil{\gamma{\sss H}}{\sss{Q}}{\sman},
\nll
\vvertil{\gamma tt}{\sss D}{\sman} &=&
  \vvertil{\gamma{\sss Z}}{\sss D}{\sman}
+ \vvertil{\gamma{\sss W}}{\sss D}{\sman}
+ \vvertil{\gamma{\sss H}}{\sss D}{\sman},
\nll
\vvertil{ztt}{\sss{L}}{\sman}  &=&
  \vvertil{z{\sss Z}}{\sss{L}}{\sman} 
+ \vvertil{z{\sss W}}{\sss{L}}{\sman}
+ \vvertil{z{\sss H}}{\sss{L}}{\sman},
\nll
\vvertil{ztt}{\sss{Q}}{\sman}  &=&
  \vvertil{z{\sss Z}}{\sss{Q}}{\sman} 
+ \vvertil{z{\sss W}}{\sss{Q}}{\sman}
+ \vvertil{z{\sss H}}{\sss{Q}}{\sman},
\nll
\vvertil{ztt}{\sss D}{\sman} &=&
  \vvertil{z{\sss Z}}{\sss D}{\sman}
+ \vvertil{z{\sss W}}{\sss D}{\sman}
+ \vvertil{z{\sss H}}{\sss D}{\sman}.
\eqa
The quantities $\vvertil{\ph(z){\sss B}}{\sss L,Q,D}{\sman}$
originate from groups of diagrams, which we will call {\it clusters}. 

\subsection{Library of form factors for $Btt$ clusters}
Here we present a complete collection of scalar form factors
$\vvertil{\ph(z){\sss B}}{\sss L,Q,D}{\sman}$ originating 
from a vertex diagram with a virtual vector boson, contribution of 
a scalar partner of this vector boson, and relevant counterterms.

Actually three gauge-invariant subsets of diagrams of this kind,
$\ab$, $\zb$ and $\hb$, appear in our calculation.
They may be termed {\it clusters}, since they are 
natural building blocks of the complete scalar form factors, which are
the aim of our calculation. 
Again, in the spirit of our presentation, we write down their pole and finite parts.
The remaining vertices with virtual $\wb$ and $\hkp,\hkm$ with relevant 
counterterms we also define as the $\wb$ cluster. However, the latter diagrams
do not form a gauge-invariant subset.
\subsubsection{Form factors of the $Z$  cluster}
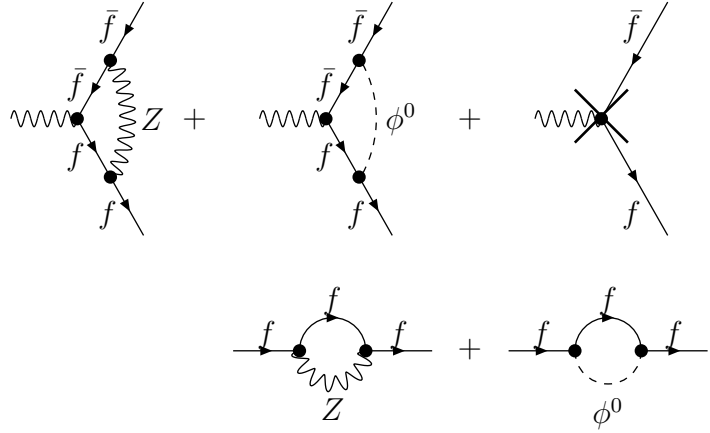
\begin{figure}[h]
\vspace{-18mm}
\[
\baa{ccccccccc}
&&
\begin{picture}(55,88)(0,41)
  \Photon(0,44)(25,44){3}{5}
  \Vertex(25,44){2.5}
  \PhotonArc(0,44)(44,-28,28){3}{9}
  \Vertex(37.5,22){2.5}
  \Vertex(37.5,66){2.5}
  \ArrowLine(50,88)(37.5,66)
  \ArrowLine(37.5,66)(25,44)
  \ArrowLine(25,44)(37.5,22)
  \ArrowLine(37.5,22)(50,0)
  \Text(33,75)[lb]{$\fbf$}
  \Text(21,53)[lb]{$\fbf$}
  \Text(49,44)[lc]{$\zb$}
  \Text(21,35)[lt]{$\ff$}
  \Text(33,13)[lt]{$\ff$}
  \Text(1,1)[lb]{}
\end{picture}
&+&
\begin{picture}(55,88)(0,41)
  \Photon(0,44)(25,44){3}{5}
  \Vertex(25,44){2.5}
  \DashCArc(0,44)(44,-28,28){3}
  \Vertex(37.5,22){2.5}
  \Vertex(37.5,66){2.5}
  \ArrowLine(50,88)(37.5,66)
  \ArrowLine(37.5,66)(25,44)
  \ArrowLine(25,44)(37.5,22)
  \ArrowLine(37.5,22)(50,0)
  \Text(33,75)[lb]{$\fbf$}
  \Text(21,53)[lb]{$\fbf$}
  \Text(47.5,44)[lc]{$\hkn$}
  \Text(21,35)[lt]{$\ff$}
  \Text(33,13)[lt]{$\ff$}
  \Text(1,1)[lb]{}
\end{picture}
&+&
\begin{picture}(55,88)(0,41)
  \Photon(0,44)(25,44){3}{5}
  \Vertex(25,44){2.5}
  \ArrowLine(50,88)(25,44)
  \ArrowLine(25,44)(50,0)
  \Text(33,75)[lb]{$\fbf$}
  \Text(33,13)[lt]{$\ff$}
  \Text(1,1)[lb]{}
\SetScale{2.0}
  \Line(17.5,17.5)(7.5,27.5)
  \Line(7.5,17.5)(17.5,27.5)
\end{picture}
\nl \nl  [17mm] 
& &
& &
\begin{picture}(75,20)(0,7.5)
\Text(12.5,13)[bc]{$\ff$}
\Text(62.5,13)[bc]{$\ff$}
\Text(37.5,26)[bc]{$\ff$}
\Text(37.5,-8)[tc]{$\zb$}
\Text(0,-7)[lt]{$ $}
  \ArrowLine(0,10)(25,10)
  \ArrowArcn(37.5,10)(12.5,180,0)
  \PhotonArc(37.5,10)(12.5,180,0){3}{7}
  \Vertex(25,10){2.5}
  \Vertex(50,10){2.5}
  \ArrowLine(50,10)(75,10)
\end{picture}
&+&
\begin{picture}(75,20)(0,7.5)
\Text(12.5,13)[bc]{$\ff$}
\Text(62.5,13)[bc]{$\ff$}
\Text(37.5,26)[bc]{$\ff$}
\Text(37.5,-8)[tc]{$\hkn$}
\Text(0,-7)[lt]{$ $}
  \ArrowLine(0,10)(25,10)
  \ArrowArcn(37.5,10)(12.5,180,0)
  \DashCArc(37.5,10)(12.5,180,0){3}
  \Vertex(25,10){2.5}
  \Vertex(50,10){2.5}
  \ArrowLine(50,10)(75,10)
\end{picture}
\eaa
\]
\vspace{2mm}
\caption
[$Z$ cluster.]
{$Z$ cluster. The two fermionic self-energy diagrams in the second row
give rise to the counterterm contribution depicted by the solid cross
in the last diagram of the first row.
\label{zavert}}
\end{figure}

The diagrams shown in \fig{zavert} contribute to the $\zb$ cluster.

 Separating out pole contributions $1/{\bar\varepsilon}$, we define finite
(calligraphic) quantities. We note that, if a form factor $F_{\sss A}^{ij}(s)$
has a pole, then the corresponding finite part ${\cal F}_{\sss A}^{ij}(s)$
is $\mu$-dependent: 
\bqa
 \vvertil{\gamma{\sss Z}}{\sss{L}}{\sman} & = & 
 \cvertil{\gamma{\sss Z}}{\sss{L}}{\sman}, 
\nll 
\vvertil{\gamma{\sss Z}}{\sss{Q}}{\sman} & = &
  \cvertil{\gamma{\sss Z}}{\sss{Q}}{\sman},
\nll 
\vvertil{\gamma{\sss Z}}{\sss{D}}{\sman} & = &
  \cvertil{\gamma{\sss Z}}{\sss{D}}{\sman},
\nll 
 \vvertil{z{\sss Z}}{\sss L}{\sman} & = & 
 -\frac{1}{4} \rtw \pole 
 +\cvertil{z{\sss Z}}{\sss L}{\sman}, 
\nll
  \vvertil{z{\sss Z}}{\sss{Q}}{\sman} & = &
 -\frac{1}{16} \frac{1}{\qum \stws} \rtw  \pole
 +\cvertil{z{\sss Z}}{\sss{Q}}{\sman},
\nll 
\vvertil{z{\sss Z}}{\sss{D}}{\sman} & = &
  \cvertil{z{\sss Z}}{\sss{D}}{\sman}.
\eqa
Here the finite parts are:
\bqa
{\cal F }^{\gamma {\sss Z}}_{\sss L} \lpar \sman \rpar  &=&
 \frac{1}{\ctws} \qu \vu \Biggl\{
            2 \lpar 2+\frac{1}{\Rz} \rpar
                        \mzs \cff{-\mts}{-\mts}{-s}{\mtl}{\mzl}{\mtl}
\\  &&
           -3 \fbff{0}{-s}{\mtl}{\mtl} + 2 \fbff{0}{-\mts}{\mtl}{\mzl} - \Lmmt
\nll &&
   + \fbff{d1}{-\mts}{\mtl}{\zml} -2 \lpar 1+4 \ruz \rpar
     \frac{\mzs}{\sdtit} \Longab{\mtl}{\mtl}{\mzl}  \Biggr\},
\nonumber
\eqa
\bqa
\label{b0d_containing}
 {\cal F }^{\gamma {\sss Z}}_{\sss Q}\lpar \sman \rpar
&=&
 \frac{1}{4 \cows} \Bigg\{
\delta_t^2 \Biggl[ 2 \biggl(2 \lpar 1-\ruz \rpar+\frac{1}{\Rz} \biggr) \mzs 
                \cff{-\mts}{-\mts}{-s}{\mtl}{\mzl}{\mtl}
\nll &&
                -3 \fbff{0}{-s}{\mtl}{\mtl}+4 \fbff{0}{-\mts}{\mtl}{\mzl}+\Lmmz
                -\ruz \fbff{d2}{-\mts}{\mtl}{\zml}
\nll &&
                -2 \lpar 1+2 \ruz \rpar 
                 \mzs \bff{0p}{-\mts}{\mtl}{\mzl}-\frac{5}{2} \Biggr]
\nll &&
 + 2 \vu \au \ruz \Biggl[-4 \mzs \cff{-\mts}{-\mts}{-s}{\mtl}{\mzl}{\mtl}
       +2\lpar \Lmmt - \Lmmz \rpar
\nll &&
               - 2 \lpar 1 - \ruz \rpar \fbff{d2}{-\mts}{\mtl}{\mzl}+1
\nll &&
               - \frac{2}{\ruz} \biggl(  \lpar 1 + 2 \ruz\rpar\mzs 
                 \bff{0p}{-\mts}{\mtl}{\mzl} + \frac{1}{2} \biggr) \Biggr]
\nll &&
 +2 \au^2 \ruz   \Biggl[ \fbff{0}{-s}{\mtl}{\mtl}  + \Lmmz
               - \ruz \fbff{d2}{-\mts}{\uml}{\zml}
\nll &&
                +6\mzs \bff{0p}{-\mts}{\uml}{\zml} - \frac{5}{2} \Biggr]
\nll && 
 -4 \lrbr \frac{ \delta_t^2}{2}-\lpar 4\vu \au-\au^2 \rpar \ruz \rrbr
  \frac{\mzs}{\sdtit} \Longab{\mtl}{\mtl}{\mzl}
                  \Biggr\}, 
\eqa
\bqa 
{\cal F}^{\gamma{\sss Z}}_{\sss D} \lpar \sman \rpar  & = &
  -\frac{2 Q_t}{\tcit\ctws}
\frac{1}{\sdtit} \Biggl\{
      \frac{v_t^2+a_t^2}{2} \Bigg[ -4 \mzs 
 \cff{-\mts}{-\mts}{-s}{\mtl}{\mzl}{\mtl}
\nll &&
 + \fbff{0}{-\sman}{\mtl}{\mtl}-2 \fbff{0}{-\mts}{\mtl}{\mzl}-\Lmmt
 + \fbff{d1}{-\mts}{\mtl}{\zml} 
\nll &&
           +2  +6 \frac{\mzs}{\sdtit} \Longab{\mtl}{\mtl}{\mzl}
                 \Bigg]
\nll &&
         +a_t^2 \Bigg[ 2 \lpar 3 \rtz-\frac{1}{\Rz} \rpar \mzs 
 \cff{-\mts}{-\mts}{-s}{\mtl}{\mzl}{\mtl}
\nll &&
                +\fbff{0}{-\mts}{\mtl}{\mzl} + \Lmmz - 1
                -\rtz \biggl[\fbff{0}{-\sman}{\mtl}{\mtl} + \Lmmt - 2 \biggr]
\nll &&
          -2 \lpar 2 - 3 \frac{\mts}{\sdtit} \rpar \Longab{\mtl}{\mtl}{\mzl}
               \Bigg]
                       \Biggr\},
\eqa
\bqa
{\cal F}^{z{\sss Z}}_{\sss L}\lpar \sman \rpar
&=&
\frac{1}{4 \cows} \Biggl\{
\frac{3\vu^2+\au^2}{3} \Bigg[ 2 \Bigg( 3 \lpar 2+\frac{1}{\Rz} \rpar-2 \ruz \Bigg) \mzs 
     \cff{-\mts}{-\mts}{-s}{\mtl}{\mzl}{\mtl}
\nll[1mm] &&
     -9 \fbff{0}{-s}{\mtl}{\mtl}+8 \fbff{0}{-\mts}{\mtl}{\mzl}-\Lmmt
\nll[1mm] &&
 + \fbff{d1}{-\mts}{\mtl}{\zml} 
    -2 \lpar 1+2 \ruz \rpar \mzs \bff{0p}{-\mts}{\mtl}{\mzl}-2 \Bigg]
\nll &&
    -\frac{2}{3} \au^2 \Bigg[ 4 \mts \cff{-\mts}{-\mts}{-s}{\mtl}{\mzl}{\mtl}
     +3 \ruz \bigg[ \fbff{0}{-s}{\mtl}{\mtl} + \Lmmt \bigg]
\nll[2mm]&& 
     +\fbff{0}{-\mts}{\mtl}{\mzl} + 3 \Lmmz + 2 \Lmmt
\nll[2mm] &&
     + 2\fbff{d1}{-\mts}{\mtl}{\zml}
     + 2 \lpar 1-7 \ruz \rpar \mzs \bff{0p}{-\mts}{\mtl}{\mzl} - 1 \Bigg] 
\nll &&
     -2 \lrbr \lpar 3\vu^2+\au^2\rpar \lpar 1 + 4 \ruz \rpar-2 \au^2 \ruz \rrbr 
 \frac{\mzs} {\sdtit} \Longab{\mtl}{\mtl}{\mzl}  
                \Biggr\}, 
\eqa
\bqa
{\cal F}^{z{\sss Z}}_{\sss Q}\lpar \sman \rpar
&=&
 \frac{1}{\cows} \Biggl\{ 
 \frac{1}{4} \delta_t^2 
  \Biggl[ 2 \lpar 2-2 \ruz+\frac{1}{\Rz}\rpar
            \mzs \cff{-\mts}{-\mts}{-s}{\mtl}{\mzl}{\mtl}
\nll[2mm] &&
  - 3 \fbff{0}{-s}{\mtl}{\mtl}+4 \fbff{0}{-\mts}{\mtl}{\mzl}+\Lmmt
  - \fbff{d1}{-\mts}{\mtl}{\zml}
\nll[2mm] &&
                -2 \lpar 1+2 \ruz \rpar \mzs \bff{0p}{-\mts}{\mtl}{\mzl} -2
                -2\frac{\mzs}{\sdtit} \Longab{\mtl}{\mtl}{\mzl} \Biggr]
\nll && 
 +\au \ruz \Biggl( \vu \Biggl[-2 \mzs \cff{-\mts}{-\mts}{-s}{\mtl}{\mzl}{\mtl}
\nll &&
 +\frac{1}{\ruz} \biggl[ \fbff{0}{-\mts}{\mtl}{\mzl} + \Lmmt - 1
 -\fbff{d1}{-\mts}{\mtl}{\zml}
\nll &&
               -\lpar 1 + 2 \ruz \rpar \mzs \bff{0p}{-\mts}{\mtl}{\mzl} \biggr]
       +6 \frac{\mzs}{\sdtit} \Longab{\mtl}{\mtl}{\mzl} \Biggr]
\nll &&
       -\frac{1}{2}\au\Biggl[ 8 \mzs \cff{-\mts}{-\mts}{-s}{\mtl}{\mzl}{\mtl}
\nll[2mm] &&
       - \fbff{0}{-s}{\mtl}{\mtl} - \Lmmt + 2 
       +\fbff{d1}{-\mts}{\mtl}{\zml}
\nll[2mm] &&
       - 6 \mzs \bff{0p}{-\mts}{\mtl}{\mzl} 
       - 10 \frac{\mzs}{\sdtit} \Longab{\mtl}{\mtl}{\mzl} 
               \Biggr]
\nll && 
 -\frac{\au^2}{\delta_t} \Biggl[
                 4 \mzs \cff{-\mts}{-\mts}{-s}{\mtl}{\mzl}{\mtl}
\nll &&
                         - \fbff{0}{-s}{\mtl}{\mtl}+1 
                -6 \frac{\mzs}{\sdtit} \Longab{\mtl}{\mtl}{\mzl} \Biggr]
           \Biggr)             \Biggr\},
\eqa
\bqa
{\cal F}^{z{\sss Z}}_{\sss D}\lpar \sman \rpar  & = &
-\frac{1}{2\tcit\ctws} 
\frac{1}{\sdtit}
   \Biggl\{\lpar 3\au^2+\vu^2\rpar \vu
            \Biggl[-4 \mzs \cff{-\mts}{-\mts}{-s}{\mtl}{\mzl}{\mtl}
\nll[2mm] &&
      + \fbff{0}{-\sman}{\mtl}{\mtl}-2 \fbff{0}{-\mts}{\mtl}{\mzl}-\Lmmt
      + \fbff{d1}{-\mts}{\mtl}{\zml}
\nll[2mm] &&
       +2 +6 \frac{\mzs}{\sdtit} \Longab{\mtl}{\mtl}{\mzl}
               \Biggr]
\nll &&
    +2 v_t a_t^2 \Bigg[ 2 \lpar 7 \rtz-2 \frac{1}{\Rz} \rpar \mzs 
 \cff{-\mts}{-\mts}{-s}{\mtl}{\mzl}{\mtl}
\nll &&
                +\fbff{0}{-\mts}{\mtl}{\mzl}+\Lmmz-1
                -\rtz \biggl[ \fbff{0}{-\sman}{\mtl}{\mtl}+\Lmmt-2 \biggr] 
\nll &&
                -2 \Bigg( 4-3 \frac{\mts}{\sdtit} \Bigg) 
         \Longab{\mtl}{\mtl}{\mzl}
               \Bigg]                        
    \Biggr\}.
\eqa
\vskip 10pt
In \eqn{b0d_containing} the `once and twice subtracted' functions $\sfbff{d1}$ and $\sfbff{d2}$  
are met:
\vskip 1pt
\bqa
\fbff{d1}{-\mts}{\mtl}{\zml}&=&
 \frac{1}{\rtz} \biggl[
 \fbff{0}{-\mts}{\mtl}{\zml}+\Lmmz-1\biggr],
\\
\fbff{d2}{-\mts}{\mtl}{\zml}&=&
 \frac{1}{\rtz^2} \biggl[
 \fbff{0}{-\mts}{\mtl}{\zml}+\Lmmz-1
\nll &&\hspace{1cm}
 -\rtz \lpar\Lmmt-\Lmmz+\frac{1}{2} \rpar\biggr].
\nn
\eqa
They remain finite in the limit $\mtl\to 0$.

We note that, for the $\zb$ cluster, all the six scalar form factors
$\vvertil{\ph(z){\sss Z}}{\sss L,Q,D}{\sman}$ are {\em separately}
gauge-invariant.
\clearpage
\subsubsection{Form factors of the $H$ cluster}
\begin{figure}[h]
\vspace{-15mm}
\[
\hspace{-1cm}
\baa{cccccccc} 
& &
\begin{picture}(55,88)(0,41)
  \Photon(0,44)(25,44){3}{5}
  \Vertex(25,44){2.5}
  \DashCArc(0,44)(44,-28,28){3}
  \Vertex(37.5,22){2.5}
  \Vertex(37.5,66){2.5}
  \ArrowLine(50,88)(37.5,66)
  \ArrowLine(37.5,66)(25,44)
  \ArrowLine(25,44)(37.5,22)
  \ArrowLine(37.5,22)(50,0)
  \Text(33,75)[lb]{$\fbf$}
  \Text(21,53)[lb]{$\fbf$}
  \Text(47.5,44)[lc]{$\hb$}
  \Text(21,35)[lt]{$\ff$}
  \Text(33,13)[lt]{$\ff$}
  \Text(1,1)[lb]{}
\end{picture}
& + &
\begin{picture}(55,88)(0,41)
  \Photon(0,44)(25,44){3}{5}
  \Vertex(25,44){2.5}
  \ArrowArcn(0,44)(44,28,-28)
  \Vertex(37.5,22){2.5}
  \Vertex(37.5,66){2.5}
  \ArrowLine(50,88)(37.5,66)
  \ArrowLine(37.5,22)(50,0)
  \Photon(25,44)(37.5,66){3}{5}
  \DashLine(37.5,22)(25,44){3}
  \Text(33,75)[lb]{$\fbf$}
  \Text(-1,51)[lb]{$(\zb)$}
  \Text(17,53)[lb]{$\zb$}
  \Text(48,44)[lc]{$\ff$}
  \Text(17,35)[lt]{$\hb$}
  \Text(33,13)[lt]{$\ff$}
  \Text(1,1)[lb]{}
\end{picture}
& + &
\begin{picture}(55,88)(0,41)
  \Photon(0,44)(25,44){3}{5}
  \Vertex(25,44){2.5}
  \ArrowArcn(0,44)(44,28,-28)
  \Vertex(37.5,22){2.5}
  \Vertex(37.5,66){2.5}
  \ArrowLine(50,88)(37.5,66)
  \ArrowLine(37.5,22)(50,0)
  \DashLine(25,44)(37.5,66){3}
  \Photon(37.5,22)(25,44){3}{5}
  \Text(33,75)[lb]{$\fbf$}
  \Text(-1,51)[lb]{$(\zb)$}
  \Text(17,53)[lb]{$\hb$}
  \Text(48,44)[lc]{$\ff$}
  \Text(17,35)[lt]{$\zb$}
  \Text(33,13)[lt]{$\ff$}
  \Text(1,1)[lb]{}
\end{picture}
 \nl \nl [3mm]
& + & 
\begin{picture}(55,88)(0,41)
  \Photon(0,44)(25,44){3}{5}
  \Vertex(25,44){2.5}
  \ArrowArcn(0,44)(44,28,-28)
  \Vertex(37.5,22){2.5}
  \Vertex(37.5,66){2.5}
  \ArrowLine(50,88)(37.5,66)
  \ArrowLine(37.5,22)(50,0)
  \DashLine(25,44)(37.5,66){3}
  \DashLine(37.5,22)(25,44){3}
  \Text(33,75)[lb]{$\fbf$}
  \Text(-1,51)[lb]{$(\zb)$}
  \Text(17,53)[lb]{$\hb$}
  \Text(48,44)[lc]{$\ff$}
  \Text(17,35)[lt]{$\hkn$}
  \Text(33,13)[lt]{$\ff$}
  \Text(1,1)[lb]{}
\end{picture}
& + &
\begin{picture}(55,88)(0,41)
  \Photon(0,44)(25,44){3}{5}
  \Vertex(25,44){2.5}
  \ArrowArcn(0,44)(44,28,-28)
  \Vertex(37.5,22){2.5}
  \Vertex(37.5,66){2.5}
  \ArrowLine(50,88)(37.5,66)
  \ArrowLine(37.5,22)(50,0)
  \DashLine(25,44)(37.5,66){3}
  \DashLine(37.5,22)(25,44){3}
  \Text(33,75)[lb]{$\fbf$}
  \Text(-1,51)[lb]{$(\zb)$}
  \Text(17,53)[lb]{$\hkn$}
  \Text(48,44)[lc]{$\ff$}
  \Text(17,35)[lt]{$\hb$}
  \Text(33,13)[lt]{$\ff$}
  \Text(1,1)[lb]{}
\end{picture}
&+&
\begin{picture}(55,88)(0,41)
  \Photon(0,44)(25,44){3}{5}
  \Vertex(25,44){2.5}
  \ArrowLine(50,88)(25,44)
  \ArrowLine(25,44)(50,0)
  \Text(33,75)[lb]{$\fbf$}
  \Text(33,13)[lt]{$\ff$}
  \Text(1,1)[lb]{}
\SetScale{2.0}
  \Line(17.5,17.5)(7.5,27.5)
  \Line(7.5,17.5)(17.5,27.5)
\end{picture}
\nl 
\nl [15mm]
& &  & &  & &
\begin{picture}(75,20)(0,7.5)
\Text(12.5,13)[bc]{$\ff$}
\Text(62.5,13)[bc]{$\ff$}
\Text(37.5,26)[bc]{$\ff$}
\Text(37.5,-8)[tc]{$\hb$}
\Text(0,-7)[lt]{$$}
  \ArrowLine(0,10)(25,10)
  \ArrowArcn(37.5,10)(12.5,180,0)
  \DashCArc(37.5,10)(12.5,180,0){3}
  \Vertex(25,10){2.5}
  \Vertex(50,10){2.5}
  \ArrowLine(50,10)(75,10)
\end{picture}
\eaa
\]
\vspace*{2mm}
\caption[$\hb$ cluster.]
{$\hb$ cluster: the vertices and the counterterm. 
\label{fig:H_cluster}}
\end{figure}
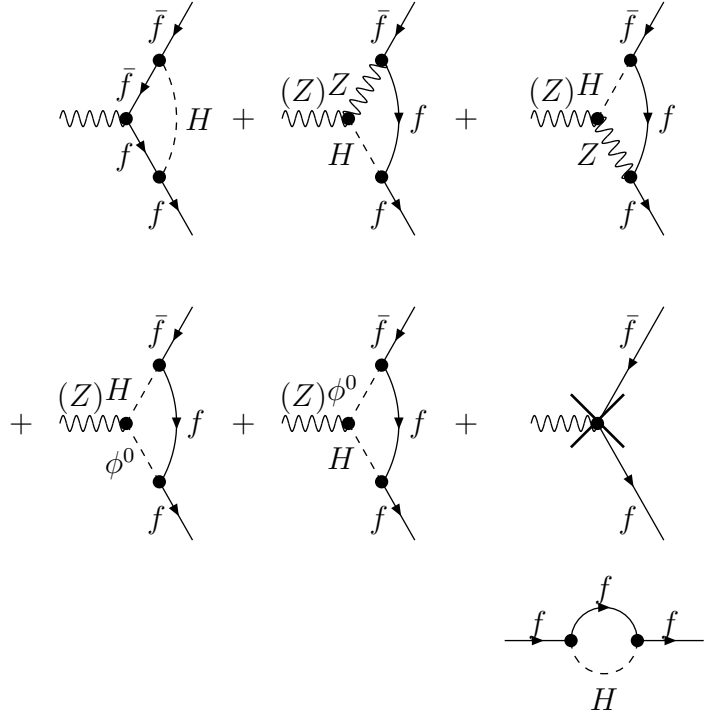

The diagrams of \fig{fig:H_cluster} contribute to the $\hb$ cluster, 

Separating UV poles, we have:
\bqa
\vvertil{\gamma{\sss H}}{\sss{Q}}{\sman} & = &
  \cvertil{\gamma{\sss H}}{\sss{Q}}{\sman},
\nll
\vvertil{\gamma{\sss H}}{\sss{D}}{\sman} & = &
  \cvertil{\gamma{\sss H}}{\sss{D}}{\sman},
\nll
 \vvertil{z{\sss H}}{\sss L}{\sman}   & = &
  \frac{1}{4} \rtw \pole
+ \cvertil{z{\sss H}}{\sss L}{\sman},
\nll
 \vvertil{z{\sss H}}{\sss{Q}}{\sman} & = & 
 \frac{1}{16} \frac{1}{\qum} \rtw  \frac{1}{\stws} \pole
+ \cvertil{z{\sss H}}{\sss{Q}}{\sman},
\nll
  \vvertil{z{\sss H}}{\sss{D}}{\sman} & = &
  \cvertil{z{\sss H}}{\sss{D}}{\sman},
\label{higgs_formfactors}
\eqa
with the finite parts:
\bqa
{\cal F}^{\gamma{\sss H}}_{\sss Q}\lpar \sman \rpar
&=&
 \frac{1}{8} \ruw \Biggl\{
   8 \mts \cff{-\mts}{-\mts}{-s}{\mtl}{\mhl}{\mtl}
\nll &&
            + \fbff{0}{-s}{\mtl}{\mtl}+\Lmmt-2 
            - \fbff{d1}{-\mts}{\mtl}{\mhl}
\nll &&
                -2 \lpar 1-4 \ruh \rpar \mhs \bff{0p}{-\mts}{\mtl}{\mhl}
                -2 \frac{\mhs}{\sdtit} \Longab{\mtl}{\mtl}{\mhl} \Biggr\}, 
\eqa
\bqa 
{\cal F}^{\gamma{\sss H}}_{\sss D}\lpar \sman \rpar  & = &
 - \frac{Q_t}{2\tcit} 
\frac{\rtw }{\sdtit} 
\Biggl\{
     -6 \mhs 
 \cff{-\mts}{-\mts}{-s}{\mtl}{\mhl}{\mtl}
\nll &&
      +3 \fbff{0}{-\sman}{\mtl}{\mtl}-4 \fbff{0}{-\mts}{\mtl}{\mhl}-\Lmmt+2    
\nll &&
      + \fbff{d1}{-\mts}{\mtl}{\mhl}
      +6 \frac{\mhs}{\sdtit} \Longab{\mtl}{\mtl}{\mhl}    
                                                \Biggr\},
\eqa 
\vspace{-3mm}
\bqa
{\cal F}^{z{\sss H}}_{\sss L}\lpar \sman \rpar
&=&
 \frac{1}{4}\ruw \Biggl\{
    4  \mts  \cff{-\mts}{-\mts}{-s}{\mtl}{\mhl}{\mtl}
\nll[1mm] &&
  +\Bigl[4\lpar 1-\ruz\rpar +(1-\rhz)^2 \Rz \Bigr]
     \mzs \cff{-\mts}{-\mts}{-s}{\mhl}{\mtl}{\mzl}
\nll[1mm] &&
   +2 \fbff{0}{-s}{\mzl}{\mhl}-\frac{1}{2} \fbff{0}{-s}{\mtl}{\mtl}
   +\frac{1}{2} \Lmmt
\nll[1mm] &&
   -\frac{1}{2} \fbff{d1}{-\mts}{\mtl}{\mhl}
   - (1-4 \ruh) {\mhs} \bff{0p}{-\mts}{\mtl}{\mhl} + 2 
\nll[1mm] &&
   +(1-\rhz) \Rz
   \Bigl[ \fbff{0}{-\mts}{\mzl}{\mtl}-\fbff{0}{-\mts}{\mtl}{\mhl} \Bigr]
\\[1mm] &&
                +\frac{\mzs}{\sdtit}
  \biggl[
                 \lpar \rhz-8 \ruz \rpar \Longab{\mtl}{\mtl}{\mhl} 
  -2 \lpar 3 - \rhz + 4\rtz \rpar 
          \LongHi{\mtl}{\mhl}{\mzl} 
  \biggr]           \Biggr\},
\nonumber
\eqa
\vspace{-3mm}
\bqa
{\cal F}^{z {\sss H}}_{\sss Q}\lpar \sman \rpar&=&\ruw\Biggl\{
  \mts \cff{-\mts}{-\mts}{-s}{\mtl}{\mhl}{\mtl}
\nll[1mm] &&
 +\Bigl[ 1 + \lpar 1-\rhz \rpar \Rz \Bigr] 
 \mzs \cff{-\mts}{-\mts}{-s}{\mhl}{\mtl}{\mzl}
\nll[1mm]&&
 +\frac{1}{8} \bigg[ \fbff{0}{-s}{\mtl}{\mtl}+\Lmmt-2 \bigg]
\nll &&
 +\Rz \Bigl[ \fbff{0}{-\mts}{\mzl}{\mtl}-\fbff{0}{-\mts}{\mtl}{\mhl} \Bigr]
 -\frac{1}{8}\fbff{d1}{-\mts}{\mtl}{\mhl}
\nll &&
 -\frac{1}{4}\lpar 1-4 \ruh \rpar\mhs \bff{0p}{-\mts}{\mtl}{\mhl}
 -\frac{1}{4} \frac{\mhs}{\sdtit} \Longab{\mtl}{\mtl}{\mhl} 
\nll &&  
 +\frac{1}{4} \frac{\au}{\delta_t} 
\Biggl(
   \Bigl[ 4 \rtz  + \lpar 3+\rhz \rpar (1-\rhz) \Rz \Bigr]
   \mzs \cff{-\mts}{-\mts}{-s}{\mhl}{\mtl}{\mzl}
\nll[2mm] &&
                + \fbff{0}{-s}{\mtl}{\mtl} - 2 \fbff{0}{-s}{\mzl}{\mhl}-3
\nll[2mm] &&
+ \lpar 3+\rhz \rpar \Rz
\Bigl[\fbff{0}{-\mts}{\mzl}{\mtl} -\fbff{0}{-\mts}{\mtl}{\mhl} \Bigr]
\\ && 
  - 2 \frac{\mzs}{\sdtit} 
         \bigg[ \lpar 1 - 4 \ruh \rpar \rhz \Longab{\mtl}{\mtl}{\mhl}
  - \lpar 3 - \rhz + 4\rtz \rpar 
          \LongHi{\mtl}{\mhl}{\mzl} 
    \bigg] 
\Biggr)
\Biggr\},
\nonumber
\eqa
\vspace{-3mm}
\bqa
{\cal F}^{z{\sss H}}_{\sss D}\lpar \sman \rpar  & = &
-\frac{v_t}{2\tcit\ctws} 
\frac{1}{\sdtit}  \Biggl\{
   -3 \rtz \mhs \cff{-\mts}{-\mts}{-s}{\mtl}{\mhl}{\mtl}
\nll &&
   + 2 \biggl[
         2 \lpar \rhz-1 \rpar \frac{\mts}{s}-\rhz+2 \rtz\biggr] \mzs  
 \cff{-\mts}{-\mts}{-s}{\mhl}{\mtl}{\mzl}
\nll &&
                -4 \frac{\mts}{s} 
    \biggl[ \fbff{0}{-\mts}{\mzl}{\mtl}-\fbff{0}{-\mts}{\mtl}{\mhl} \biggr]
 -2 \fbff{0}{-\sman}{\mzl}{\mhl}
\nll &&
 +2\fbff{0}{-\mts}{\mzl}{\mtl}
                +\frac{3}{2} \rtz 
 \biggl[\fbff{0}{-\sman}{\mtl}{\mtl}
      -\fbff{0}{-\mts}{\mtl}{\mhl} \biggr]
\nll &&
                +\frac{1}{2} \rhz \biggl[
       \fbff{0}{-\mts}{\mtl}{\mhl}+\Lmmh-1 \biggr]
\nll &&
 -\frac{1}{2} \rtz \biggl[
       \fbff{0}{-\mts}{\mtl}{\mhl}+\Lmmt-2 \biggr]
  +3 \rtz \frac{\mhs}{\sdtit} \Longab{\mtl}{\mtl}{\mhl}
             \Biggr\}.\qquad
\eqa

Again, the five (one does not exist) scalar form factors in \eqn{higgs_formfactors} 
are {\em separately} gauge-invariant. Note also that UV poles persisting
in the scalar form factors of the $\hb$ cluster cancel exactly the corresponding poles
of the $\zb$ cluster. In other words, the form factors of the `neutral sector'
cluster ($\zb+\hb$) are UV finite.

In total, we have 11 separately gauge-invariant {\em building blocks}
that originate from $\zb$ and $\hb$ clusters.
\subsubsection{Form factors of the $W$ cluster}
\begin{figure}[h]
\vspace{-18.5mm}
\[
\baa{ccccccccc}

\begin{picture}(55,88)(0,41)
  \Photon(0,44)(25,44){3}{5}
  \Vertex(25,44){2.5}
  \PhotonArc(0,44)(44,-28,28){3}{9}
  \Vertex(37.5,22){2.5}
  \Vertex(37.5,66){2.5}
  \ArrowLine(50,88)(37.5,66)
  \ArrowLine(37.5,66)(25,44)
  \ArrowLine(25,44)(37.5,22)
  \ArrowLine(37.5,22)(50,0)
  \Text(33,75)[lb]{$\fbf$}
  \Text(21,53)[lb]{$\fbfp$}
  \Text(47.5,44)[lc]{$\wb$}
  \Text(21,35)[lt]{$\ffp$}
  \Text(33,13)[lt]{$\ff$}
  \Text(1,1)[lb]{}
\end{picture}
& & 
&+&
\begin{picture}(55,88)(0,41)
  \Photon(0,44)(25,44){3}{5}
  \Vertex(25,44){2.5}
  \DashCArc(0,44)(44,-28,28){3}
  \Vertex(37.5,22){2.5}
  \Vertex(37.5,66){2.5}
  \ArrowLine(50,88)(37.5,66)
  \ArrowLine(37.5,66)(25,44)
  \ArrowLine(25,44)(37.5,22)
  \ArrowLine(37.5,22)(50,0)
  \Text(33,75)[lb]{$\fbf$}
  \Text(19,53)[lb]{$\fbfp$}
  \Text(47.5,44)[lc]{$\hkg$}
  \Text(19,35)[lt]{$\ffp$}
  \Text(33,13)[lt]{$\ff$}
  \Text(1,1)[lb]{}
\end{picture}
& &
&+&
\begin{picture}(55,88)(0,41)
  \Photon(0,44)(25,44){3}{5}
  \Vertex(25,44){2.5}
  \ArrowLine(50,88)(25,44)
  \ArrowLine(25,44)(50,0)
  \Text(33,75)[lb]{$\fbf$}
  \Text(33,13)[lt]{$\ff$}
  \Text(1,1)[lb]{}
\SetScale{2.0}
  \Line(17.5,17.5)(7.5,27.5)
  \Line(7.5,17.5)(17.5,27.5)
\end{picture}
\nl \nl [16mm]
&& && 
&\hspace*{-2.5mm}+\hspace*{-2.5mm}&
\begin{picture}(75,20)(0,7.5)
\Text(12.5,13)[bc]{$\ff$}
\Text(62.5,13)[bc]{$\ff$}
\Text(37.5,26)[bc]{$\ffp$}
\Text(37.5,-8)[tc]{$\hkg$}
\Text(0,-7)[lt]{$$}
  \ArrowLine(0,10)(25,10)
  \ArrowArcn(37.5,10)(12.5,180,0)
  \DashCArc(37.5,10)(12.5,180,0){3}
  \Vertex(25,10){2.5}
  \Vertex(50,10){2.5}
  \ArrowLine(50,10)(75,10)
\end{picture}
&\hspace*{-2.5mm}+\hspace*{-2.5mm}& 
\begin{picture}(75,20)(0,7.5)
\Text(12.5,13)[bc]{$\ff$}
\Text(62.5,13)[bc]{$\ff$}
\Text(37.5,26)[bc]{$\ffp$}
\Text(37.5,-8)[tc]{$\wb$}
\Text(0,-7)[lt]{$$}
  \ArrowLine(0,10)(25,10)
  \ArrowArcn(37.5,10)(12.5,180,0)
  \PhotonArc(37.5,10)(12.5,180,0){3}{7}
  \Vertex(25,10){2.5}
  \Vertex(50,10){2.5}
  \ArrowLine(50,10)(75,10)
\end{picture}
\nl[-5mm] 
\begin{picture}(55,88)(0,41)
  \Photon(0,44)(25,44){3}{5}
  \Vertex(25,44){2.5}
  \ArrowArcn(0,44)(44,28,-28)
  \Vertex(37.5,22){2.5}
  \Vertex(37.5,66){2.5}
  \ArrowLine(50,88)(37.5,66)
  \ArrowLine(37.5,22)(50,0)
  \Photon(25,44)(37.5,66){3}{5}
  \Photon(37.5,22)(25,44){3}{5}
  \Text(33,75)[lb]{$\fbf$}
  \Text(15,53)[lb]{$\wb$}
  \Text(48,44)[lc]{$\ffp$}
  \Text(15,35)[lt]{$\wb$}
  \Text(33,13)[lt]{$\ff$}
  \Text(1,1)[lb]{}
\end{picture}
&+&
\begin{picture}(55,88)(0,41)
  \Photon(0,44)(25,44){3}{5}
  \Vertex(25,44){2.5}
  \ArrowArcn(0,44)(44,28,-28)
  \Vertex(37.5,22){2.5}
  \Vertex(37.5,66){2.5}
  \ArrowLine(50,88)(37.5,66)
  \ArrowLine(37.5,22)(50,0)
  \Photon(25,44)(37.5,66){3}{5}
  \DashLine(37.5,22)(25,44){3}
  \Text(33,75)[lb]{$\fbf$}
  \Text(15,53)[lb]{$\wb$}
  \Text(48,44)[lc]{$\ffp$}
  \Text(19,35)[lt]{$\hkg$}
  \Text(33,13)[lt]{$\ff$}
  \Text(1,1)[lb]{}
\end{picture}
&+&
\begin{picture}(55,88)(0,41)
  \Photon(0,44)(25,44){3}{5}
  \Vertex(25,44){2.5}
  \ArrowArcn(0,44)(44,28,-28)
  \Vertex(37.5,22){2.5}
  \Vertex(37.5,66){2.5}
  \ArrowLine(50,88)(37.5,66)
  \ArrowLine(37.5,22)(50,0)
  \DashLine(25,44)(37.5,66){3}
  \Photon(37.5,22)(25,44){3}{5}
  \Text(33,75)[lb]{$\fbf$}
  \Text(19,53)[lb]{$\hkg$}
  \Text(48,44)[lc]{$\ffp$}
  \Text(15,35)[lt]{$\wb$}
  \Text(33,13)[lt]{$\ff$}
  \Text(1,1)[lb]{}
\end{picture}
&+&
\begin{picture}(55,88)(0,41)
  \Photon(0,44)(25,44){3}{5}
  \Vertex(25,44){2.5}
  \ArrowArcn(0,44)(44,28,-28)
  \Vertex(37.5,22){2.5}
  \Vertex(37.5,66){2.5}
  \ArrowLine(50,88)(37.5,66)
  \ArrowLine(37.5,22)(50,0)
  \DashLine(25,44)(37.5,66){3}
  \DashLine(37.5,22)(25,44){3}
  \Text(33,75)[lb]{$\fbf$}
  \Text(19,53)[lb]{$\hkg$}
  \Text(48,44)[lc]{$\ffp$}
  \Text(19,35)[lt]{$\hkg$}
  \Text(33,13)[lt]{$\ff$}
  \Text(1,1)[lb]{}
\end{picture}
&+&
\begin{picture}(55,88)(0,41)
  \Photon(0,44)(25,44){3}{5}
  \Vertex(25,44){2.5}
  \ArrowLine(50,88)(25,44)
  \ArrowLine(25,44)(50,0)
  \Text(33,75)[lb]{$\fbf$}
  \Text(33,13)[lt]{$\ff$}
  \Text(1,1)[lb]{}
\SetScale{2.0}
  \Line(17.5,17.5)(7.5,27.5)
  \Line(7.5,17.5)(17.5,27.5)
\end{picture}
\eaa
\]
\vspace{10mm}
\caption
[$\wb$ cluster.]
{$\wb$ cluster: the first row shows the abelian diagrams of the cluster,
the last row the non-abelian diagrams; the second row shows diagrams that contribute
to both counterterm crosses (last diagrams in first and third rows).
\label{fig:W_cluster} }
\end{figure}
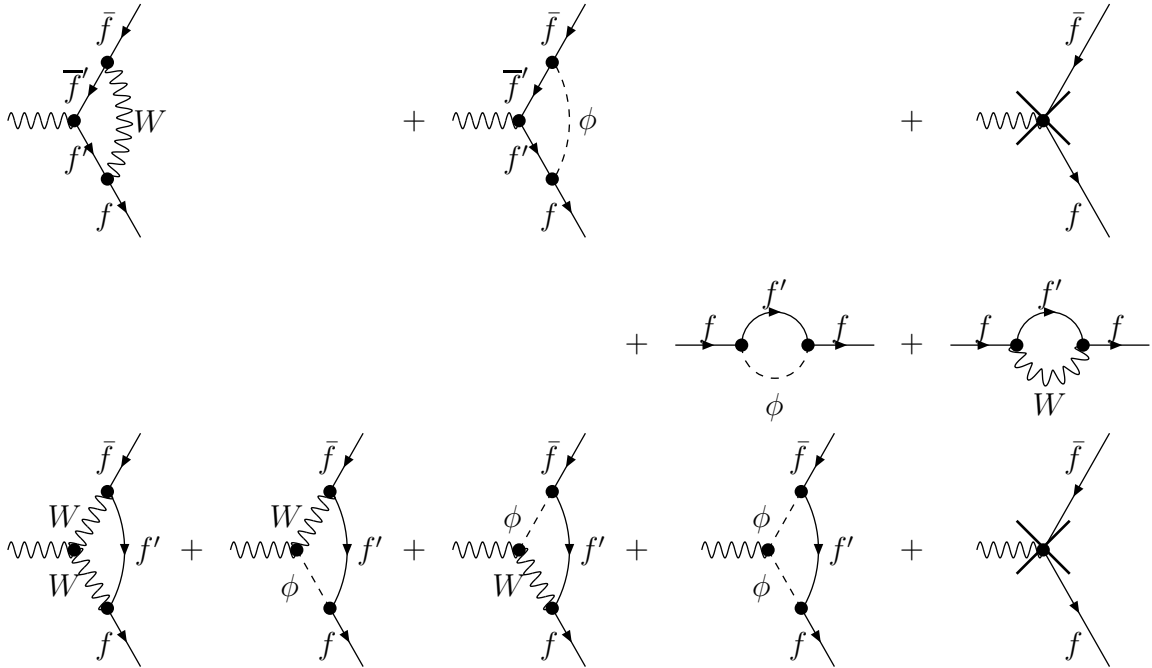

Finally, the $\wb$ cluster is made of the diagrams shown in 
Fig.~\ref{fig:W_cluster}.

In the formulae below, we present contributions to scalar form factors from all 
the diagrams of the $\wb$ cluster, not subdividing them into abelian and
non-abelian contributions. To some extent two sub-clusters are automatically 
marked by the type of arguments of $\scff{0}$ functions and typical coupling
constants. Separating poles, we have:
\bqa
 {\cal F}^{\gamma W}_{\sss L} &=& 
\frac{\qd}{2\tcit}
\Biggl\{ \lrbr 3+\lpar 1 + \rD \rpar^2+\frac{2}{\Rw} \rrbr
              \mws \cff{-\mts}{-\mts}{-s}{\mbl}{\mwl}{\mbl} 
\nll &&
      -\frac{1}{2} \lpar 6 + \rD \rpar 
            \bigg[ \fbff{0}{-s}{\mbl}{\mbl}-\fbff{0}{-\mts}{\mbl}{\mwl} \bigg]
\nll &&
      +\frac{1}{2} \lpar 2-\rD\rpar \biggl[\fbff{d\sss{W}}{-\mts}{\mbl}{\mwl}-1 \bigg]
\nll &&
      +\Bigg[ 4-\lpar 2+\rD \rpar \lpar 3 + 3\rtw + \rbw \rpar \Bigg]
       \frac{ \mws}{\sdtit} \Longab{\mtl}{\mbl}{\mwl}
\Biggr\}
\nll &&
 -\rbw \lpar 2-\rD  \rpar \mws \cff{-\mts}{-\mts}{-s}{\mwl}{\mbl}{\mwl}
\nll &&
 +\frac{1}{2} \lpar  6-\rD \rpar 
     \bigg[ \fbff{0}{-s}{\mwl}{\mwl}-\fbff{0}{-\mts}{\mbl}{\mwl} \bigg]
\nll &&
 + 2 \fbff{0}{-\mts}{\mbl}{\mwl}
 + \frac{1}{2} \lpar 2-3\rD \rpar \bigg[ \fbff{d\sss{W}}{-\mts}{\mbl}{\mwl} + 1 \bigg]
\nll &&
 -\Bigg(4-\bigg[ 2 + \rtw \lpar 13+\rtw\rpar - \rbw \lpar 1+\rbw \rpar \bigg]
      \frac{\mws}{\sdtit} \Bigg) \Longna{\mtl}{\mbl}{\mwl} 
\nll &&
 - \frac{\qt}{4\tcit}
   \bigg(\Delta_{{\cal F}_{\rm Im}}-3\biggr) i\, {\rm Im} \fbff{0}{-\mts}{\mwl}{\mbl},
\label{gammaWL}
\eqa
where
\vspace{-6mm}
\bqa
\Delta_{{\cal F}_{\rm Im}}=\lpar 1-\rbw \rpar \frac{\lpar 2+\rbw\rpar}{\rtw}+\rtw\,.
\eqa
The last term in \eqn{gammaWL} is due to a non-cancellation of the imaginary part of the
function
$\fbff{0}{-\mts}{\mwl}{\mbl}$ which appear in
real counterterms and complex-valued vertices.
\bqa
{\cal F}^{\gamma W}_{\sss Q} &=&
   \frac{\rtw}{4\qu} \Biggl\{
   \qd \Biggl[-4 \mws \cff{-\mts}{-\mts}{-s}{\mbl}{\mwl}{\mbl}
\nll &&
   +\fbff{0}{-s}{\mbl}{\mbl}-\fbff{0}{-\mts}{\mbl}{\mwl}-1
   -\frac{2+\rbw}{\rtw}  \fbff{d\sss{W}}{-\mts}{\mbl}{\mwl}
\nll &&
   -\Biggl( \frac{\lpar 1-\rbw\rpar \lpar 2+\rbw\rpar}{\rtw}
   -1 - \rtw + 2\rbw \Biggr) \mws \bff{0p}{-\mts}{\mbl}{\wml}
\nll &&
   +2 \lpar 3+\rD \rpar\frac{\mws}{\sdtit}\Longab{\mtl}{\mbl}{\mwl}\Biggr]
\nll [2mm] &&
   -2 \tcit \bigg[ 2 \mbs \cff{-\mts}{-\mts}{-s}{\wml}{\mbl}{\wml}
\nll [2mm] &&
   -\fbff{0}{-s}{\wml}{\wml}+\fbff{0}{-\mts}{\mbl}{\mwl}-1
   +\frac{2+\rbw}{\rtw} \fbff{d\sss{W}}{-\mts}{\mbl}{\mwl}
\nll &&
   +\Biggl( \frac{ 2-\rbw \lpar 1+\rbw \rpar}{\rtw}
     -1-\rtw+2\rbw \Biggr) \mws \bff{0p}{-\mts}{\mbl}{\wml}
\nll &&                                    
   +2 \bigg( 7+\rP \bigg) \frac{ \mws}{\sdtit} \Longna{\mtl}{\mbl}{\mwl}
 \bigg] \Biggr\}
 + \frac{1}{4} \Delta_{{\cal F}_{\rm Im}} i\,{\rm Im} \fbff{0}{-\mts}{\mwl}{\mbl}\,,
\eqa
\bqa
{\cal F}^{\gamma{\sss W}}_{\sss D} &=& -\frac{1}{\sdtit } \Bigg\{
   \frac{\qd}{2\tcit} \Biggl( -2 \lrbr 8+ {\rD}^2-6 \rbw+\frac{2}{\Rw} \rrbr 
   \mws \cff{-\mts}{-\mts}{-s}{\mbl}{\mwl}{\mbl}
\nll [2mm] &&
 + \lpar 10+\rtw-3 \rbw \rpar 
         \bigg[\fbff{0}{-s}{\mbl}{\mbl}-\fbff{0}{-\mts}{\mbl}{\mwl} \bigg]
\nll &&
 + \lpar 2+\rP \rpar
       \biggl[ \fbff{d\sss{W}}{-\mts}{\mbl}{\mwl} + 1 \biggr]
\nll &&
 + 6 \bigg[ \lpar 1+\rtw \rpar  \lpar 2+\rtw\rpar-\rbw \lpar 1+\rbw \rpar \bigg]
             \frac{\mws}{\sdtit}  \Longab{\mtl}{\mbl}{\mwl} \Biggr)
\nll &&
 + 2 \lrbr \lpar 1-\rP \rpar \lpar 2-\rtw \rpar-\rbw 
                    \lpar 1-4 \rbw-\frac{1}{\Rw} \rpar \rrbr
\nll [2mm] &&
   \times \mws \cff{-\mts}{-\mts}{-s}{\mwl}{\mbl}{\mwl}
\nll [2mm] &&    
 + \lpar 6-\rtw-5 \rbw \rpar \bigg[ \fbff{0}{-s}{\mwl}{\mwl}
                                  - \fbff{0}{-\mts}{\mbl}{\mwl} \bigg]
\nll &&
 + \lpar 2+\rP \rpar \bigg[ \fbff{d\sss{W}}{-\mts}{\mbl}{\mwl} - 1 \bigg]
\nll &&
 - 6 \bigg[ \lpar 1-\rtw \rpar \lpar 2+  \rtw      \rpar-\rbw 
                              \lpar 1+2 \rtw+\rbw \rpar \bigg]
     \frac{\mws}{\sdtit} \Longna{\mtl}{\mbl}{\mwl} \Bigg\},
\qquad
\eqa
\bqa
{\cal F}^{z{\sss W}}_{\sss L} &=& \frac{\vpa{d}{}}{4 \tcit} 
\Biggl\{
  \lrbr 3 + \lpar 1 + \rD \rpar^2 + \frac{2}{\Rw} \rrbr 
      \mws \cff{-\mts}{-\mts}{-s}{\mbl}{\mwl}{\mbl}
\nll &&
  - \frac{1}{2}\lpar 6+\rD \rpar
     \bigg[\fbff{0}{-s}{\mbl}{\mbl}-\fbff{0}{-\mts}{\mbl}{\mwl} \bigg]
\nll &&
  + \frac{1}{2}\lpar 2-\rD \rpar 
    \bigg[ \fbff{d\sss{W}}{-\mts}{\mbl}{\mwl} - 1 \bigg]
\nll &&
  + \bigg[ 4-\lpar 2+\rD\rpar \lpar 3+3\rtw+\rbw \rpar \bigg]
      \frac{\mws}{\sdtit} \Longab{\mtl}{\mbl}{\mwl} 
\Biggr\}
\nll[1mm]&&
  + \rbw\mws \cff{-\mts}{-\mts}{-s}{\mbl}{\mwl}{\mbl} 
\nll[2mm]&&
  + \frac{1}{4} \Biggl\{ \rD\fbff{0}{-s}{\mwl}{\wml}
               -\rtw\bigg[\fbff{0}{-\mts}{\mbl}{\mwl} - 1\bigg]
\nll &&
               +\rbw\bigg[\fbff{0}{-s}{\mbl}{\mbl} - 2 \bigg]
 - \lpar 2+\rbw\rpar \fbff{d\sss{W}}{-\mts}{\mbl}{\mwl}
\nll &&
 - \bigg[ \lpar 1-\rtw \rpar \lpar 2+\rtw \rpar
 - \rbw \lpar 1-2 \rtw+\rbw \rpar \bigg] \mws \bff{0p}{-\mts}{\mbl}{\wml}
\nll &&
  -2\rbw \lpar 1+\rD \rpar \frac{\mws}{\sdtit} \Longab{\mtl}{\mbl}{\mwl} 
\Biggr\}
\nll &&
  - \ctws \Biggl\{ \Bigg( 4 \lpar 1-\rtw \rpar 
  + \frac{1}{2} \rbw \lrbr 4+\frac{\stws-\ctws}{\ctws} 
      \lpar 4+\rD \rpar \rrbr\Bigg)
\nll [2mm] &&
  \times \mws \cff{-\mts}{-\mts}{-s}{\mwl}{\mbl}{\mwl}
\nll [2mm] &&
 + \frac{1}{2} \lpar 2+\rD \rpar
   \bigg[  \fbff{0}{-\mts}{\mwl}{\mwl}-\fbff{0}{-\mts}{\mbl}{\mwl} \bigg]
\nll &&
 - \frac{1}{2} \lpar 2-\rD \rpar
   \bigg[ \fbff{d\sss{W}}{-\mts}{\mbl}{\mwl} + 1 \bigg]
 - 2 \fbff{0}{-\mts}{\mbl}{\mwl}
\nll &&
 - \frac{1}{2} \Biggl( 4 + 12 \rtw-\frac{\stws-\ctws}{\ctws} 
                                \rtw \lpar 7+\rtw \rpar 
\nll &&
 - \rbw \lrbr 4 + \frac{\stws-\ctws}{\ctws}\lpar 1 - \rbw \rpar 
\rrbr \Biggr)
    \frac{\mws}{\sdtit}  \Longna{\mtl}{\mbl}{\mwl}\Biggr\}
\nll &&
 + \frac{1}{8\tcit}
\bigg(3 \vpau-\vmau\Delta_{{\cal F}_{\rm Im}}  \bigg) i\, {\rm Im} \fbff{0}{-\mts}{\mwl}{\mbl}\,,
\eqa
\bqa
{\cal F}^{z{\sss W}}_{\sss Q} &=& - \frac{\rtw}{4}\frac{\vpa{d}{}}{\vma{t}{}}
\Bigg\{4 \mws 
\cff{-\mts}{-\mts}{-s}{\mbl}{\mwl}{\mbl}
\nll &&
  -\fbff{0}{-s}{\mbl}{\mbl}+\fbff{0}{-\mts}{\mbl}{\mwl}+1
  -2\lpar 3+\rD \rpar \frac{\mws}{\sdtit} \Longab{\mtl}{\mbl}{\mwl}
\Bigg\}
\nll &&
           -\frac{1}{4}\lpar 2+\rbw\rpar\fbff{d\sss{W}}{-\mts}{\mbl}{\mwl}
\nll &&
           -\frac{1}{4}
           \bigg[ \lpar 1-\rtw \rpar\lpar 2+\rD\rpar+\rbw \rD\bigg] 
      \mws \bff{0p}{-\mts}{\mbl}{\wml}
\nll &&
      +\tcit\ctws \frac{\rtw}{\vma{t}{}} 
        \Biggl\{\frac{\stws-\ctws}{\ctws} 
    \bigg(\rbw\mws \cff{-\mts}{-\mts}{-s}{\mwl}{\mbl}{\mwl}
\nll &&
   -\frac{1}{2} \bigg[\fbff{0}{-s}{\mwl}{\mwl} - \fbff{0}{-\mts}{\mbl}{\mwl} + 1 \bigg] \bigg)
\nll &&
  -\bigg[8-\frac{\stws-\ctws}{\ctws} \lpar 3+\rtw+\rbw \rpar \bigg] \frac{\mws}{\sdtit} 
 \Longna{\mtl}{\mbl}{\mwl} \Biggr\}
\nll &&
 + \frac{1}{4}\Delta_{{\cal F}_{\rm Im}}i\,{\rm Im} \fbff{0}{-\mts}{\mwl}{\mbl},
\eqa
\bqa
{\cal F}^{z{\sss W}}_{\sss D} &=& \frac{1}{\sdtit}\Biggr\{
  \frac{\vd+\ad}{2\tcit}  \Biggr( \lrbr  8 + \lpar{\rD}\rpar^2 - 6 \rbw + \frac{2}{\Rw} \rrbr
\mws \cff{-\mts}{-\mts}{-s}{\mbl}{\mwl}{\mbl}
\nll [2mm] &&
   -\frac{1}{2}\lpar 10+\rtw-3 \rbw \rpar 
      \bigg[\fbff{0}{-s}{\mbl}{\mbl}-\fbff{0}{-\mts}{\mbl}{\mwl}\bigg]
\nll && 
   -\frac{1}{2}\lpar 2+\rP \rpar
               \bigg[\fbff{d\sss{W}}{-\mts}{\mbl}{\mwl} + 1 \bigg]
\nll &&
   -3 \bigg[ \lpar 1+\rtw \rpar \lpar 2+\rtw \rpar
                               -\rbw \lpar 1+\rbw \rpar \bigg]
                   \frac{\mws}{\sdtit}  \Longab{\mtl}{\mbl}{\mwl}
                           \Biggr)
\nll &&
  + \rbw \Biggl(\mws \cff{-\mts}{-\mts}{-s}{\mbl}{\mwl}{\mbl}
\nll &&
  +\frac{1}{2}\bigg[\fbff{0}{-s}{\mbl}{\mbl}-\fbff{0}{-\mts}{\mbl}{\mwl}-1\bigg]
  -\frac{1}{2}\fbff{d\sss{W}}{-\mts}{\mbl}{\mwl}
\nll &&
  -3 \lpar 1+\rD \rpar \frac{\mws}{\sdtit} \Longab{\mtl}{\mbl}{\mwl}
                         \Biggr)
\nll &&
  -\ctws\Biggl(
\Bigg[ 4\rbw-\frac{\stws-\ctws}{\ctws}
             \bigg[ \lpar 1-\rtw\rpar \lpar 2-\rtw \rpar
             -\rbw  \lpar 5-\rtw-4\rbw-\frac{1}{\Rz} \rpar \bigg]
  \Bigg]
\nll [2mm] &&
\times  \mws \cff{-\mts}{-\mts}{-s}{\mwl}{\mbl}{\mwl}
\nll [2mm] &&
 +\frac{1}{2}\lrbr 4 - \frac{\stws-\ctws}{\ctws} \lpar 4 - \rtw - 5 \rbw \rpar  \rrbr
  \bigg[ \fbff{0}{-s}{\mwl}{\mwl}-\fbff{0}{-\mts}{\mbl}{\mwl} \bigg]
\nll &&
 +\frac{1}{2}\lpar 4 - \frac{\stws-\ctws}{\ctws} r_{tb}^+ \rpar
  \Bigg[ \fbff{d\sss{W}}{-\mts}{\mbl}{\mwl} - 1 
\nll &&
  -6 \lpar  1 - r_{tb}^+ \rpar  
     \frac{\mws}{\sdtit}  \Longna{\mtl}{\mbl}{\mwl}\Biggr] \Biggl) \Biggr\}.  
\eqa
\vspace*{1mm}

Here we introduce more symbols, 
which were not given in ~\eqn{abbrev-old}:
\vspace*{-1mm}
\bqa 
 r_{tb}^{\pm}=  \rtw\pm\rbw\,,\qquad
\sdtit       =  4\mts-\sman\,.
\label{moredefinitions}
\eqa
\vspace*{-5mm}

\noindent
Furthermore, we used one more `subtracted' function:
\bqa
\fbff{d\sss{W}}{-\mts}{\mbl}{\mwl}&=&\frac{1}{\rtw}
\biggl\{\lpar 1-\rbw\rpar\Bigl[\fbff{0}{-\mts}{\mbl}{\mwl}
+\Lmmw-1\Bigr]
\nll&&
-\rbw\Bigl[\Lmmb-\Lmmw\Bigr]\biggl\},
\eqa
and the three auxiliary functions:
\bqa
L_{ab}( M_1,M_2,M_3) & = &
\lpar M^2_3 + M^2_1 - M^2_2 \rpar 
        \cff{-\mts}{-\mts}{-\sman}{M_2}{M_3}{M_2}
\nll &&
 -\fbff{0}{-\sman}{M_2}{M_2} + \fbff{0}{-\mts }{M_2}{M_3},
\eqa
\vspace*{-5mm}
\bqa
L_{na}(M_1,M_2,M_3)  & = &
 \lpar M^2_3 - M^2_1 - M^2_2 \rpar \cff{-\mts}{-\mts}{-\sman}{M_3}{M_2}{M_3}
\nll &&
 +\fbff{0}{-\sman}{M_3}{M_3} - \fbff{0}{-\mts}{M_3}{M_2},
\eqa
\vspace*{-5mm}
\bqa
\LongHi{M_1}{M_2}{M_3} &=&
     \lrbr \frac{1}{2} \lpar M_2^2+M_3^2 \rpar
                           -2 M_1^2  \rrbr
         \cff{-\mts}{-\mts}{-\sman}{M_2}{M_1}{M_3}
\\ &&
        +           \fbff{0}{-\sman}{M_3}{M_2}
        -\frac{1}{2}\fbff{0}{-\mts}{M_1}{M_2} 
        -\frac{1}{2}\fbff{0}{-\mts}{M_3}{M_1}.
\nonumber
\eqa
Four scalar form factors, ${\cal F}^{\gamma(z){\sss W}}_{\sss{Q,D}}$, as 
follows from calculations, are both gauge-invariant and finite,
thus enlarging the number of gauge-invariant {\em building blocks} to 15.
On the contrary, two form factors, ${\cal F}^{\gamma(z){\sss W}}_{\sss{L}}$,
are neither gauge-invariant nor finite. Gauge dependence on $\gpar$, 
as well as UV poles, of $L$ form factors cancel in the sum with the $\wb\wb$ box 
and the self-energy contributions. 
\clearpage
\subsection{Library of scalar form factors for electron vertex}
Besides $B\ft\ft$ clusters, we need also $Bee$ clusters, which can, in principle,
be taken from \cite{Bardin:1999yd} or derived from the $B\ft\ft$ case 
in the $\mtl\to0$ limit. Here we simply list the results:
\bqa
\cvetril{\gamma ee}{\sss{L}}{\sman,\tman}   &=&
 -\frac{2}{\cows} \qe v_e a_e
  \cvertil{\sss Z,e}{}{\sman}  
 +\cvertil{\sss Wna,e}{}{\sman},
\nll
\cvetril{\gamma ee}{\sss{Q}}{\sman,\tman}  &=&
  \frac{1}{4 \cows} \delta_e^2
    \cvertil{\sss Z,e}{}{\sman},
\nll
\cvetril{zee}{\sss{L}}{\sman,\tman}  &=&
- \frac{1}{2 \cows} A_e
  \cvertil{\sss Z,e}{}{\sman}
+ \cvertil{\sss W,e}{}{\sman},
\nll
\cvetril{zee}{\sss{Q}}{\sman}  &=&
  \frac{1}{4 \cows} \delta_e^2
    \cvertil{\sss Z,e}{}{\sman},
\label{Beeffs}
\eqa
with
\bq
 \cvertil{\sss W,e}{}{\sman}=
 -\cvertil{\sss Wab,e}{}{\sman}
+ \cows
  \cvertil{\sss Wna,e}{}{\sman}.
\eq
In \eqn{Beeffs} we use three more auxiliary functions:
\bqa
{\cal F}^{\sss Z,e}\equiv{\cal F}^{\sss Zab,e}
&=&
 2 \frac{\lpar 1+\Rz \rpar^2}{\Rz}  \mzs \cff{0}{0}{-s}{0}{\mzl}{0}
 -3 \lrbr \fbff{0}{-s}{0}{0} + \Lmmz \rrbr
\nll &&
+\frac{5}{2} 
  -2 \Rz \lrbr \fbff{0}{-s}{0}{0} + \Lmmz - 1 \rrbr,
\eqa
\bqa
{\cal F}^{\sss Wna,e}
&=&
 -2 \Rw 
\lrbr\mws \cff{0}{0}{-s}{\mwl}{0}{\mwl}
         +\fbff{0}{-s}{\mwl}{\mwl}+\Lmmw-1 \rrbr  
\nll &&
  -4 \mws \cff{0}{0}{-s}{\mwl}{0}{\mwl}
      -\fbff{0}{-s}{\mwl}{\mwl}-3\Lmmw+\frac{9}{2}\,,\qquad
\eqa
\bqa
{\cal  F}^{\sss Wab,e}
&=&  
   \sigma_\nu
 \Biggl\{   
 \frac{\lpar 1 + \Rw \rpar^2}{\Rw} \mws \cff{0}{0}{-s}{0}{\mwl}{0}
      -\frac{3}{2} \lrbr \fbff{0}{-s}{0}{0} + \Lmmw \rrbr
\nll &&
+\frac{5}{4}
      -\Rw \lrbr \fbff{0}{-s}{0}{0} + \Lmmw - 1 \rrbr \Biggr\}.
\eqa

\subsection{The $\wb\wb$ box}
\begin{figure}[h!]
\vspace{-30mm}
\begin{center}
\[
\begin{picture}(125,130)(0,40)
  \ArrowLine(0,15)(37.5,15)
  \ArrowLine(37.5,15)(37.5,90)
  \ArrowLine(37.5,90)(0,90)
  \Photon(37.5,15)(112.5,90){3}{10}
  \Photon(37.5,90)(112.5,15){3}{10}
  \Vertex(37.5,15){2.5}
  \Vertex(37.5,90){2.5}
  \Vertex(112.5,15){2.5}
  \Vertex(112.5,90){2.5}
  \ArrowLine(150,90)(112.5,90)
  \ArrowLine(112.5,90)(112.5,15)
  \ArrowLine(112.5,15)(150,15)
\Text(18.75,105)[tc]{$e^+$}
  \Text(75,108.75)[tc]{$W$}
  \Text(131.25,112.5)[tc]{$\bar t$}
  \Text(18.75,52.5)[lc]{$\nu_e$}
  \Text(138,52.5)[lc]{$b,s,d$}
  \Text(18.75,0)[cb]{$e^-$}
  \Text(75,-3.75)[bc]{$W$}
  \Text(131.25,-3)[cb]{$t$}
\end{picture}
\]
\vspace*{5mm}
\caption{Crossed $WW$ box. \label{WW_box_c} }
\end{center}
\end{figure}
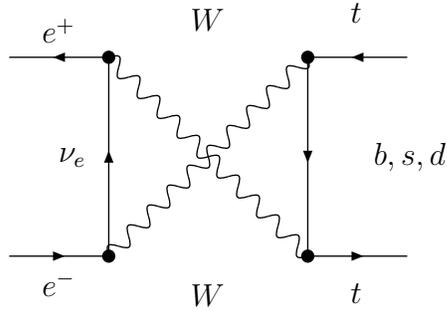

There is only one, {\it crossed}, $\wb\wb$ diagram contributing to our 
process, see \fig{WW_box_c}.

 Here we give the contribution of this diagram
to the scalar form factor $LL$:
\bq
\bigg( {\cal B}^{\sss WW} \bigg)^{c} = (2\pi)^4\ib\frac{g^4}{16\pi^2}\frac{1}{\sman}
\gadi{\mu}\gap\otimes\gadi{\mu}\gap
{\cal F}_{\sss {LL}}^{\sss WW}\lpar\sman,\uman \rpar,
\eq
where
\bqa
{\cal F}_{\sss {LL}}^{\sss WW}\lpar \sman, \uman \rpar &=&
 \frac{s}{8} \bigg[ -\lpar\uman-\mbs\rpar
   \dffp{0}{0}{-\mts}{-\mts}{-\sman}{-\uman} \dffm{\mwl}{0}{\mwl}{\mbl} 
\nll &&
+\cff{-\mts}{-\mts}{-\sman}{\mwl}{0}{\mwl}
+\cff{0}{0}{-\sman}{\mwl}{0}{\mwl}
             \bigg],\quad
\label{WW_box_1}
\eqa
with $\sman$, $\tman$, and $\uman$ being the usual Mandelstamm variables
satisfying
\bq
\sman+\tman+\uman=2\mts\,.
\eq
\subsection{ The $\zb\zb$ box}
There are four $\zb\zb$ diagrams, which form a gauge-invariant and UV finite
subset. Its contribution is originally presented in terms of 
six structures  $(L,R) \otimes (L,R,D)$ (i.e. here we used initially 
the $L,R,D$ basis):
\bqa
\bigg( {\cal B}^{\sss ZZ} \bigg)^{d+c} & = &
\frac{1}{32} \frac{g^4}{\ctwf}\frac{1}{s}
\Biggl[
    \lrbr  \gdmu \gdp \otimes  \gdmu \gdp \rrbr
{\cal F}^{\sss ZZ}_{\sss LL} \lpar s,t,u \rpar
  + \lrbr  \gdmu \gdp \otimes  \gdmu \gdm  \rrbr
{\cal F}^{\sss ZZ}_{\sss LR}\lpar s,t,u \rpar 
\nll[1mm] & &
  + \lrbr  \gdmu \gdm \otimes  \gdmu \gdp \rrbr 
{\cal F}^{\sss ZZ}_{\sss RL}\lpar s,t,u \rpar
  + \lrbr  \gdmu \gdm   \otimes  \gdmu \gdm \rrbr
{\cal F}^{\sss ZZ}_{\sss RR} \lpar s,t,u \rpar
\\[1mm] & &
  + \lrbr \gdmu \gdp  \otimes \lpar -i \mtl I D_\mu \rpar \rrbr
{\cal F}^{\sss ZZ}_{\sss LD}\lpar s,t,u \rpar
  + \lrbr \gdmu \gdm  \otimes \lpar -i \mtl I D_\mu \rpar \rrbr 
{\cal F}^{\sss ZZ}_{\sss RD}\lpar s,t,u \rpar
\Biggr].
\nonumber
\eqa
\begin{figure}[!h]
\vspace{-20mm}
\[
\begin{array}{ccc}
\begin{picture}(100,105)(0,49.5)
  \ArrowLine(0,15)(37.5,15)
  \ArrowLine(37.5,15)(37.5,90)
  \ArrowLine(37.5,90)(0,90)
  \Photon(37.5,15)(112.5,90){3}{10}
  \Photon(37.5,90)(112.5,15){3}{10}
  \Vertex(37.5,15){2.5}
  \Vertex(37.5,90){2.5}
  \Vertex(112.5,15){2.5}
  \Vertex(112.5,90){2.5}
  \ArrowLine(150,90)(112.5,90)
  \ArrowLine(112.5,90)(112.5,15)
  \ArrowLine(112.5,15)(150,15)
\Text(18.75,105)[tc]{$e^+$}
  \Text(75,108.75)[tc]{$Z$}
  \Text(131.25,112.5)[tc]{$\bar t$}
  \Text(18.75,52.5)[lc]{$e$}
  \Text(131.8,52.5)[lc]{$t$}
  \Text(18.75,0)[cb]{$e^-$}
  \Text(75,-3.75)[bc]{$Z$}
  \Text(131.25,-3)[cb]{$t$}
\end{picture}
\qquad \qquad \qquad
&+&
\qquad 
\begin{picture}(100,105)(0,49.5)
  \ArrowLine(0,15)(37.5,15)
  \ArrowLine(37.5,15)(37.5,90)
  \ArrowLine(37.5,90)(0,90)
  \Photon(37.5,15)(112.5,15){3}{10}
  \Photon(37.5,90)(112.5,90){3}{10}
  \Vertex(37.5,15){2.5}
  \Vertex(37.5,90){2.5}
  \Vertex(112.5,15){2.5}
  \Vertex(112.5,90){2.5}
  \ArrowLine(150,90)(112.5,90)
  \ArrowLine(112.5,90)(112.5,15)
  \ArrowLine(112.5,15)(150,15)
\Text(18.75,105)[tc]{$e^+$}
  \Text(75,108.75)[tc]{$Z$}
  \Text(131.25,112.5)[tc]{$\bar t$}
  \Text(18.75,52.5)[lc]{$e$}
  \Text(131.8,52.5)[lc]{$t$}
  \Text(18.75,0)[cb]{$e^-$}
  \Text(75,-3.75)[bc]{$Z$}
  \Text(131.25,-3)[cb]{$t$}
\end{picture}
\end{array}
\]
\vspace{12mm}
\caption{Crossed and direct $ZZ$ boxes. \label{ZZ_box_dc} }
\end{figure}
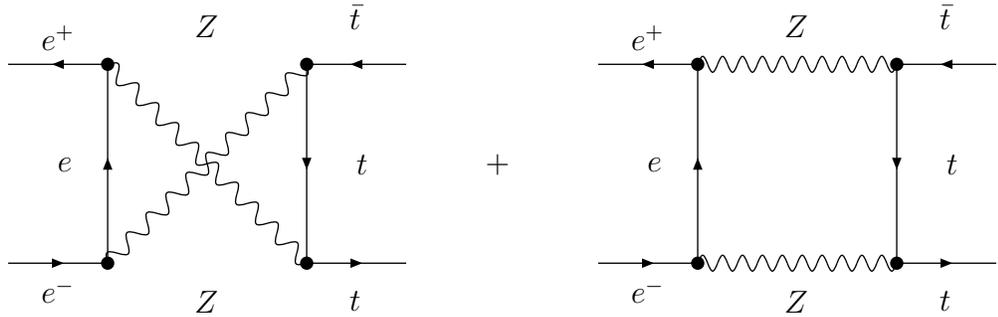

 Moreover, we used three auxiliary functions ${\cal{F,H,G}}$:
\bqa
{\cal F}^{\sss ZZ}_{\sss LD}\lpar s,t,u \rpar  &=&
 {\cal F}\lpar \vpae,\vpau,\vmau,s,t \rpar  
-{\cal F}\lpar \vpae,\vmau,\vpau,s,u \rpar, 
\nll
{\cal F}^{\sss ZZ}_{\sss RD} \lpar s,t,u \rpar &=&
 {\cal F}\lpar \vmae,\vmau,\vpau,s,t \rpar
-{\cal F}\lpar \vmae,\vpau,\vmau,s,u \rpar,   
\nll
{\cal F}^{\sss ZZ}_{\sss LR}\lpar s,t,u \rpar  &=&
 {\cal H}\lpar \vpae,\vpau,\vmau,s,t \rpar
-{\cal G}\lpar \vpae,\vmau,\vpau,s,u \rpar,  
\nll
{\cal F}^{\sss ZZ}_{\sss RL}\lpar s,t,u \rpar  &=&
 {\cal H}\lpar \vmae,\vmau,\vpau,s,t \rpar 
-{\cal G}\lpar \vmae,\vpau,\vmau,s,u \rpar,  
\nll
{\cal F}^{\sss ZZ}_{\sss LL} \lpar s,t,u \rpar &=&
  {\cal G}\lpar \vpae,\vpau,\vmau,s,t \rpar 
- {\cal H}\lpar \vpae,\vmau,\vpau,s,u \rpar, 
\nll
{\cal F}^{\sss ZZ}_{\sss RR}\lpar s,t,u \rpar  &=&
  {\cal G}\lpar \vmae,\vmau,\vpau,s,t \rpar 
- {\cal H}\lpar \vmae,\vpau,\vmau,s,u \rpar. 
\eqa

Separating out $\zb$ fermion coupling constants and some common factors,
we introduce more auxiliary functions. For ${\cal F}\lpar \vpae,\vpau,\vmau,s,t \rpar$
defined as
\bqa
{\cal F}\lpar \vpae,\vpau,\vmau,s,t \rpar &=&
 \frac{s}{\sdfit} 
 \Bigl[   \vpae^2 \vpau^2     {\cal F}_1\lpar s,t \rpar 
       +  \vpae^2 \vpau \vmau {\cal F}_2\lpar s,t \rpar   \Bigr],
\eqa 
there are two
\bqa
{\cal F}_1(s,t) &=&
  \Bigg(   \tpl \mzs+\tmis 
           +\lrbr \lpar-\sz+2 \mzs \rpar \tmis + 4 t \mzf \rrbr
            \frac{\tmi}{ \sdfit}
  \Bigg) 
\nll &&
\times 
 \dffp{0}{0}{-\mts}{-\mts}{-s}{-t}
 \dffm{\mzl}{0}{\mzl}{\mtl}
\nll &&
  -\Bigg[ 
      \mts - \mzs 
 - \frac{ \tmi \lpar \tmi t + 2 \mzs \tpl \rpar + s \mtq}{\sdfit}
                 +\frac{ 2  \tpl \lpar \mzs-2 \mts \rpar}{\sdtit}    
   \Bigg]
\nll &&
\times \cff{-\mts}{-\mts}{s}{\mzl}{\mtl}{\mzl}
\nll &&
  -  2\Bigg[
                   t \lpar 1+\frac{\mzs}{\tmi}\rpar
      +\frac{ t^3 + 2 t \tmi \mzs 
      -\mts \lpar 3 t\tmi + \mtq \rpar }{\sdfit}
     \Bigg]
\,\cff{0}{-\mts}{t}{0}{\mzl}{\mtl}
\nll &&
  -  \Bigg[
      \mzs- \mts  + 
       \frac{ \lpar t-2 \tmi-2 \mzs \rpar \tmis + s \mtq} 
            { \sdfit}
 \Bigg]
\,\cff{0}{0}{s}{\mzl}{0}{\mzl} 
\nll &&
-\lpar 1 + \frac{s + 2 \tmi}{ \sdtit} \rpar
\lrbr
 \bff{0}{-s}{\mzl}{\mzl}
-\bff{0}{-\mts}{\mzl}{\mtl}
\rrbr
\nll &&
  + \frac{2 t}{\tmi}
\lrbr  \bff{0}{t}{\mtl}{0} - \bff{0}{-\mts}{\mzl}{\mtl} \rrbr
    \Bigg\},
\eqa
and
\bqa
{\cal F}_2\lpar s,t \rpar &=&
  -\lpar \tmis + 2 \mzs  t  \rpar
 \dffp{0}{0}{-\mts}{-\mts}{-s}{-t}
 \dffm{\mzl}{0}{\mzl}{\mtl}
\nll [1mm] &&
   - \tpl \cff{-\mts}{-\mts}{s}{\mzl}{\mtl}{\mzl}
\nll[2mm] &&
   + 2 t   \cff{0}{-\mts}{t}{0}{\mzl}{\mtl}
   - \tmi   \cff{0}{0}{s}{\mzl}{0}{\mzl}.   
\eqa
For ${\cal H}$ written out as 
\bqa
{\cal H}\lpar \vpae,\vpau,\vmau,s,t \rpar &=&
\frac{ \mts s}{ \sdfit}
\Bigl[ 
   \vpae^2 \vpau^2       {\cal  H}_1 \lpar s,t \rpar 
+  \vpae^2 \vpau \vmau   {\cal  H}_2 \lpar s,t \rpar \Bigr]
+  \vpae^2 \vmau^2       {\cal  H}_3 \lpar s,t \rpar,  
\eqa
we need three auxiliary subfunctions:
\bqa
{\cal H}_1\lpar s,t \rpar 
 &=& 
  \Bigg[
 \frac{ s\tmi }{2}-(\tmi+\mzs)^2-2 \mzs t-\frac{\szs \tmi \tpl}{2\sdfit}
 \Bigg]
\nll &&  
\times 
\dffp{0}{0}{-\mts}{-\mts}{-s}{-t}
 \dffm{\mzl}{0}{\mzl}{\mtl}
\nll &&
 - \Bigg(
  \tpl-\frac{\sz \lrbr \tpls+\sdtit \mts \rrbr}{2\sdfit}
   \Bigg) 
\cff{-\mts}{-\mts}{s}{\mzl}{\mtl}{\mzl}
\nll &&
 + \Bigg( 
 2 t+\tmi+\frac{2\mzs t}{\tmi}-\frac{\sz \tmi \tpl }{\sdfit}
  \Bigg)
\cff{0}{0}{s}{\mzl}{0}{\mzl}
\nll &&
 -  \Bigg[
\tmi +  \frac{\sz \lpar s \mts-\tmis \rpar}{2 \sdfit}
    \Bigg] 
\cff{0}{-\mts}{t}{0}{\mzl}{\mtl}
\\ &&
 -2 \frac{\mts}{\tmi} 
 \Biggl( \bff{0}{t}{\mtl}{0}-\bff{0}{-\mts}{\mzl}{\mtl} \Biggr)
 - \bff{0}{t}{\mtl}{0}+\bff{0}{s}{\mzl}{\mzl},
\nn \nll
{\cal H}_2\lpar s,t \rpar &=&
  \lrbr \tmi \lpar s + \tmi \rpar + 2 \mts \mzs \rrbr
\dffp{0}{0}{-\mts}{-\mts}{-s}{-t}
\dffm{\mzl}{0}{\mzl}{\mtl}
\nll [1mm] &&
   - \lpar s + \tmi - 2 \mts\rpar \cff{-\mts}{-\mts}{s}{\mzl}{\mtl}{\mzl} 
\\ [1mm] &&
   - 2 \mts\cff{-\mts}{-\mts}{s}{0}{\mzl}{\mtl} 
   + \lpar s+\tmi \rpar\cff{0}{0}{s}{\mzl}{0}{\mzl}, 
\nonumber
\eqa  
and
\bqa
{\cal H}_3\lpar s,t \rpar &=&-
 s\biggl[  \tmi 
\dffp{0}{0}{-\mts}{-\mts}{-s}{-t}
\dffm{\mzl}{0}{\mzl}{\mtl}
\nll[1mm] &&
   + \cff{-\mts}{-\mts}{s}{\mzl}{\mtl}{\mzl} 
   + \cff{0}{-\mts}{t}{\mzl}{0}{\mzl} \biggr].      
\eqa
Finally, ${\cal G}$ also, defined as follows:
\bq
{\cal G} \lpar \vpae,\vpau,\vmau,s,t \rpar = \frac{s}{\sdfit}
\Bigl[
  \vpae^2 \vpau^2 {\cal G}_1\lpar s,t \rpar  
+ \vpae^2 \vpau \vmau {\cal G}_2\lpar s,t \rpar  \Bigr],
\eq
needs only two additional functions:
\bqa
{\cal G}_1\lpar s,t \rpar
&=&-
   \Biggl[
     \tmi \tpl \lpar 2 \mzs + \frac{\szs t}{2\sdfit}\rpar 
      -\tmi   \lpar \frac{s \mts}{2}-\tmi t\rpar 
      + t\mzs \lpar 2 \mts-\mzs \rpar
   \Biggr]
\nll &&
 \times 
\dffp{0}{0}{-\mts}{-\mts}{-s}{-t}
\dffm{\mzl}{0}{\mzl}{\mtl}
\nll &&
 + \Bigg[    
\tpl t+ \frac{\sz}{2} \lpar t - \frac{\tmi\tpls}{\sdfit}\rpar
   \Bigg]
 \cff{-\mts}{-\mts}{s}{\mzl}{\mtl}{\mzl}
\nll &&
 -  \Bigg[
  \tpl \lpar t+\tmi \rpar + 2 \mzs \mts \frac{t}{\tmi}
 -\frac{\tpl \tmi t \sz}{\sdfit}
    \Bigg]
 \cff{0}{-\mts}{-t}{0}{\mzl}{\mtl}
\nll &&
 - \Bigg[ \tmi t + \frac{\sz}{2} 
            \bigg( t - \frac{ \tpl \tmis}{\sdfit} \bigg)
   \Bigg] 
\cff{0}{0}{s}{\mzl}{0}{\mzl} 
\nll &&
 + 2 \mts \lpar 1 + \frac{\mts}{\tmi} \rpar
 \bigg[ \bff{0}{t}{\mtl}{0} -\bff{0}{-\mts}{\mzl}{\mtl} \bigg]
\nll [1mm] &&
 -  t \bigg[ \bff{0}{s}{\mzl}{\mzl}-\bff{0}{t}{\mtl}{0} \bigg]\,,
\eqa
and
\bqa
{\cal G}_2\lpar s,t \rpar &=&
 \mts \Bigg[     \big( \tmis + 2 \mzs t \big) 
\dffp{0}{0}{-\mts}{-\mts}{-s}{-t}\dffm{\mzl}{0}{\mzl}{\mtl} 
\nll [1mm] &&
                  + \tpl \cff{-\mts}{-\mts}{s}{\mzl}{\mtl}{\mzl}
                  - 2 t     \cff{0}{-\mts}{t}{0}{\mzl}{\mtl}
\nll [1mm] &&  
                + \tmi  \cff{0}{0}{s}{\zml}{0}{\mzl}\Biggr].
\eqa
In this section we used the notation:
\bqa
 \sdfit &=& - t u + \mtq\,;
\eqa
together with $\sdtit$ of \eqn{moredefinitions},
this denotes remnants of Gram determinants that remained after cancellation
of factors $\sman$ and $4$, leading to a simplification of the expressions.

\subsubsection{Transition to the $\it L,Q,D$ basis}
Since the $\zb\zb$ box contribution is given in the ${\it L,R,D}$ basis, while all the rest
is in the ${\it L,Q,D}$ basis, we should transfer one of them to a chosen basis.
At this phase of the calculations there is not much difference which basis is 
chosen. For definiteness we choose the ${\it L,Q,D}$ basis and transfer the $\zb\zb$ box 
contribution to it.The transition formulae are simple:
\bqa
{\tilde{\cal F}}^{\sss ZZ}_{\sss LL}\lpar s,t,u \rpar &=&   
 {\cal F}^{\sss ZZ}_{\sss LL}\lpar s,t,u \rpar
+{\cal F}^{\sss ZZ}_{\sss RR}\lpar s,t,u \rpar
-{\cal F}^{\sss ZZ}_{\sss LR}\lpar s,t,u \rpar
-{\cal F}^{\sss ZZ}_{\sss RL}\lpar s,t,u \rpar,
\nll 
{\tilde{\cal F}}^{\sss ZZ}_{\sss QL} \lpar s,t,u \rpar &=& 
 2\Bigl[ {\cal F}^{\sss ZZ}_{\sss RL}\lpar s,t,u \rpar
       -{\cal F}^{\sss ZZ}_{\sss RR}\lpar s,t,u \rpar \Bigr], 
\nll
{\tilde{\cal F}}^{\sss ZZ}_{\sss LQ} \lpar s,t,u \rpar  &=&  
 2\Bigl[{\cal F}^{\sss ZZ}_{\sss LR}\lpar s,t,u \rpar
             -{\cal F}^{\sss ZZ}_{\sss RR}\lpar s,t,u \rpar  \Bigr],
\nll
{\tilde{\cal F}}^{\sss ZZ}_{\sss QQ} \lpar s,t,u \rpar &=&  
 4{\cal F}^{\sss ZZ}_{\sss RR}\lpar s,t,u \rpar,
\nll[1mm]
{\tilde{\cal F}}^{\sss ZZ}_{\sss LD} \lpar s,t,u \rpar &=&   
        {\cal F}^{\sss ZZ}_{\sss LD}\lpar s,t,u \rpar
       -{\cal F}^{\sss ZZ}_{\sss RD}\lpar s,t,u \rpar,
\nll[1mm]
{\tilde{\cal F}}^{\sss ZZ}_{\sss QD} \lpar s,t,u \rpar &=& 
 2{\cal F}^{\sss ZZ}_{\sss RD}\lpar s,t,u \rpar.   
\eqa

\clearpage

\section{Scalar form factors for electroweak amplitude}
Having all the building blocks, it is time to construct
{\it complete} electroweak scalar form factors.
\subsection{Vertices scalar form factors}
We begin with two vertex contributions:
\begin{figure}[th]
\vspace*{-15mm}
\[
\begin{array}{ccc}
\begin{picture}(125,86)(0,40)
    \Photon(25,43)(100,43){3}{15}
    \Vertex(25,43){12.5}
    \ArrowLine(0,0)(25,43)
    \ArrowLine(25,43)(0,86)
      \ArrowLine(125,86)(100,43)
      \Vertex(100,43){2.5}
      \ArrowLine(100,43)(125,0)
\Text(14,74)[lb]{$\fbe$}
\Text(108,74)[lb]{$\bar t$}
\Text(62.5,50)[bc]{$(\gamma,\zb)$}
\Text(14,12)[lt]{$\fe$}
\Text(108,12)[lt]{$t$}
\end{picture}
\qquad
&&
\qquad
\begin{picture}(125,86)(0,40)
 \Photon(25,43)(100,43){3}{15}
 \Vertex(100,43){12.5}
 \ArrowLine(125,86)(100,43)
 \ArrowLine(100,43)(125,0)
   \ArrowLine(0,0)(25,43)
   \Vertex(25,43){2.5}
   \ArrowLine(25,43)(0,86)
\Text(14,74)[lb]{$\fbe$}
\Text(108,74)[lb]{$\bar t$}
\Text(62.5,50)[bc]{$(\gamma,\zb)$}
\Text(14,12)[lt]{$\fe$}
\Text(108,12)[lt]{$t$}
\end{picture}
\end{array}
\]
\vspace{15mm}
\caption[Electron and final fermion vertices 
in $\fe\fbe\to(\gamma,\zb)\to\ff\fbf$.]
{\it
Electron (a) and final fermion (b) vertices 
in $\fe\fbe\to(\gamma,\zb)\to\ff\fbf$.
\label{zavert6}}
\end{figure}

 In the same way as described in~\cite{Bardin:1999yd}, we reduce two vertex 
contributions to our six form factors:
\bqa
\vvertil{}{\sss{LL}}{\sman}&=&
    \vvertil{zee}{\sss{L}}{\sman} 
 +  \vvertil{ztt}{\sss{L}}{\sman} - 4\ctws \Delta(\mwl),  
\nll [2mm]
\vvertil{}{\sss{QL}}{\sman}&=&
    \vvertil{zee}{\sss{Q}}{\sman}
 +  \vvertil{ztt}{\sss{L}}{\sman} - 2\ctws \Delta(\mwl)            
 +  k
 \lrbr \vvertil{\gamma tt}{\sss{L}}{\sman}- 2\Delta(\mwl) \rrbr,
\nll [2mm]
\vvertil{}{\sss{LQ}}{\sman}&=&
    \vvertil{zee}{\sss{L}}{\sman} - 2\ctws \Delta(\mwl)
  + \vvertil{ztt}{\sss{Q}}{\sman}    
 +  k
 \lrbr \vvertil{\gamma ee}{\sss{L}}{\sman}- 2\Delta(\mwl) \rrbr,
\nll [1mm]
\vvertil{}{\sss{QQ}}{\sman}&=&
    \vvertil{zee}{\sss{Q}}{\sman}
 +  \vvertil{ztt}{\sss{Q}}{\sman}                   
 -  \frac{k}{\stws} 
    \lrbr \vvertil{\gamma ee}{\sss{Q}}{\sman}
 +  \vvertil{\gamma tt}{\sss{Q}}{\sman} \rrbr,
\nll [1mm]
\vvertil{}{\sss{LD}}{\sman}&=&
    \vvertil{ztt}{\sss{D}}{\sman},
\nll [3mm]
\vvertil{}{\sss{QD}}{\sman}&=&
    \vvertil{ztt}{\sss{D}}{\sman}
 +  k
    \vvertil{\gamma tt}{\sss{D}}{\sman},
\eqa
where 
\bqa
k = \cows \lpar \Rz - 1 \rpar.
\label{koeff_k}
\eqa

With the term containing $\Delta(\mwl)$,
\bqa
\Delta(\mwl) = \pole - \ln \frac{\mws}{\mu^2}\,,
\eqa    
we explicitly show the contribution of the  
so-called {\em special} vertices \cite{Passarino:1991b}.
Note that they accompany every $L$ form factor.
The poles $1/{\bar{\varepsilon}}$ originating from special vertices
will be canceled in the sum of all contributions, including self-energies 
and boxes.
\newpage
\subsection{Bosonic self-energies and bosonic counterterms}
 The contributions to form factors from bosonic self-energy diagrams and 
counterterms, originating from bosonic self-energy diagrams,
come from four classes of diagrams; their sum is depicted by a black
circle in \fig{zavert8}.

\begin{figure}[thbp]
\vspace{-17mm}
\[
\begin{array}{ccc}
\begin{picture}(125,86)(0,40)
    \Photon(25,43)(50,43){3}{5}
    \Vertex(62.5,43){12.5}
    \Photon(75,43)(100,43){3}{5}
  \ArrowLine(125,86)(100,43)
  \Vertex(100,43){2.5}
  \ArrowLine(100,43)(125,0)
    \ArrowLine(0,0)(25,43)
    \Vertex(25,43){2.5}
    \ArrowLine(25,43)(0,86)
\Text(14,74)[lb]{$\fbe$}
\Text(108,74)[lb]{$\fbf$}
\Text(37.5,50)[bc]{$(\ph,\zb)$}
\Text(87.5,50)[bc]{$(\ph,\zb)$}
\Text(14,12)[lt]{$\fe$}
\Text(108,12)[lt]{$\ff$}
\end{picture}
\qquad
 &=&
\nll  \nll 
\begin{picture}(125,86)(0,40)
    \Photon(25,43)(50,43){3}{5}
    \GCirc(62.5,43){12.5}{0.5}
    \Photon(75,43)(100,43){3}{5}
  \ArrowLine(125,86)(100,43)
  \Vertex(100,43){2.5}
  \ArrowLine(100,43)(125,0)
    \ArrowLine(0,0)(25,43)
    \Vertex(25,43){2.5}
    \ArrowLine(25,43)(0,86)
\Text(14,74)[lb]{$\fbe$}
\Text(108,74)[lb]{$\fbf$}
\Text(37.5,50)[bc]{$(\ph,\zb)$}
\Text(87.5,50)[bc]{$(\ph,\zb)$}
\Text(14,12)[lt]{$\fe$}
\Text(108,12)[lt]{$\ff$}
\end{picture}
\qquad
&+&
\qquad
\begin{picture}(125,86)(0,40)
     \Photon(25,43)(100,43){3}{15}
\SetScale{2.0}
     \Line(26.5,16.75)(28,18.25)
     \Line(29,19.25)(34,24.25)
     \Line(35,25.25)(36.5,26.75)
      \Line(26.5,26.75)(28,25.25)
      \Line(29,24.25)(34,19.25)
      \Line(35,18.25)(36.5,16.75)
\SetScale{1.0}
     \ArrowLine(125,86)(100,43)
     \Vertex(100,43){2.5}
     \ArrowLine(100,43)(125,0)
  \ArrowLine(0,0)(25,43)
  \Vertex(25,43){2.5}
  \ArrowLine(25,43)(0,86)
\Text(14,74)[lb]{$\fbe$}
\Text(108,74)[lb]{$\fbf$}
\Text(37.5,50)[bc]{$(\ph,\zb)$}
\Text(87.5,50)[bc]{$(\ph,\zb)$}
\Text(14,12)[lt]{$\fe$}
\Text(108,12)[lt]{$\ff$}
\end{picture} \nll \nll 
\begin{picture}(125,86)(0,40)
  \Photon(25,43)(100,43){3}{15}
\SetScale{2.0}
    \Line(45,16.5)(46.5,18)
    \Line(47.5,19)(52.5,24)
    \Line(53.5,25)(55,26.5)
    \Line(45,26.5)(46.5,25)
    \Line(47.5,24)(52.5,19)
    \Line(53.5,18)(55,16.5)
\SetScale{1.0}
  \ArrowLine(125,86)(100,43)
  \Vertex(100,43){2.5}
  \ArrowLine(100,43)(125,0)
    \ArrowLine(0,0)(25,43)
    \Vertex(25,43){2.5}
    \ArrowLine(25,43)(0,86)
\Text(14,74)[lb]{$\fbe$}
\Text(108,74)[lb]{$\fbf$}
\Text(62.5,50)[bc]{$(\ph,\zb)$}
\Text(14,12)[lt]{$\fe$}
\Text(108,12)[lt]{$\ff$}
\end{picture}
\qquad
&+&
\qquad
\begin{picture}(125,86)(0,40)
  \Photon(25,43)(100,43){3}{15}
\SetScale{2.0}
    \Line(7.5,16.5)(9,18)
    \Line(10,19)(15,24)
    \Line(16,25)(17.5,26.5)
    \Line(7.5,26.5)(9,25)
    \Line(10,24)(15,19)
    \Line(16,18)(17.5,16.5)
\SetScale{1.0}
    \Vertex(25,43){2.5}
    \ArrowLine(0,0)(25,43)
    \ArrowLine(25,43)(0,86)
 \ArrowLine(125,86)(100,43)
 \Vertex(100,43){2.5}
 \ArrowLine(100,43)(125,0)
\Text(14,74)[lb]{$\fbe$}
\Text(108,74)[lb]{$\fbf$}
\Text(62.5,50)[bc]{$(\ph,\zb)$}
\Text(14,12)[lt]{$\fe$}
\Text(108,12)[lt]{$\ff$}
\end{picture}
\end{array}
\]
\vspace{10mm}
\caption[Bosonic self-energies
and bosonic counterterms for $\fe\fbe\to(\zb,\ph)\to\ff\fbf$.]
{\it
Bosonic self-energies
and bosonic counterterms for $\fe\fbe\to(\zb,\ph)\to\ff\fbf$.
\label{zavert8}} 
\end{figure}

The contribution of these diagrams to the four scalar form factors is derived
straightforwardly ~\cite{Bardin:1999ak}, ~\cite{Bardin:1999yd}:
\bqa
\vvertil{ct}{\sss{LL}}{\sman} &=& 
      \Dz{}{\sman}-\stws\Pgg^{}(0)
     +\frac{\cows-\siws}{\siws} \lpar \delrho{} + \bdelrho{\bos} \rpar,   
\label{Fct_b_LL}
\\ [2mm]
\vvertil{ct}{\QL(\LQ)}{\sman}&=& 
      \Dz{}{\sman}-
      \lpar \Pzg^{}(\sman) + \bPzga{\bos}{\sman} \rpar
     -\stws \Pgg^{}(0)-\lpar\delrho{}+\bdelrho{\bos} \rpar, 
\label{Fct_b_LQ}
\\ [2mm]
\vvertil{ct,\bos}{\sss{QQ}}{\sman}&=&
      \Dz{\bos}{\sman}-2\lpar \Pzg^{\bos}(\sman) + \bPzga{\bos}{\sman} \rpar
     + k \Bigl[\Pgg^{\bos}\lpar s \rpar 
     - \Pgg^{\bos}\lpar 0 \rpar \Bigr]                  
\nll &&
-\stws \Pgg^{\bos}\lpar 0 \rpar 
     -\frac{1}{\siws} \lpar \delrho{\bos} + \bdelrho{\bos} \rpar,    
\label{Fct_b_QQ}
\\ [2mm]
\vvertil{ct,\fer}{\sss{QQ}}{\sman}&=&
      \Dz{\fer}{\sman}-2\Pzg^{\fer}(\sman)
     -\stws \Pgg^{\fer}\lpar 0 \rpar 
     -\frac{1}{\siws} \delrho{\fer}.
\label{Fct_f_QQ}
\eqa
 We note that the term $ k \bigl[\Pgg^{\fer}\lpar\sman\rpar
-\Pzg^{\fer}\lpar\sman\rpar\bigr]$ is conventionally extracted from 
$\vvertil{ct,\fer}{\sss{QQ}}{\sman}$. 
This contribution is shifted to $A_{\gamma}^{\sss{\rm{IBA}}}$, \eqn{Born_modulo-old}.

 In \eqns{Fct_b_LL}{Fct_f_QQ} $\bdelrho{\bos}$ and $\bPzga{\bos}{\sman}$
stand for shifts of bosonic self-energies. They have the same origin as 
special vertices and they are equal to:
\bqa
\bdelrho{\bos}      &=& 4\stws\Delta(\mwl),
\\[2mm]
\bPzga{\bos}{\sman} &=& -2\Rw \Delta(\mwl),
\eqa
see Eqs.~(6.137) and ~(6.139) of ~\cite{Bardin:1999ak}.
These poles also cancel in the sum of all contributions.
\subsection{Complete scalar form factors of the one-loop amplitude}
 Adding all contributions together, we observe the cancellation of all poles.
The ultraviolet-finite results for six scalar form factors are:
\bqa
\vvertil{}{\sss{LL}}{\sman,\tman,\uman}&=&
    \cvetril{zee}{\sss{L}}{\sman}
 +  \cvetril{ztt}{\sss{L}}{\sman}  
 +  \cvertil{ct}{\sss{LL}}{\sman}  
 + k^{\sss WW}~\cvetril{\sss WW}{\sss LL}{\sman,\uman}
 + k^{\sss ZZ}~\cvetril{\sss ZZ}{\sss LL}{\sman,\tman,\uman},
\nll [2mm]
\vvertil{}{\sss{QL}}{\sman,\tman,\uman}&=&
    \cvetril{zee}{\sss{Q}}{\sman}
 +  \cvetril{ztt}{\sss{L}}{\sman}             
 + k~\cvetril{\gamma tt}{\sss{L}}{\sman}
 +  \cvertil{ct}{\sss{QL}}{\sman}
 + k^{\sss ZZ}~\cvetril{\sss ZZ}{\sss QL}{\sman,\tman,\uman},
\nll [2mm]
\vvertil{}{\sss{LQ}}{\sman,\tman,\uman}&=&
     \cvetril{zee}{\sss{L}}{\sman} + \cvetril{ztt}{\sss{Q}}{\sman}    
+  k~\cvetril{\gamma ee}{\sss{L}}{\sman} + \cvertil{ct}{\sss{LQ}}{\sman}
 + k^{\sss ZZ}~\cvetril{\sss ZZ}{\sss LQ}{\sman,\tman,\uman},
\nll [1mm]
\vvertil{}{\sss{QQ}}{\sman,\tman,\uman}&=&
    \cvetril{zee}{\sss{Q}}{\sman}
 +  \cvetril{ztt}{\sss{Q}}{\sman}                   
 - \frac{k}{\stws}~\lrbr \cvetril{\gamma ee}{\sss{Q}}{\sman}
 +  \cvetril{\gamma tt}{\sss{Q}}{\sman} \rrbr
 +  \cvertil{ct}{\sss{QQ}}{\sman}
 + k^{\sss ZZ}~\cvetril{\sss ZZ}{\sss QQ}{\sman,\tman,\uman},
\nll [1mm]
\vvertil{}{\sss{LD}}{\sman,\tman,\uman}&=&
    \cvetril{ztt}{\sss{D}}{\sman}
 + k^{\sss ZZ}~\cvetril{\sss ZZ}{\sss LD}{\sman,\tman,\uman},
\nll [4mm]
\vvertil{}{\sss{QD}}{\sman,\tman,\uman}&=&
    \cvetril{ztt}{\sss{D}}{\sman}
+  k~\cvetril{\gamma tt}{\sss{D}}{\sman}
 + k^{\sss ZZ}~\cvetril{\sss ZZ}{\sss QD}{\sman,\tman,\uman},
\label{soular_ff}
\eqa
where
\bqa
k^{\sss{WW}} = 16 k\,,
\eqa
\bqa
k^{\sss ZZ} =\frac{\lpar \Rz - 1 \rpar}{2 \ctws}\,.
\eqa
In \eqn{soular_ff}, the quantities $\cvetril{ct}{\sss{AB}}{\sman}$
denote finite parts of the counterterm 
contributions, see \eqns{Fct_b_LL}{Fct_f_QQ}. 

The formulae of Sections 2 and 3 put together present the one-loop core of 
the {\tt eeffLib} code.

\clearpage

\section{Improved  Born Approximation cross-section}
In this section we give the improved  Born approximation (IBA) differential in the
scattering angle cross-section. It is derived by simple squaring the $(\ph+\zb)$ exchange
IBA amplitude, \eqns{Born_modulo-old}{structures-old}, and accounting for proper 
normalization factors. We simply give the result:
\bq
\frac{d\sigma^{\sss{\rm{IBA}}}}{d\cos\vartheta}=\frac{\pi\alpha^2}{s^3}\beta_{\ft} N_{c}
\bigl(\sigma_{\sss \gamma\gamma}^{\sss IBA}
     +\sigma_{\sss \gamma Z    }^{\sss IBA}
     +\sigma_{\sss ZZ          }^{\sss IBA}
\bigr),
\label{differential-cs}
\eq
where $\beta_{\ft}=\sqrt{1-{4\mts}/{\sman}}$ and 
\bqa
\sigma_{\sss \gamma\gamma}^{\sss IBA} &=&
      \qt^2 \lpar s^2+ 2st+2\tmis \rpar \Bigl| \alpha\lpar s \rpar \Bigr|^2,
\nll 
\sigma_{\sss \gamma Z}^{\sss IBA} &=& 2 \qt{\rm Re} \biggl\{ \chi \biggl(
            2\lrbr \lpar s+\tmi\rpar^2+s \mts \rrbr\FLLt 
\nll &&
            +\lpar s^2+2st+2\tmis\rpar\biggl[
 \FQLt + \FLQt  + \FQQt   \biggr]
\nll &&
           - 4 \mts \lpar st+\tmis \rpar \biggl[   \FLDt   
 +  \FQDt   \biggr] \biggr)
\alpha^*\lpar s \rpar \biggr\},
\nll
\sigma_{\sss ZZ}^{\sss IBA} &=&  |\chi|^2 {\rm Re} \biggl\{
   8 \lpar s+\tmi\rpar^2 \bigg[ \Bigl| \FLLt \Bigr|^2+ \FLLt \FQLtc \bigg]
\nll[2mm] &&
 + 2 \lrbr \lpar s+\tmi \rpar^2+\tmis \rrbr \Bigl| \FQLt \Bigr|^2
\nll[2mm] &&
 + 4 \lrbr\lpar s+\tmi\rpar^2+s\mts \rrbr\bigg[ 2  \FLLt \FLQtc 
\nll &&
 +  \FLLt \FQQtc + \FQLt \FLQtc  \bigg]
\nll &&
 + \lrbr s^2+2 \lpar s t+ \tmis \rpar\rrbr \biggl[ 2 \Bigl| \FLQt \Bigl|^2 
                          +\Bigl| \FQQt \Bigr|^2
\nll &&
 + 2 \bigg( \FQLt + \FLQt \bigg) \FQQtc  \biggr]
\nll && 
 - 8 \mts \lpar st+\tmis \rpar \bigg[ \bigg( 2\FLDt + \FQDt \bigg) \FLLtc 
\nll && 
 + \bigg( \FLDt + \FQDt \bigg) \FQLtc 
\nll && 
 + \bigg( 2 \FLDt + \FQDt \bigg) \FLQtc 
\nll && 
 + \bigg( \FLDt + \FQDt \bigg)\FQQtc  \bigg] 
\nll &&
 - 2\mts \lpar s t + \tmis \rpar \sdtit 
       \bigg[ 2 \Bigl| \FLDt \Bigr|^2 
\nll &&
 + 2 \FLDt \FQDtc  + \Bigl| \FQDt \Bigr|^2 \bigg] \biggr\}.
\eqa
\clearpage
\section{Numerical results and discussion}
All the formulae derived in this article are realized in a {\tt FORTRAN} code 
with a tentative name {\tt eeffLib}. All the numbers are produced with December 2000
version of the code~\cite{eeffLib:2000}.
In this section we present several examples of numerical results.

We will show several examples of comparison with {\tt ZFITTER v6.30}~\cite{zfitterv6.30:2000}. 
In the present realization,
{\tt eeffLib} does not calculate $\mwl$ from $\mu$ decay and does not precompute either
Sirlin's parameter $\Delta r$ or total $\zb$ width, which enters the $\zb$ boson propagator.
For this reason, the three parameters: $\mwl\,,\;\Delta r\,,\;\gz$ were being taken 
from
{\tt ZFITTER} and used as {\tt INPUT} for {\tt eeffLib}. Moreover, present {\tt eeffLib}
is a purely one-loop code, while in {\tt ZFITTER} it was not foreseen to access one-loop
form factors with users flags. To accomplish the goals of comparison at the one-loop level, 
we had to modify a little the {\tt DIZET} electroweak library. The most important change 
was an addition to the {\tt SUBROUTINE ROKANC}:
\vspace*{-5mm}

\begin{verbatim}
*
* For eett
*  
      FLL=(XROK(1)-1D0+DR )*R1/AL4PI
      FQL=FLL+(XROK(2)-1D0)*R1/AL4PI
      FLQ=FLL+(XROK(3)-1D0)*R1/AL4PI
      FQQ=FLL+(XROK(4)-1D0)*R1/AL4PI 
\end{verbatim}
with the aid of which we reconstruct four form factors from {\tt ZFITTER}'s effective
couplings $\rho$ and $\kappa$'s ($F_{\sss{LD}}$ and $F_{\sss{QD}}$ do not contribute in massless
approximation).
\subsection{Flags of {\tt eeffLib}\label{t-flaggs}}
Here we give a very brief description of flags (user options) of {\tt eeffLib}. While creating 
the code, we followed the  principle to preserve as much as possible the meaning of flags as 
described in the {\tt ZFITTER} description~\cite{Bardin:1999yd}. In the list below, 
a comment `{\tt as in ZFD}' means that the flag has exactly the same meaning as in
~\cite{Bardin:1999yd}.
\begin{itemize}
\item \verb+ ALEM=3  ! as in ZFD +
\vspace*{-2mm}

\item \verb+ ALE2=3  ! as in ZFD +
\vspace*{-2mm}

\item \verb+ VPOL=0  ! =0 \alpha(0); =1,=2 as in ZFD; =3 is reserved for later use+ \\
Note that the flag is extended to {\tt VPOL=0} to allow calculations `without running of 
$\alpha$'.
\vspace*{-2mm}

\item \verb+ QCDC=0  ! as in ZFD +
\vspace*{-2mm}

\item \verb+ ITOP=1  ! as in DIZET (internal flag) +
\vspace*{-2mm}

\item \verb+ GAMS=1  ! as in ZFD +
\vspace*{-2mm}

\item \verb+ WEAK=1  ! as in ZFD (use WEAK=2 in v6.30 to throw away some HO-terms) +
\vspace*{-2mm}

\item \verb+ IMOMS=1 ! =0 \alpha-scheme; =1 GFermi-scheme + \\
New meaning of an old flag: switches between two renormalization schemes;
\vspace*{-2mm}

\item \verb+ BOXD=0  ! =0 + without any boxes  \\
      \verb+         ! =1 + with  $\ph\ph$ box \\
      \verb+         ! =2 + with  $\zb\ph$ box \\
      \verb+         ! =3 + with  $\ph\ph$ and $\zb\ph$ boxes; 1,~2,~3 are used together with 
{\tt WEAK=0} \\
      \verb+         ! =4 + with  $\wb\wb$ box \\
      \verb+         ! =5 + with  $\wb\wb$ and $\zb\zb$ boxes; 4,~5 are used together with 
{\tt WEAK=1}
\vspace*{-2mm}

\item \verb+ GAMZTR=1! =0 GAMZ=0; =1 GAMZ.NE.0 + \\
Treatment of $\Gamma_{\sss{Z}}$. The option is implemented for the sake of comparison with 
{\tt FeynArts}.
\vspace*{-7mm}

\item \verb+ EWFFTR=0! =0 EWFFs ; =1 RHO-KAPPAS+ \\
Treatment of EW form factors; switches between form factors and effective
{\tt ZFITTER} couplings $\rho$ and $\kappa$'s. The option is implemented for
comparison with {\tt ZFITTER}.
\vspace*{-2mm}

\item \verb+ FERMTR=1! =1 a `standard' set of fermions masses; =2,3 `modified' + \\ 
Treatment of fermionic masses; switches between three different
sets of `effective quark masses'.
\vspace*{-20mm}

\end{itemize}

\begin{table}[!b]
\vspace*{-1cm}
\caption[EWFF for the process $\fep\fem\to\fu\bar{u}$. {\tt eeffLib}--{\tt ZFITTER} 
comparison without and with $\wb\wb$ boxes.]
{EWFF for the process $\fep\fem\to\fu\bar{u}$. {\tt eeffLib}--{\tt ZFITTER} comparison.
\label{t_com1}}
\vspace*{3mm}
\begin{tabular} {||c|c||l|l|l||}
\hline
\multicolumn{5}{||c||}{Without EW boxes}                                              \\
\hline
\hline
\multicolumn{2}{||c||}{Quantity}&\multicolumn{3}{|c||}{$E_{\rm{cm}}$}                 \\
\hline
\multicolumn{1}{||c}{$~~~~~~~~~\sqrt s$}\hspace{-5mm}&
\multicolumn{1}{c||}{}&~~~~~~~~100 GeV    &~~~~~~~~200 GeV    &~~~~~~~~300 GeV        \\
\hline
\hline
           &$\wml/10$&$13.47777~-~i1.84781$&$16.22034~-i10.49408$&$23.75241~-i11.27464$\\
$F_{\sss{LL}}$&$\wml$&$13.47777~-~i1.84781$&$16.22034~-i10.49408$&$23.75241~-i11.27464$\\
           &$10\wml $&$13.47777~-~i1.84781$&$16.22034~-i10.49408$&$23.75241~-i11.27464$\\
\hline
\multicolumn{2}{||c||}{\tt ZFITTER}
                     &$13.47771~-~i1.84786$&$16.22031~-i10.49405$&$23.75237~-i11.27464$\\
\hline
\hline
           &$\wml/10$&$29.34725~+~i3.67334$&$30.33892~+~i3.34535$&$31.64554~+~i2.75260$\\
$F_{\sss{QL}}$&$\wml$&$29.34725~+~i3.67334$&$30.33891~+~i3.34535$&$31.64554~+~i2.75260$\\
           &$10\wml $&$29.34725~+~i3.67334$&$30.33891~+~i3.34535$&$31.64554~+~i2.75260$\\
\hline
\multicolumn{2}{||c||}{\tt ZFITTER}
                     &$29.34720~+~i3.67330$&$30.33889~+~i3.34535$&$31.64552~+~i2.75259$\\
\hline
\hline
           &$\wml/10$&$29.13302~+~i3.26972$&$30.03854~+~i1.54158$&$31.68636~-~i0.22635$\\
$F_{\sss{LQ}}$&$\wml$&$29.13302~+~i3.26972$&$30.03854~+~i1.54158$&$31.68636~-~i0.22635$\\
           &$10\wml $&$29.13302~+~i3.26972$&$30.03854~+~i1.54158$&$31.68636~-~i0.22635$\\
\hline
\multicolumn{2}{||c||}{\tt ZFITTER}
                     &$29.13304~+~i3.26973$&$30.03855~+~i1.54163$&$31.68635~-~i0.22634$\\
\hline
\hline
           &$\wml/10$&$44.90390~+~i8.85688$&$43.80287~+i10.02412$&$44.21224~+i10.83899$\\
$F_{\sss{QQ}}$&$\wml$&$44.90389~+~i8.85688$&$43.80285~+i10.02412$&$44.21222~+i10.83899$\\
           &$10\wml $&$44.90390~+~i8.85688$&$43.80286~+i10.02412$&$44.21223~+i10.83899$\\
\hline
\multicolumn{2}{||c||}{\tt ZFITTER}
                     &$44.90392~+i8.85688$&$43.80285~+i10.02411$&$44.21224~+i10.83894$\\
\hline
\hline
\multicolumn{5}{||c||}{$\wb\wb$ is added}                                             \\
\hline
\hline
           &$\wml/10$&$12.94471~-~i1.84781$&$~9.34003~-~i9.42493$&$~9.03774~-i11.56004$\\
$F_{\sss{LL}}$&$\wml$&$12.94471~-~i1.84781$&$~9.34003~-~i9.42493$&$~9.03774~-i11.56004$\\
           &$10\wml $&$12.94471~-~i1.84781$&$~9.34003~-~i9.42493$&$~9.03774~-i11.56004$\\
\hline
\multicolumn{2}{||c||}{\tt ZFITTER}
                     &$12.94468~-~i1.84786$&$~9.34065~-~i9.42467$&$~9.03903~-i11.55958$\\
\hline
\hline
\end{tabular}
\end{table}

\clearpage

\begin{table}[ht]
\caption[EWFF for the process $\fep\fem\to\fu\bar{u}$. {\tt eeffLib}--{\tt ZFITTER} comparison
with $\zb\zb$ boxes.]
{EWFF for the process $\fep\fem\to\fu\bar{u}$. {\tt eeffLib}--{\tt ZFITTER} comparison.
\label{t_com2}}
\vspace*{3mm}
\begin{tabular} {||c|c||l|l|l||}
\hline
\multicolumn{5}{||c||}{With $\zb\zb$ boxes}                                              \\
\hline
\hline
\multicolumn{2}{||c||}{Quantity}&\multicolumn{3}{|c||}{$E_{\rm{cm}}$}                    \\
\hline
\multicolumn{1}{||c}{$~~~~~~~~~\sqrt s$}\hspace{-5mm}&\multicolumn{1}{c||}{}
&~~~~~~~~100 GeV    &~~~~~~~~200 GeV   &~~~~~~~~300 GeV \\
\hline
\hline
           &$\wml/10$&$12.89587~-~i1.84781$&$~8.24674~-i10.64677$&$~8.98241~-i12.88512$\\
$F_{\sss{LL}}$&$\wml$&$12.89586~-~i1.84781$&$~8.24673~-i10.64677$&$~8.98241~-i12.88512$\\
           &$10\wml $&$12.89587~-~i1.84781$&$~8.24673~-i10.64677$&$~8.98241~-i12.88512$\\
\hline
\multicolumn{2}{||c||}{\tt ZFITTER}
                     &$12.89583~-~i1.84786$&$~8.24736~-i10.64651$&$~8.98370~-i12.88466$\\
\hline
\hline
           &$\wml/10$&$29.30451~+~i3.67334$&$29.38219~+~i2.27613$&$31.59712~+~i1.59304$\\
$F_{\sss{QL}}$&$\wml$&$29.30451~+~i3.67334$&$29.38218~+~i2.27613$&$31.59712~+~i1.59304$\\
           &$10\wml $&$29.30451~+~i3.67334$&$29.38219~+~i2.27613$&$31.59712~+~i1.59304$\\
\hline
\multicolumn{2}{||c||}{\tt ZFITTER}
                     &$29.30445~+~i3.67330$&$29.38216~+~i2.27613$&$31.59710~+~i1.59304$\\
\hline
\hline
           &$\wml/10$&$29.10829~+~i3.26972$&$29.48510~+~i0.92306$&$31.65836~-~i0.89713$\\
$F_{\sss{LQ}}$&$\wml$&$29.10829~+~i3.26972$&$29.48509~+~i0.92306$&$31.65835~-~i0.89713$\\
           &$10\wml $&$29.10829~+~i3.26972$&$29.48509~+~i0.92306$&$31.65835~-~i0.89713$\\
\hline
\multicolumn{2}{||c||}{\tt ZFITTER}
                     &$29.10832~+~i3.26973$&$29.48512~+~i0.92312$&$31.65835~-~i0.89711$\\
\hline
\hline
           &$\wml/10$&$44.88226~+~i8.85688$&$43.31856~+~i9.48287$&$44.18773~+i10.25200$\\
$F_{\sss{QQ}}$&$\wml$&$44.88226~+~i8.85688$&$43.31854~+~i9.48287$&$44.18771~+i10.25200$\\
           &$10\wml $&$44.88226~+~i8.85688$&$43.31855~+~i9.48287$&$44.18772~+i10.25200$\\
\hline
\multicolumn{2}{||c||}{\tt ZFITTER}
                     &$44.88228~+~i8.85688$&$43.31854~+~i9.48286$&$44.18773~+i10.25196$\\
\hline
\hline
\end{tabular}
\end{table}

\subsection{{\tt eeffLib}--{\tt ZFITTER} comparison of scalar form factors}
First of all we discuss the results of a computation of the four scalar form factors, 
\bqa
\vvertil{}{\sss{LL}}{s,t},\quad
\vvertil{}{\sss{QL}}{s,t},\quad
\vvertil{}{\sss{LQ}}{s,t},\quad
\vvertil{}{\sss{QQ}}{s,t},
\eqa
for three variants:\\
1) without EW boxes, i.e. without gauge-invariant contribution of $\zb\zb$ boxes, 
and without 
$\gpar=1$ part of the $\wb\wb$ box, \eqn{WW_box_1};\\
2) without $\zb\zb$ boxes;\\
3) with full content of EWRC.

In this comparison we use flags as in subsection \ref{t-flaggs} and, moreover,
\bqa
\mwl    &=&80.4514958\;\mbox{GeV},
\nl
\Delta r&=&0.0284190602\,,
\nl
\Gamma_{\sss Z}&=&2.499 776\;\mbox{GeV}.
\label{firstPOs}
\eqa

In \tbn{t_com1} we show an example of comparison of four form factors 
$\vvertil{}{\sss{LL,QL,LQ,QQ}}{\sman,\tman}$ between the {\tt eeffLib}, where we set 
$\mtl=0.2$
 GeV and {\tt ZFITTER} (the latter is able to deliver only massless results).
The form factors are shown as complex numbers for the three c.m.s. energies 
(for $\tman = \mts-\sman/2$)
and for the three values of scale $\tHs=\wml/10,\;\wml,\;10\wml$. The table demonstrates 
scale independence and very good agreement with {\tt ZFITTER} results (6 or 7 digits).
One should stress that total agreement with {\tt ZFITTER} is not expected because 
in the {\tt eeffLib} code we use massive expressions to compute the nearly massless case.
Certain numerical cancellations leading to losing some numerical precision are expected.  
We should conclude that the agreement is very good and uniquely demonstrates that our 
formulae have the correct $\mtl\to 0$ limit.

In \tbn{t_com2} we show a similar comparison with {\tt ZFITTER} when $\zb\zb$ boxes are added.
As seen, the agreement has not deteriorated.
\subsection{{\tt eeffLib}--{\tt ZFITTER} comparison of IBA cross-section}
As the next step of the comparison of {\tt eeffLib} with calculations from the 
literature, we present a comparison of the IBA cross-section.

In \tbn{IBA_table1} we show the differential cross-section \eqn{differential-cs} in pb
for three values of $\cos\vartheta=-0.9,\,0,\,+0.9$, with IPS of \eqn{firstPOs} and
without running e.m. coupling, i.e. $\alpha\lpar\sman\rpar\to\alpha$.
\begin{table}[h]
\caption[{\tt eeffLib}--{\tt ZFITTER} comparison of the differential cross-section
without running $\alpha$.]
{IBA, First  row -- {\tt ZFITTER} ($u\bar{u}$ channel); second 
row -- {\tt eeffLib} ($\mtl=0.1$ GeV); 
 third  row -- {\tt eeffLib} ($\mtl=173.8$ GeV). }
\label{IBA_table1}
\vspace*{5mm}
\centering
\begin{tabular} {||c|c|l|l|l|l|l|l||}
\hline
\hline
${\sqrt s}$         & 100$GeV$ & 200$GeV$ & 300$GeV$ & 400$GeV$ & 700$GeV$ & 1000$GeV$\\
\hline
                    & 47.664652& 0.291823 & 0.169510 &          &          &          \\ 
\cline{2-7}
$\cos\vartheta=-0.9$& 47.661401& 0.291827 & 0.169515 & 0.103284 & 0.035318 & 0.017203 \\
\cline{2-7}
                    &          &          &          & 0.162579 & 0.043974 & 0.018850 \\
\hline
\hline
                    & 59.768387& 1.718830 & 0.695061 &          &          &          \\
\cline{2-7}
$\cos\vartheta= 0 $ & 59.770715& 1.718870 & 0.695072 & 0.376868 & 0.117276 & 0.055870 \\
\cline{2-7}
                    &          &          &          & 0.264874 & 0.112923 & 0.054211 \\
\hline
\hline
                    &168.981978& 5.954048 & 2.292260 &          &          &          \\
\cline{2-7}
$\cos\vartheta= 0.9$&168.991272& 5.954166 & 2.292289 & 1.222343 & 0.372903 & 0.176030 \\
\cline{2-7}
                    &          &          &          & 0.438952 & 0.293415 & 0.154784 \\
\hline
\hline
\end{tabular}
\end{table}

Next, we present the same comparison as in \tbn{IBA_table1}, but now with running e.m. 
coupling.
Since the flags setting {\tt VPOL=1}, which is  relevant to this case, affects {\tt ZFITTER}
numbers, we now use, instead of \eqn{firstPOs}, the new {\tt INPUT} set:
\bqa
\mwl    &=&80.4467671\;\mbox{GeV},
\nl
\Delta r&=&0.0284495385\,,
\nl
\Gamma_{\sss Z}&=&2.499 538\;\mbox{GeV}.
\label{secondPOs}
\eqa
The numbers, collected in \tbn{IBA_table2}, exhibit good level of agreement.
\begin{table}[h]
\caption[{\tt eeffLib}--{\tt ZFITTER} comparison of the differential cross-section
with running $\alpha$.]
{IBA, First  row -- {\tt ZFITTER} ($u\bar{u}$ channel); second row -- {\tt eeffLib} 
($\mtl=0.1$ GeV); third  row -- {\tt eeffLib} ($\mtl=173.8$ GeV). }
\label{IBA_table2}
\vspace*{5mm}
\centering
\begin{tabular} {||c|c|l|l|l|l|l|l||}
\hline
\hline
${\sqrt s}$       & 100$GeV$ & 200$GeV$ & 300$GeV$ & 400$GeV$ & 700$GeV$ & 1000$GeV$\\
\hline
                  & 45.404742& 0.386966 & 0.225923 &          &          &          \\
$\cos\vartheta=-0.9$ & 45.404593& 0.386966 & 0.225923 & 0.138065 & 0.048621 & 0.024155 \\
                  &          &          &          & 0.195069 & 0.057892 & 0.025877 \\
\hline
\hline
                  & 60.382423& 1.882835 & 0.771939 &          &          &          \\
$\cos\vartheta= 0 $  & 60.382553& 1.882835 & 0.771938 & 0.421409 & 0.133474 & 0.064244 \\
                  &          &          &          & 0.303984 & 0.130208 & 0.062853 \\
\hline
\hline
                  &173.467517& 6.450000 & 2.510881 &          &          &          \\
$\cos\vartheta= 0.9$ &173.467515& 6.449995 & 2.510877 & 1.346616 & 0.417292 & 0.198839 \\
                  &          &          &          & 0.493006 & 0.330480 & 0.175598 \\
\hline
\hline
\end{tabular}
\end{table}

\vspace*{1mm}

Finally, in \tbn{IBA_table3}, we give a comparison of the cross-section integrated within  
the angular interval $|\cos\vartheta| \leq 0.999$.
(Flags setting is the same as for Table 4.)

\begin{table}[h]
\caption[{\tt eeffLib}--{\tt ZFITTER} comparison of the total cross-section.]
{{\tt eeffLib}--{\tt ZFITTER} comparison of the total cross-section.
Cross-sections are given in picobarns:
the first row  -- $\sigma^{tf}_{\rm{tot}}$, i.e. {\tt eeffLib ($m_t=0.1$ GeV)}; 
the second row -- $\sigma^{ZF}_{\rm{tot}}$, i.e. {\tt ZFITTER} ($u\bar{u}$ channel); 
the last entry shows the deviation $\displaystyle{\lpar\sigma^{tf}_{\rm{tot}}
                                 -\sigma^{ZF}_{\rm{tot}}\rpar/\sigma^{ZF}_{\rm{tot}}}$ 
in per mill.
          }
\label{IBA_table3}
\vspace*{5mm}
\centering
\begin{tabular}{||c|c||c|c||c|c||}
\hline
\hline
\multicolumn{2}{||c||}{$100$ GeV}   &\multicolumn{2}{|c||}{200 GeV}  
                                 &\multicolumn{2}{|c||}{300 GeV}       \\
\hline
$\sigma_{\rm{tot}}$&$\sigma_{\sss{\rm{FB}}}$&$\sigma_{\rm{tot}}$&$\sigma_{\sss{\rm{FB}}}$
                                  &$\sigma_{\rm{tot}}$&$\sigma_{\sss{\rm{FB}}}$\\
\hline
\hline
    160.8981   &   70.98419   &    5.021797   &    3.360836   &    2.031750   &    1.269552  \\
    160.8980   &   70.98406   &    5.021808   &    3.360848   &    2.031754   &    1.269556  \\
\hline 
     +0.001    &   +0.002     &   -0.002      &   -0.004      &   -0.002      &   -0.003     \\
\hline
\hline
\end{tabular}
\vspace*{-2mm}
\end{table}
A typical deviation between {\tt eeffLib} and {\tt ZFITTER} is of the order $\sim 10^{-6}$, i.e. 
of the order of the required precision of the numerical integration over $\cos\vartheta$.
Examples of numbers obtained with {\tt eeffLib}, which were shown in this section, 
demonstrate that {\tt ZFITTER} numbers are recovered for light $\mtl$.

We conclude this subsection with a comment about technical precision of our calculations
(modulo bugs, of course). We do not use {\tt looptools} package~\cite{Hahn:2000jm}.
For all PV functions, but one, namely $D_0$ function, we use our own coding where we can
control precision internally and, typically, we can guarantee 9-10 digits precision.
For $D_0$ function we use, instead, {\tt REAL*8 TOPAZ0} coding~\cite{Montagna:1999kp} and 
the only accessible for us way to control the precision is to compare results with 
those computed with {\tt REAL*16 TOPAZ0} coding. 
This was done for a typical $D_0$ function entering $WW$ box contribution.
We got an agreement within 9 digits between these two versions for all 
$\sqrt{s}=400,\,700,\,1000$ GeV and $\cos\theta=0.9,\,0,\,-0.9$.

\subsection{About a comparison with the other codes}
As is well known, the one-loop differential cross-section of $\fep\fem\to\ft\bar{t}$
may be generated with the aid of the FeynArts system~\cite{Hahn:2000jm}.
FeynArts-generated versions of the code with and without QED contributions are 
available~\cite{Code:1998}, and an attempt to compare results was undertaken.
We compared $d\sigma/d\cos\theta$ without QED contributions
at $\sqrt{\sman}=400,700,1000$ GeV and three values of $\cos\theta=0.9,\;,0,\;-0.9$.
An agreement of numbers for the Born cross-section within 6 digits was found,
while for the one-loop corrected cross-section we managed to reach an agreement
within $1-3\%$ only.

\begin{figure}[!h]
\centering
\setlength{\unitlength}{0.1mm}
\begin{picture}(1800,1800)
\put(-220,0){\makebox(0,0)[lb]{\epsfig{file=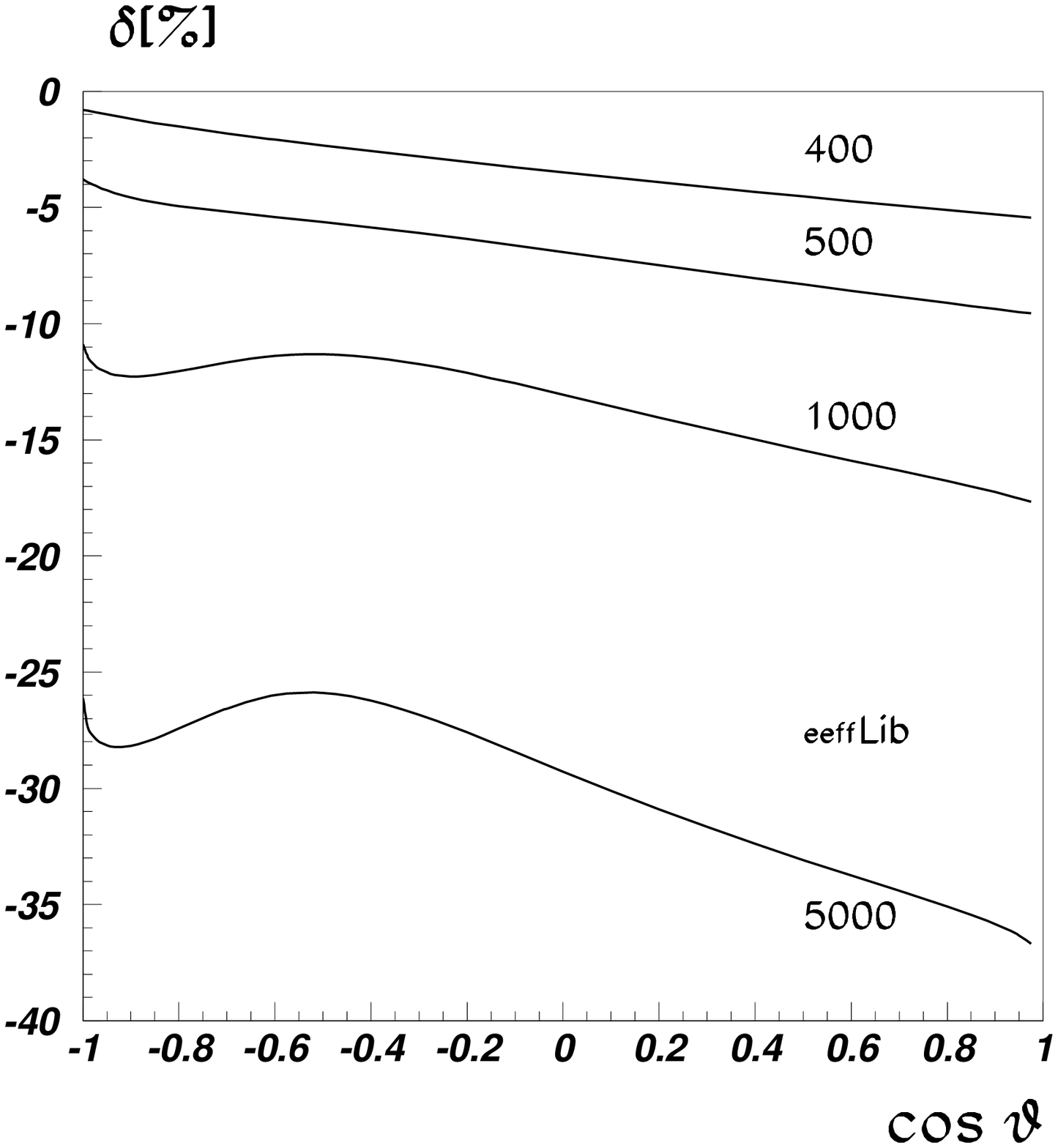,width=175mm,height=180mm}}}
\end{picture}
\caption[Relative EWRC to the $e^+e^-\to\ft\bar{t}$ differential cross-section.] 
{Relative EWRC $\delta$ [\eqn{delta-double}] to the $e^+e^-\to\ft\bar{t}$ differential 
cross-section.
Numbers near the curves show $\sqrt{\sman}$ in GeV.\label{fig:delta-double}}
\end{figure}

There is another {\tt FORTRAN} code for $\ft\bar{t}$ production.
It was originally written for the MSSM \cite{Hollik:1998md}, but it has also been tailored 
for the SM. 
So far we managed to completely agree with this code only for the Born cross-section;
while we turned to the one-loop one, we realized that no separation between QED 
and EW corrections is implemented into this code. On the contrary, in the {\tt eeffLib} version
used to produce numbers for this paper, we coded only the EW part of the cross-section.
A present, the QED part is also available in our code~\cite{part2:2001}.
Moreover, this code produces cross-section integrated over the angle, 
while it would be more informative to compare the differential quantity:
\bq
\delta(\sqrt{s},\cos\vartheta)=
\ds\frac{d\sigma^{(1)}(\sman,\tman)/d\tman}{d\sigma^{(0)}(\sman)/d\tman}-1.
\label{delta-double}
\eq
For the time being we limit ourselves by presenting {\tt eeffLib} results for 
$\delta(\sqrt{s},\cos\vartheta)$ of \linebreak
\eqn{delta-double},  
which are shown in \fig{fig:delta-double}. (Flags setting is the same as for Table 3.)

\begin{figure}[!h]
\centering
\setlength{\unitlength}{0.1mm}
\begin{picture}(1800,1800)
\put(-220,0){\makebox(0,0)[lb]{\epsfig{file=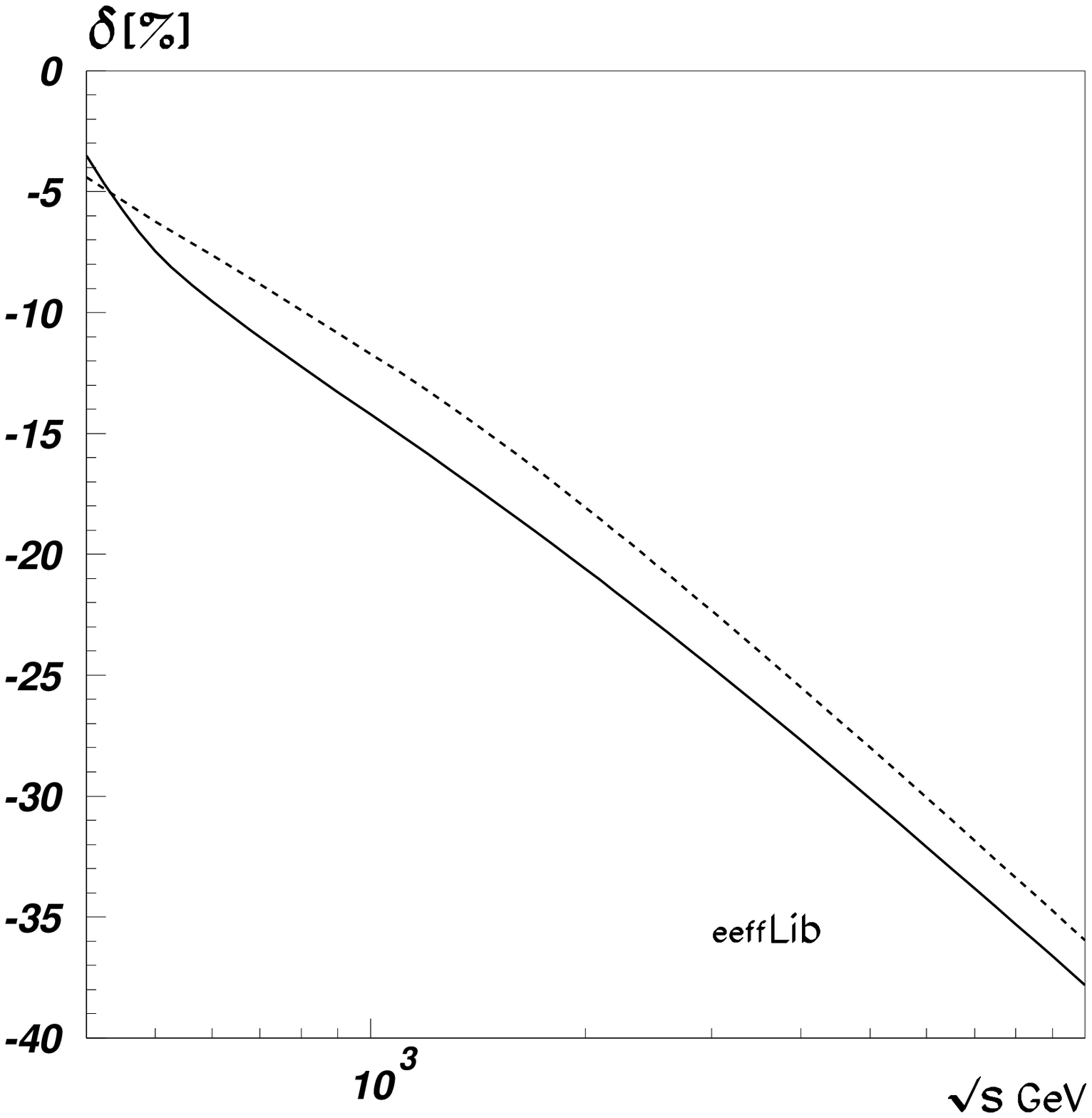,width=175mm,height=180mm}}}
\end{picture}
\caption[
 Relative EWRC to $e^+e^-\to\ft\bar{t}$ total cross-section.]
{Relative EWRC to $e^+e^-\to\ft\bar{t}$ for $\mhl=100$ GeV (solid line) and $\mhl=1000$ GeV
(dashed line).\label{fig:delta-integrated}}
\end{figure}

Furthermore, in Fig. 3 of paper \cite{Driesen:1996tn}, an interesting result is presented. 
We tried to reproduce it with the aid of {\tt eeffLib}.
The results are shown in \fig{fig:delta-integrated}.
As might be seen from a comparison of two figures, there is nearly ideal agreement
for $\sqrt{\sman}$ in the interval $[500$--$3000]$ GeV, while above 3000 GeV the {\tt eeffLib} 
curve goes a bit higher than the curve shown in~\cite{Driesen:1996tn}. 
Note, that both curves show a very similar $\mhl$ dependence.
It is difficult to expect more from such a pilot comparison, because even input parameters 
and various options were not tuned.  


Meantime, a Bielefeld--Zeuthen team~\cite{Zeuthen:2001}
started alternative calculations using the DIANA system~\cite{DIANA}.
A comparison of results was undertaken. It showed good agreement of numbers.

Recently, we were provided with the numbers computed with the FeynArts 
system~\cite{FynArts:2000} for $d\sigma/d\cos\theta$ without QED contributions;
they showed better agreement than we managed to reach ourselves.

The results of latest comparisons will be presented in more detail elsewhere.

\addcontentsline{toc}{section}{Acknowledgments}
\section*{Acknowledgements}
 We would like to thank our colleagues P.~Christova and A.~Andonov for fruitful discussions.
 We are indebted to W.~Hollik and C.~Schappacher for a discussion of issues of the 
comparison with FeynArts.
 We acknowledge a common work on numerical comparison with J.~Fleischer, A.~Leike, T.~Riemann, 
and A.~Werthenbach which helped us to debug our code.
 We are grateful to T.~Riemann for a critical reading of selected sections of this text and useful 
comments.
 We also wish to thank G.~Altarelli for extending to us the hospitality of the CERN TH Division 
at various stages of this work.

\end{document}